\documentclass[aps,prc,superscriptaddress,showpacs,nofootinbib,floatfix,twocolumn]{revtex4}
\usepackage{graphics}
\usepackage{epsfig}

\newcommand{\pp} {\mbox{$p+p$}}

\newcommand{\dA} {\mbox{$d+Au$}}
\newcommand{\pt} {\mbox{$p_T$}}

\newcommand{\rabo}[1]{\raisebox{1.5ex}[1.5ex]{#1}}

\def\mean#1{\left<{#1}\right>}

\begin{document}

\title{Jet Structure from Dihadron Correlations in d+Au collisions at $\sqrt{s_{NN}}$=200 GeV}

\newcommand{\abilene}{Abilene Christian University, Abilene, TX 79699, USA}
\newcommand{\acadsin}{Institute of Physics, Academia Sinica, Taipei 11529, Taiwan}
\newcommand{\banaras}{Department of Physics, Banaras Hindu University, Varanasi 221005, India}
\newcommand{\barc}{Bhabha Atomic Research Centre, Bombay 400 085, India}
\newcommand{\bnl}{Brookhaven National Laboratory, Upton, NY 11973-5000, USA}
\newcommand{\caucr}{University of California - Riverside, Riverside, CA 92521, USA}
\newcommand{\ciae}{China Institute of Atomic Energy (CIAE), Beijing, People's Republic of China}
\newcommand{\cns}{Center for Nuclear Study, Graduate School of Science, University of Tokyo, 7-3-1 Hongo, Bunkyo, Tokyo 113-0033, Japan}
\newcommand{\colorado}{University of Colorado, Boulder, CO 80309, USA}
\newcommand{\columbia}{Columbia University, New York, NY 10027 and Nevis Laboratories, Irvington, NY 10533, USA}
\newcommand{\dapnia}{Dapnia, CEA Saclay, F-91191, Gif-sur-Yvette, France}
\newcommand{\debrecen}{Debrecen University, H-4010 Debrecen, Egyetem t{\'e}r 1, Hungary}
\newcommand{\elte}{ELTE, E{\"o}tv{\"o}s Lor{\'a}nd University, H - 1117 Budapest, P{\'a}zm{\'a}ny P. s. 1/A, Hungary}
\newcommand{\fsu}{Florida State University, Tallahassee, FL 32306, USA}
\newcommand{\gsu}{Georgia State University, Atlanta, GA 30303, USA}
\newcommand{\hiroshima}{Hiroshima University, Kagamiyama, Higashi-Hiroshima 739-8526, Japan}
\newcommand{\ihepprot}{IHEP Protvino, State Research Center of Russian Federation, Institute for High Energy Physics, Protvino, 142281, Russia}
\newcommand{\illuiuc}{University of Illinois at Urbana-Champaign, Urbana, IL 61801, USA}
\newcommand{\isu}{Iowa State University, Ames, IA 50011, USA}
\newcommand{\jinrdubna}{Joint Institute for Nuclear Research, 141980 Dubna, Moscow Region, Russia}
\newcommand{\kek}{KEK, High Energy Accelerator Research Organization, Tsukuba, Ibaraki 305-0801, Japan}
\newcommand{\kfki}{KFKI Research Institute for Particle and Nuclear Physics of the Hungarian Academy of Sciences (MTA KFKI RMKI), H-1525 Budapest 114, POBox 49, Budapest, Hungary}
\newcommand{\korea}{Korea University, Seoul, 136-701, Korea}
\newcommand{\kurchatov}{Russian Research Center ``Kurchatov Institute", Moscow, Russia}
\newcommand{\kyoto}{Kyoto University, Kyoto 606-8502, Japan}
\newcommand{\labllr}{Laboratoire Leprince-Ringuet, Ecole Polytechnique, CNRS-IN2P3, Route de Saclay, F-91128, Palaiseau, France}
\newcommand{\lawllnl}{Lawrence Livermore National Laboratory, Livermore, CA 94550, USA}
\newcommand{\losalamos}{Los Alamos National Laboratory, Los Alamos, NM 87545, USA}
\newcommand{\lpc}{LPC, Universit{\'e} Blaise Pascal, CNRS-IN2P3, Clermont-Fd, 63177 Aubiere Cedex, France}
\newcommand{\lund}{Department of Physics, Lund University, Box 118, SE-221 00 Lund, Sweden}
\newcommand{\muenster}{Institut f\"ur Kernphysik, University of Muenster, D-48149 Muenster, Germany}
\newcommand{\myongji}{Myongji University, Yongin, Kyonggido 449-728, Korea}
\newcommand{\nagasaki}{Nagasaki Institute of Applied Science, Nagasaki-shi, Nagasaki 851-0193, Japan}
\newcommand{\newmex}{University of New Mexico, Albuquerque, NM 87131, USA }
\newcommand{\nmsu}{New Mexico State University, Las Cruces, NM 88003, USA}
\newcommand{\ornl}{Oak Ridge National Laboratory, Oak Ridge, TN 37831, USA}
\newcommand{\orsay}{IPN-Orsay, Universite Paris Sud, CNRS-IN2P3, BP1, F-91406, Orsay, France}
\newcommand{\peking}{Peking University, Beijing, People's Republic of China}
\newcommand{\pnpi}{PNPI, Petersburg Nuclear Physics Institute, Gatchina, Leningrad region, 188300, Russia}
\newcommand{\riken}{RIKEN (The Institute of Physical and Chemical Research), Wako, Saitama 351-0198, JAPAN}
\newcommand{\rikjrbrc}{RIKEN BNL Research Center, Brookhaven National Laboratory, Upton, NY 11973-5000, USA}
\newcommand{\saopaulo}{Universidade de S{\~a}o Paulo, Instituto de F\'{\i}sica, Caixa Postal 66318, S{\~a}o Paulo CEP05315-970, Brazil}
\newcommand{\seoulnat}{System Electronics Laboratory, Seoul National University, Seoul, South Korea}
\newcommand{\stonybrkc}{Chemistry Department, Stony Brook University, Stony Brook, SUNY, NY 11794-3400, USA}
\newcommand{\stonycrkp}{Department of Physics and Astronomy, Stony Brook University, SUNY, Stony Brook, NY 11794, USA}
\newcommand{\subatech}{SUBATECH (Ecole des Mines de Nantes, CNRS-IN2P3, Universit{\'e} de Nantes) BP 20722 - 44307, Nantes, France}
\newcommand{\tenn}{University of Tennessee, Knoxville, TN 37996, USA}
\newcommand{\titech}{Department of Physics, Tokyo Institute of Technology, Oh-okayama, Meguro, Tokyo 152-8551, Japan}
\newcommand{\tsukuba}{Institute of Physics, University of Tsukuba, Tsukuba, Ibaraki 305, Japan}
\newcommand{\vandy}{Vanderbilt University, Nashville, TN 37235, USA}
\newcommand{\waseda}{Waseda University, Advanced Research Institute for Science and Engineering, 17 Kikui-cho, Shinjuku-ku, Tokyo 162-0044, Japan}
\newcommand{\weizmann}{Weizmann Institute, Rehovot 76100, Israel}
\newcommand{\yonsei}{Yonsei University, IPAP, Seoul 120-749, Korea}
\newcommand{\deceased}{\dagger}
\affiliation{\abilene}
\affiliation{\acadsin}
\affiliation{\banaras}
\affiliation{\barc}
\affiliation{\bnl}
\affiliation{\caucr}
\affiliation{\ciae}
\affiliation{\cns}
\affiliation{\colorado}
\affiliation{\columbia}
\affiliation{\dapnia}
\affiliation{\debrecen}
\affiliation{\elte}
\affiliation{\fsu}
\affiliation{\gsu}
\affiliation{\hiroshima}
\affiliation{\ihepprot}
\affiliation{\illuiuc}
\affiliation{\isu}
\affiliation{\jinrdubna}
\affiliation{\kek}
\affiliation{\kfki}
\affiliation{\korea}
\affiliation{\kurchatov}
\affiliation{\kyoto}
\affiliation{\labllr}
\affiliation{\lawllnl}
\affiliation{\losalamos}
\affiliation{\lpc}
\affiliation{\lund}
\affiliation{\muenster}
\affiliation{\myongji}
\affiliation{\nagasaki}
\affiliation{\newmex}
\affiliation{\nmsu}
\affiliation{\ornl}
\affiliation{\orsay}
\affiliation{\peking}
\affiliation{\pnpi}
\affiliation{\riken}
\affiliation{\rikjrbrc}
\affiliation{\saopaulo}
\affiliation{\seoulnat}
\affiliation{\stonybrkc}
\affiliation{\stonycrkp}
\affiliation{\subatech}
\affiliation{\tenn}
\affiliation{\titech}
\affiliation{\tsukuba}
\affiliation{\vandy}
\affiliation{\waseda}
\affiliation{\weizmann}
\affiliation{\yonsei}
\author{S.S.~Adler}	\affiliation{\bnl}
\author{S.~Afanasiev}	\affiliation{\jinrdubna}
\author{C.~Aidala}	\affiliation{\columbia}
\author{N.N.~Ajitanand}	\affiliation{\stonybrkc}
\author{Y.~Akiba}	\affiliation{\kek} \affiliation{\riken}
\author{A.~Al-Jamel}	\affiliation{\nmsu}
\author{J.~Alexander}	\affiliation{\stonybrkc}
\author{K.~Aoki}	\affiliation{\kyoto}
\author{L.~Aphecetche}	\affiliation{\subatech}
\author{R.~Armendariz}	\affiliation{\nmsu}
% ...
\collaboration{PHENIX Collaboration} \noaffiliation

\date{\today}

\begin{abstract}


Dihadron correlations at high transverse momentum in $d$ + Au~collisions at
$\sqrt{s_{\rm{NN}}}$ = 200 GeV at midrapidity are measured by the
PHENIX experiment at the Relativistic Heavy Ion Collider (RHIC).
From these correlations we extract several structural
characteristics of jets; the root-mean-squared  
transverse momentum of
fragmenting hadrons with respect to the jet $\sqrt{\langle j_T^2 \rangle}$,
the mean sine-squared of the azimuthal
angle between the jet axes 
$\langle \sin^2(\phi_{jj}) \rangle$, 
and the number of particles produced within
the dijet that are associated with a high-$p_T$ particle ($dN/dx_E$
distributions). We observe that the fragmentation characteristics
of jets in $d$ + Au~collisions are very similar to those in
$p+p$~collisions and that there is also little dependence on the
centrality of the $d$ + Au~collision. This is consistent with the
nuclear medium having little influence on the fragmentation
process. Furthermore, there is no statistically significant increase in
the value of $\langle \sin^2(\phi_{jj}) \rangle$ 
from $p+p$~to $d$ + Au~collisions. This constrains the
effect of multiple scattering that partons undergo in the cold
nuclear medium before and after a hard-collision.

\end{abstract}



\maketitle




\section{Introduction\label{sec:intro}}

Jet production in high energy collisions is a useful tool to study
the passage of scattered partons through a nuclear medium. A
dominant hard-scattering process is two partons scattering to
produce two high-$p_T$ partons which then fragment to produce a
dijet. In a nuclear environment the partons that participate in
the collision can undergo multiple scattering within the nucleus,
potentially changing the structure of the dijet. Such changes can
provide information on the interaction of colored partons with the
cold nuclear medium.

Some information on this interaction is already available at RHIC
energies via the Cronin enhancement of the $p_T$
spectra~\cite{Cronin:zm}. In $d$ + Au~collisions at RHIC
\cite{Adler:2003ii} \cite{Adams:2003im} \cite{Back:2003ns} the
cross-section for high-$p_T$ particle production in $d$ +
Au~collisions is enhanced compared to $p+p$ collisions, consistent
with multiple scattering in the cold-nuclear medium increasing the
transverse momentum of the partons. In this paper we report on a
complimentary observable to the Cronin effect: the broadening of
dijet distributions. Such broadening is directly related to the
additional transverse momentum delivered to the partons during
multiple scattering, and hence provides a complementary handle for
comparison between experiment and theory.

Interpreting both $d$ + Au~and Au + Au~collisions requires solid
knowledge of baseline $p+p$ collisions, especially those dijet
events at midrapidity that contain two, nearly back-to-back, jets
produced from a hard (large $Q^{2}$) parton-parton interaction.
Experimentally the jets are not exactly back-to-back and the
acoplanarity momentum vector, $\vec{k_{T}}$, was measured in $p+p$
collisions at ISR energies to have a magnitude, $k_{T}$, on the
order of 1 GeV/c\cite{DellaNegra:1977sk}. This was much larger
than the expectation that $k_{T}$ was due to intrinsic parton
transverse momentum governed by the hadron size, which
would lead to
$k_{T}\sim$ 300 MeV/c. It was realized early~\cite{Feynman:1978dt}
that additional gluon radiation either before or after the
hard scattering will increase the value of $k_T$ and the dijet
acoplanarity.

In collisions involving nuclei, multiple scattering within the
nucleus increases the parton transverse momentum.
E557~\cite{Stewart:1990wa},
E609~\cite{Corcoran:1990vq}, and E683~\cite{Naples:1994uz} have
all measured an increase in the dijet acoplanarity with atomic
mass of the target.  In the case of E683, they measured an
$A^{\frac{1}{3}}$ dependence of $\langle k_{T}^{2} \rangle$ for
both $\gamma+A$ and $\pi+A$ collisions. This dependence is
expected since the number of scatterings should be proportional to
the length traversed in the nucleus ($L \sim A^{\frac{1}{3}}$).
For large $A$ the extracted $\langle k_{T}^{2} \rangle$ values 
are about 50\% above that for collisions with the hydrogen target,
implying that the multiple-scattering effects are as important to the
broadening of the dijets as are the initial state effects at that
energy. In the case of $p+A$ reactions, the measured $\langle
k_{T}^{2} \rangle$ values increase more slowly than $A^{\frac{1}{3}}$
\cite{Corcoran:1990vq}. Since the $\langle k_{T}^{2} \rangle$ values show
a strong energy dependence~\cite{Naples:1994uz}, we need
to establish the initial and multiple-scattering contributions to
$\langle k_{T}^{2} \rangle$ for p+A reactions at RHIC energies. 
The $\langle k_{T}^{2} \rangle$ values are also known to be dependent on the
$Q^{2}$ of the parton-parton interaction, increasing with rising
$Q^{2}$ \cite{jet:ccor1,Adams:1994}.

No model is currently available that can reproduce all data on the
Cronin effect and dijet broadening although most include
multiple scattering as the underlying mechanism. A recent
review~\cite{Accardi:2002ik} considered two large classes of
models, 1) soft or Glauber scattering where the multiple
scattering is either at the hadronic or partonic level and  2)
semi-hard multiple scattering where the multiple scattering is at
the partonic level.

In both the soft and hard scattering models, the
increase $\Delta\langle k_{T}^{2} \rangle = \langle
k_{T}^{2}\rangle_{p+A} - \langle k_{T}^{2}\rangle_{p+p}$ is
proportional to the product of the scattering cross section and the
nuclear thickness function,
\begin{equation}
\Delta\langle k_{T}^{2} \rangle \propto \nu(b,\sqrt{s})- 1 =
\sigma_{\rm{MS}}(\sqrt{s})T_{A}(b)
\end{equation}
where $\nu(b,\sqrt{s})$ is the number of interactions, $b$ is the
impact parameter of the collision, $\sigma_{\rm{MS}}$ is the
multiple scattering cross-section, and $T_{A}(b)$ is the nuclear
thickness function. For the soft scattering models,
$\sigma_{\rm{MS}}$ is defined to be $\sigma_{\rm{NN}}(\sqrt{s})$,
the nucleon-nucleon scattering cross section, while for the
semi-hard models $\sigma_{\rm{MS}}$ is
$\sigma_{i|H}^{N}(\sqrt{s})$, the parton-nucleon semi-hard
cross-section. In the specific case of hard sphere nucleon
scattering~\cite{Wang:1998ww} $\nu(b,\sqrt{s}) =
\sigma_{\rm{NN}}(\sqrt{s}) \frac{3A}{2\pi
R^{2}}\sqrt{1-\frac{b^{2}}{R^{2}}}$, where $R$ is the nuclear radius,
which gives an
$A^{\frac{1}{3}}$ increase in $\Delta\langle k_{T}^{2} \rangle$.

Both types of
these models give the same trend in centrality and the same dependence
on the target's atomic mass. The difference between them
is in the strength of the increase with respect to
$T_A(b)$ and how this changes with beam energy.
We will compare the data in this paper to two specific implementations
of the hard-scattering models from Qiu and
Vitev \cite{Qiu:2004da}, and Barnafoldi {\it et al.}~\cite{Barnafoldi:2004kh}.

An alternative view of the Cronin effect was recently proposed by
Hwa and Yang~\cite{Hwa:2004zd}. These authors calculate the
recombination of hard partons with soft partons released during
the multiple collisions. Because this model reproduces the
measured Cronin effect at RHIC without imparting successive
transverse momentum kicks to the scattered partons, the authors
suggest that there may be little to no increase in $k_T$ from
$p+p$ to $d$ + Au~collisions.

We also use jet-fragmentation observables to probe
multiple scattering in cold nuclei, in particular
$\sqrt{\langle j_T^2 \rangle}$, the
RMS of the mean transverse momentum of hadrons with respect to the
fragmenting parton, and the fragmentation function of the parton,
$D(z, Q^2)$, where $z$ is the fraction of the parton's momentum
that a hadron carries. If the parton suffers semi-hard inelastic
collisions within the nuclear environment, the parton will lose
energy and its subsequent hadronization will produce fewer
high-$z$ fragments and more low-$z$ fragments. We cannot directly
measure fragmentation functions via dihadron correlations, but we
do measure the distribution of hadrons produced in association
with a high-$p_T$ trigger particle. We plot these distributions as
a function of $x_{E}$, where $x_{E}$ is defined as
\begin{equation}\label{eqn:xEdef}
x_{E} = \frac{\vec{p}_{T,\rm{trig}} \cdot
\vec{p}_{T,\rm{assoc}}}{|\vec{p}_{T,\rm{trig}}|^{2}}
\end{equation}
The motivation for the variable $x_{E}$ can be most easily seen in
the simple case where $\langle z \rangle = 1$ for the trigger
particle and the two hadrons are emitted back-to-back. In this
case, $p_{T,\rm{trig}}$ is the transverse momentum of the
scattered parton ($q_{T,\rm{parton}}$) and for the far-side $x_{E}
= z_{\rm{assoc}} = p_{T,\rm{assoc}}/q_{T,\rm{parton}}$. Relaxing
the assumption on $z_{trig}$, there is still a simple relation
between $x_{E}$ and $z$ for back-to-back jets at high-$p_T$, where
$x_E \simeq z_{\rm{assoc}}/z_{\rm{trig}}$. Hence the $dN/dx_E$
distribution for hadrons emitted back-to-back from the trigger
hadron can be related to the fragmentation function: for more
details see the end of Section \ref{sec:jets}.

There is considerable information on $x_{E}$ distributions from
$p+p$ collisions. The CCHK collaboration \cite{DellaNegra:1977sk}
demonstrated that the $x_{E}$ distribution scaled, {\it i.e.} the
distribution was approximately independent of $p_{T,\rm{trig}}$.
Scaling at higher $p_{T,\rm{trig}}$ was also established by Fisk
{\it et al.} \cite{Fisk:1978en}, and the CCOR
collaboration~\cite{Jacob:1979hk, Angelis:1978uv}, providing
support for the idea that fragmentation of high-$p_T$ partons is
independent of the momentum of the parton.

This scaling is however approximate and scaling violation was
understood by Feynman {\it et al.}~\cite{Feynman:1978dt} to be caused by
the radiation of semi-hard gluons. Scaling violation of the
fragmentation function $D(z, Q^2)$ is now well established
experimentally (\cite{Abbiendi:2004pr} and references therein).
For the $Q^2$ range considered in this paper 
($10 < Q^2 < 1000$~GeV/c$^2$),
the fragmentation functions used in NLO
calculations~\cite{Kniehl:2000fe} drop by 25\% for $z = 0.6$ over
the range $10 < Q^2 < 100$~GeV/c$^2$. At higher $Q^2$ the
fragmentation functions are less dependent on $Q^2$, e.g. the
fragmentation drops by less than 20\% at $z = 0.6$ over the much
larger range of $100 < Q^2 < 1500$~GeV/c$^2$.

In this paper we quantify the extent to which our measured $x_E$
distributions in $d$ + Au~collision scale and compare the
$x_E$ distributions to
those from $\pp$~collisions at RHIC. The goal is
to establish whether
inelastic scattering in the cold medium or the recombination
mechanism changes the effective fragmentation function. 
The $x_{E}$ distributions provide a stringent test the
recombination model from Hwa and Wang~\cite{Hwa:2004}. This model
reproduces the Cronin effect in $d$ + Au collisions through shower
thermal recombination mechanism and predicts an
increase in jet associated multiplicity~\cite{Hwa:2004,Hwa:2005ui},
{\it i.e.} an increase in the near-angle $dN/dx_E$, in $d$ + Au relative to
$\pp$ collisions.

The
measured $x_{E}$ distributions in $d$ + Au~also serve as a
critical baseline for Au + Au~collisions, where the strong
energy-loss in the dense, hot medium is expected to dramatically
change the shape of these distributions.

Our three goals for this paper are 1) to report the characteristics
of jet structures in $d$ + Au~collisions at RHIC energies, 2) to
establish the extent to which multiple scattering changes these
structures as a function of centrality and by comparison with data
from $p+p$ collisions, and 3) to establish the baseline for
jet-structure measurements in heavy ion reactions. Any difference
between jet properties in Au + Au~and $d$ + Au~collisions should
be attributable to the hot, dense nuclear matter created in the
heavy ion collisions. The main results in this paper are presented
in Section~\ref{sec:results} which details the measured values of
$\langle j_T^2 \rangle$,
$\langle \sin^2(\phi_{jj}) \rangle$, and the \pt, and $x_{E}$
distributions from $d$ + Au~collisions at $\sqrt{s_{\rm{NN}}}$=200
GeV. These results are derived from the fitted widths and yields
of two-particle azimuthal correlations which are reported
in Section~\ref{sec:results}. The experimental methods
used to obtain these correlations are described in Section
\ref{sec:expt} and the jet quantities we use throughout the paper
are fully defined in Section \ref{sec:jets}.

\section{Jet Angular (Azimuthal) Correlations\label{sec:jets}}

\def \ktrms {$\sqrt{\left<k_{T}^{2}\right>}$}
\def \vkt {$\vec{k}_{T}$}
\def \kt {$k_{T}$}

\subsection{Two particle correlation}
The defining characteristic of a jet is the production of a large
number of particles clustered in a cone in the direction of the
fragmenting parton. Traditionally, energetic jets are identified
directly using standard jet reconstruction
algorithms~\cite{jet:conemethod,jet:ktmethod}. In heavy-ion
collisions, due to the large amount of soft background, direct jet
reconstruction is difficult. Even in $p+A$ or $p+p$~collisions,
the range of energy accessible to direct jet reconstruction is
probably limited to $p_T>5-10$ GeV/c, below which the jet cone
becomes too broad and contamination from the `underlying event'
background is significant. Jet identification is even more
complicated for detectors with limited acceptance, such as the
PHENIX central arms, due to leakage of the jet cone outside the
acceptance.

\begin{figure}[th]
\includegraphics[width=\linewidth]{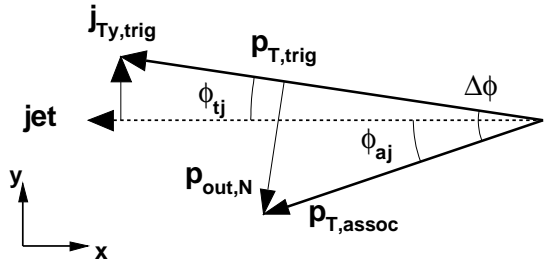}
\caption{Near-side or single jet fragmentation kinematics in the azimuthal
plane perpendicular to the beam axis.  The angles $\phi_{tj}$ and $\phi_{aj}$
are the angles (in the plane transverse to the beam axis) between the
trigger-jet and associated-jet axes, respectively.  Also shown are the $p_T$
vectors for the trigger and associated particles, as well as $p_{out}$ and
$j_{T_y}$ of the trigger particle.}
\label{fig:nearSideKine}
\end{figure}

The two-particle azimuthal angle correlation technique provides an alternative
way to access the properties of jets. It is based on the fact that the
fragments are strongly correlated in azimuth ($\phi$) and pseudo-rapidity
($\eta$). Thus, the jet signal manifests itself as a narrow peak in
$\Delta\phi$ and $\Delta\eta$ space. Jet properties can be extracted on a
statistical basis by accumulating many events to build a $\Delta\phi$
distribution or a $\Delta\phi$ correlation function.  Furthermore, we assume
that the soft background is isotropic.  Distributions in $\Delta\phi$ were
initially used in the 70's to search for jet signals in $p+p$~collisions at
CERN's ISR facility~\cite{jet:ccor1,jet:ccor2,DellaNegra:1977sk}. More
recently, $\Delta\phi$ distributions and correlation functions have been
exploited for analysis of jet correlations at
RHIC~\cite{Adler:2002ct,Adler:2002tq,Rak:2004gk,Adler:2004zd,Ajitanand:2002qd}.
A detailed discussion of the two-particle correlation method can be found
in~\cite{jet:method}. These approaches overcome problems due to background and
limited acceptance, and extend the study of jet observables to lower $p_T$.

In the correlation method, two classes of particles are correlated
with each other -- {\it trigger} particles, and {\it associated}
particles. Although the distinction between these two classes is
artificial, trigger particles are typically selected from a higher
$\pt$ range.
 In this work, we distinguish between two primary categories of
correlations:
\begin{description}
\item[Assorted-$p_T$ correlation (AC)], where the $\pt$ ranges of trigger
and associated particle do not overlap.
\item[Fixed-$p_T$ correlation (FC)], where the $\pt$ ranges for trigger and
associated
particles are identical.
\end{description}

In this paper, correlations are further categorized via a scheme which uses the
identity of the trigger and associated particles.  Four different types of
such correlations are presented. Denoting the trigger particle $\pt$ as
$p_{T,\rm{trig}}$ and associated particle $\pt$ as $p_{T,\rm{assoc}}$,
\begin{itemize}
\item {\bf $h^{\pm} - h^{\pm}$ FC}\\
The $\pt$ range of both the trigger and associated particle is
$1<p_{T}<5$ GeV/c.
\item {\bf $h^{\pm} - h^{\pm}$ AC}\\
There are three different selections,
    \begin{enumerate}
    \item $2.5<p_{T,\rm{trig}}<4$ GeV/c with $0.5<p_{T,\rm{assoc}}<2.5$ GeV/c
    \item $4<p_{T,\rm{trig}}<6$ GeV/c with $0.5<p_{T,\rm{assoc}}<4$ GeV/c
    \item $3<p_{T,\rm{trig}}<5$ GeV/c with $0.5<p_{T,\rm{assoc}}<3$ GeV/c
    \end{enumerate}
\item {\bf $\pi^0 - h^{\pm}$ AC}\\
The trigger particle is a neutral pion and the associated particle is a
charged hadron, where $5<p_{T,\rm{trig}}<10$ GeV/c with
$1<p_{T,\rm{assoc}}<5$ GeV/c.
\item {\bf $\pi^{\pm} - h^{\pm}$ AC}\\
The trigger particle is a charged pion and the associated particle
is a charged hadrons, where $5<p_{T,\rm{trig}}<16$ GeV/c with
$1<p_{T,\rm{assoc}}<5$ GeV/c.
\end{itemize}
For each type of correlation, we study jet structures as
a function of centrality and the momentum of the trigger and associated
particles.

\begin{figure}[th]
\includegraphics[width=\linewidth]{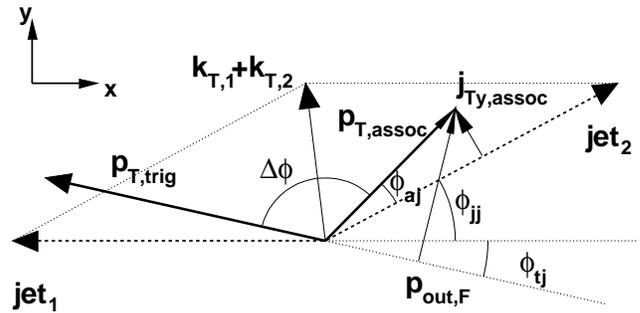}
\caption{Far-side jet fragmentation kinematics in the azimuthal
plane perpendicular to the beam axis.  Similar to Fig.\ref{fig:nearSideKine},
the $\phi$ indices $aj$, $jj$, and $tj$ denote the angles between the
associated-jet, jet-jet, and trigger-jet axes.}
\label{fig:farSideKine}
\end{figure}

\subsection{Extraction of $j_T$, $\sin^2(\phi_{jj})$ 
from the Correlation Function}
\label{subsection:eqndef}

In this section, we discuss the framework for the two particle
correlation method. Figures~\ref{fig:nearSideKine} and
\ref{fig:farSideKine} illustrate the relation between the two
particles and their parent jets, for the case when the parents are the same
jet or the dijet, respectively. The figures also show the relationship
between $j_T$, $\sin^2(\phi_{jj})$ 
and the kinematic variables describing the trigger and
associated particle. $j_T$ is the component of the particle momentum
perpendicular to the jet momentum. Its projection into the azimuthal plane is
$j_{T_y}$.  The quantity $p_{\rm{out}}$ (denoted with $N$ or $F$ for near- or
far-side, respectively) is the component of the associated particle's $p_{T}$
that is perpendicular to the trigger particle's $p_{T}$.  The vector
sum of $k_{T,1}$ and $k_{T,2}$ produces the dijet acoplanarity and
the azimuthal angle between the jet axes is $\phi_{jj}$.

The RMS value of $j_{T_y}$ can be derived from the
correlation functions. For the single jet fragmentation of
Fig.~\ref{fig:nearSideKine}, if we denote $\Delta\phi$,
$\phi_{tj}$, and $\phi_{aj}$ as the angles between
trigger-associated, trigger-jet and associated-jet, respectively,
then the following relations are true:
\begin{eqnarray}
\label{eq:sinnear} \sin(\phi_{tj}) &=&
\frac{j_{T_y,\rm{trig}}}{p_{T,\rm{trig}}}\equiv
x_{j,\rm{trig}}\quad\\\nonumber \sin(\phi_{aj}) &=&
\frac{j_{T_y,\rm{assoc}}}{p_{T,\rm{assoc}}}\equiv
x_{j,\rm{assoc}}\quad\\\nonumber \sin(\Delta\phi) &=&
\frac{p_{\rm{out},N}}{p_{T,\rm{assoc}}}\quad
\\\nonumber \Delta\phi &=& \phi_{tj} + \phi_{aj}\quad.
\end{eqnarray}
Assuming $\phi_{tj}$ and $\phi_{aj}$ are statistically
independent, we have (cross terms average to 0) for the near-side,
\begin{eqnarray}
\mean{\sin^2\Delta\phi_N}=
\mean{\sin^2\phi_{tj}\cos^2\phi_{aj}} +
\mean{\sin^2\phi_{aj}\cos^2\phi_{tj}}
\label{eq:jt1}
\end{eqnarray}

Substituting the $\sin$ and $\cos$ terms from Eq.~\ref{eq:sinnear}
into Eq.~\ref{eq:jt1}, we
obtain the equation for the RMS value of $j_{T_y}$ 
\begin{equation}
\label{eq:jt2} 
\sqrt{\langle j_{T_y}^2 \rangle}  = 
\sqrt{\left\langle p_{\rm{out},N}^2  \right\rangle  /
 (1 + \left\langle {x_h^2 } \right\rangle  - 2\left\langle
{x_{j,\rm{trig}}^2 } \right\rangle )}.
\end{equation}
where $x_h = p_{T,\rm{assoc}}/p_{T,\rm{trig}}$.

In the gaussian approximation for the near-side
azimuthal distributions, a
simple Taylor expansion connects $p_{\rm{out}}$ with the jet
width, $\sigma$:
\begin{eqnarray}
\label{eq:pout}\nonumber \mean{p_{\rm{out}}^2} &=&
\mean{p_{T,\rm{assoc}}^2\sin^2\Delta\phi}\\
&\approx&\mean{p_{T,\rm{assoc}}^2}[\sin\mean{\Delta\phi^2}-\frac{\mean{\Delta\phi^4}}{3}]\\\nonumber
&\approx&\mean{p_{T,\rm{assoc}}^2}[\sin\sigma^2-\sigma^4]
\end{eqnarray}

Since Eq.~\ref{eq:jt2} contains the variable
$x_{j,\rm{trig}}$ that depends on $j_{T_y}$, 
we should calculate $\sqrt{\langle j_{T_y}^2 \rangle}$
iteratively. In cases when trigger and
associated particle $\pt$ are much larger than the typical $j_T$
value, the near-side jet width $\sigma_{N}$ is small and
$x_{j,\rm{trig}}\approx 0$.  Hence Eq.~\ref{eq:jt2} can be
simplified as,
\begin{eqnarray}
\label{eq:jt3} \sqrt{\langle j_{T_y}^2 \rangle} & \simeq &
\frac{\sigma_N\mean{p_{T,\rm{assoc}}}}{\sqrt{1+\mean{x_h^2}}}\\\nonumber
&\simeq&\sigma_N \frac{\langle p_{T,\rm{trig}}\rangle\langle
p_{T,\rm{assoc}}\rangle} {\sqrt{\langle p_{T,\rm{trig}}\rangle^2
+\langle p_{T,\rm{assoc}}\rangle^2}}
\end{eqnarray}

Since $j_{T_y}$ is the projection of hadron $p_T$ perpendicular to
$p_{T,\rm{jet}}$, $j_{T_y}$ is necessarily less than $p_T$.  So,
for any given $p_T$ range, there is always a upper kinematic cut
off on the $j_{T_y}$ distribution. This effect, known as the {\it
seagull effect}, leads to a reduction on the observed
$\sqrt{\langle j_{T_y}^2 \rangle}$ from the expected value. It is important at
low $p_{T,\rm{trig}}$ and becomes negligible once
$p_{T,\rm{trig}}\gg\sqrt{\langle j_{T_y}^2 \rangle}$. The seagull effect can
be parameterized and removed from the $j_{T_y}$ values
\cite{constantin:2004th}.

For the far-side correlation from Fig.~\ref{fig:farSideKine}
we have
\begin{eqnarray}
\label{eq:sinfar}
\pi - \Delta\phi_F &=& \phi_{tj} + \phi_{aj} + \phi_{jj} \\
\sin(\Delta\phi_F) &=&
\frac{p_{\rm{out},F}}{p_{T,\rm{assoc}}}\nonumber
\end{eqnarray}
where $\phi_{jj}$ is the azimuthal angle between the two jet axes.
Expanding $\sin^2\Delta\phi_F$ and dropping all cross terms (which average
to 0), we get
\begin{eqnarray}
\label{eq:kt1} \nonumber\mean{\sin^2\Delta\phi_F} &=&
\mean{\left(\sin\phi_{tj}\cos\phi_{aj}\cos\phi_{jj}\right)^2}+\\\nonumber
&&\mean{\left(\sin\phi_{aj}\cos\phi_{tj}\cos\phi_{jj}\right)^2}+\\\nonumber
&&\mean{\left(\sin\phi_{jj}\cos\phi_{aj}\cos\phi_{tj}\right)^2}+\\
&&\mean{\left(\sin\phi_{tj}\sin\phi_{aj}\sin\phi_{jj}\right)^2}
\end{eqnarray}

We substitute Eq.~\ref{eq:jt1} to get
\begin{widetext}
\begin{eqnarray}
\left<\sin^{2}\Delta\phi_{F}\right> & = &
\left<\sin^{2}\Delta\phi_{N}\right>\left<\cos^{2}\phi_{jj}\right> +
\left<\cos^{2}\Delta\phi_{N}\right>\left<\sin^{2}\phi_{jj}\right>
\nonumber \\
& = & \left<\sin^{2}\Delta\phi_{N}\right>\left<1 -
\sin^{2}\phi_{jj}\right> + \left<1 -
\sin^{2}\Delta\phi_{N}\right>\left<\sin^{2}\phi_{jj}\right>
\nonumber \\
& = & \left<\sin^{2}\Delta\phi_{N}\right> +
\left<\sin^{2}\phi_{jj}\right> -
2\left<\sin^{2}\Delta\phi_{N}\right>\left<\sin^{2}\phi_{jj}\right>
\end{eqnarray}
\end{widetext}
Collecting terms in $\phi_{jj}$ produces
\begin{equation}
\left<\sin^{2}\phi_{jj}\right> =
\frac{\left<\sin^{2}\Delta\phi_{F}\right> -
\left<\sin^{2}\Delta\phi_{N}\right>} {1 -
2\left<\sin^{2}\Delta\phi_{N}\right>}
\end{equation}

Note that since $\phi_{jj}$ is the azimuthal angle between
the jet axes, $\sin^2(\phi_{jj})$ is one measure of
the extent to which the jets are not back-to-back, and is hence
a quantity that is sensitive to any additional scattering in
$d$ + Au~collisions. 
We express the right side in terms of the observables $\sigma_N$ and
$\sigma_F$,
the rms widths of distribution that we measure by expanding the
sine term 
\begin{equation}
\left<\sin^{2}\Delta\phi\right> = \sigma^{2} - \sigma^{4} +
2/3\sigma^{6}
\end{equation}
which is good to 2\% for rms widths at 0.5 rad and good to 0.6\%
for rms widths of 0.2 rad.  Therefore 
\begin{equation}
\left<\sin^{2}\phi_{jj}\right> = \frac{\left(\sigma_{F}^{2} -
\sigma_{F}^{4} + 2/3\sigma_{F}^{6}\right) - \left(\sigma_{N}^{2} -
\sigma_{N}^{4} + 2/3\sigma_{N}^{6}\right)}{1-2\left(\sigma_{N}^{2}
- \sigma_{N}^{4} + 2/3\sigma_{N}^{6}\right)}
\label{eq:LHSRHS}
\end{equation}
The right hand side is now in terms of
experimental observables which we will use to extract
$\sin(\phi_{jj})$. 

We have attempted to extract
$k_T$ from $\sin^2(\phi_{jj})$. This requires assumptions
on the scattered quark distribution, 
the magnitude of the momentum
asymmetry between the partons due to 
the kt-kick, as well as the detailed shape
of the fragmentation function. 
The current paper 
is focussed on the comparison between $p+p$~and $d$ + Au~collisions,
which can be made with $\sin^2(\phi_{jj})$. Hence we leave
the extraction of $k_T$ to future work.

In this paper we report the RMS values of $j_{T}$ and $\sin^2(\phi_{jj})$, 
where
$\sqrt{\langle j_{T}^2 \rangle} = \sqrt{2}\,\sqrt{\langle j_{T_y}^2 
\rangle}$
In the literature, a $j_{T}$ value is sometimes
reported as the geometrical mean, $\mean{|j_{T_y}|}$. 
The relation to the RMS value is
$\mean{|j_{T_y}|} = \sqrt{2/\pi}\,\sqrt{\langle j_{T_y}^2 \rangle}$.

\subsection{Conditional Yields}
\label{subsection:CY}
We also present in this paper the associated yield per trigger
particle, referred to as the conditional yield (CY),
as a function of $p_T$ and
$x_E$. The CY is the number of particles produced in the same or
opposite jet associated with a trigger particle
\begin{eqnarray}
\label{eq:cond6}
CY(p_T ) &=& \frac{1}{{N_{\rm{trig}}}}\frac{{dN_h}}{{dp_T }}\\
CY(x_E ) &=& \frac{1}{{N_{\rm{trig}} }}\frac{{dN_h}}{{dx_E }}
\end{eqnarray}
and can be directly extracted from the measured Gaussian yields
in the correlation functions.  

To emphasize the importance of the $CY$, we
note that it is related to the single- and two-particle cross sections:
\begin{eqnarray}
CY = {
 { \frac{d^2\sigma}{dp_adp_b} }
\mathord{\left/
 {\vphantom { { \frac{d^2\sigma}{dp_adp_b} } { \frac{d\sigma}{dp_T} }}} \right.
 \kern-\nulldelimiterspace} { \frac{d\sigma}{dp_T} } }\quad
.
\end{eqnarray}
The interpretation for the two-particle cross section depends on
whether one is studying the near- or far-side jet correlations.
The conditional yield for particles from the near-side jet depends
on the dihadron fragmentation function, while the conditional
yield from the far-side jet depends on two independent
fragmentation functions: one parton fragments to produce a hadron
with $p_{T,\rm{trig}}$, while the other scattered parton on the
far-side fragments to produce a hadron with $p_{T,\rm{assoc}}$.
For the far-side conditional yield at high-$p_T$, $x_E \simeq
z_{\rm{assoc}}/z_{\rm{trig}}$ (see Eq.~\ref{eqn:xEdef}). Hence,
$d(x_E) \simeq d(z_{\rm{assoc}})/z_{\rm{trig}}$ and the slope of
the far-side $CY(x_E)$ is $z_{\rm{trig}}$ times the slope of the
fragmentation function $D(z)$.

\section{Experiment and Data Analysis\label{sec:expt}}

\subsection{Data Collection\label{sec:exptCollect}}

The data presented in this paper were collected by the PHENIX
experiment at the Relativistic Heavy Ion Collider during the $d$ +
Au~and $p+p$~run of January -- May 2003.  During that time
integrated luminosities were recorded 
of 2.7~nb$^{-1}$ for $d$ + Au~collisions
and $0.35$~pb$^{-1}$ for $p+p$~collisions each at 
$\sqrt{s_{\rm{NN}}}$=200~GeV.

\begin{figure}[t!]
\includegraphics[width=1.0\linewidth]{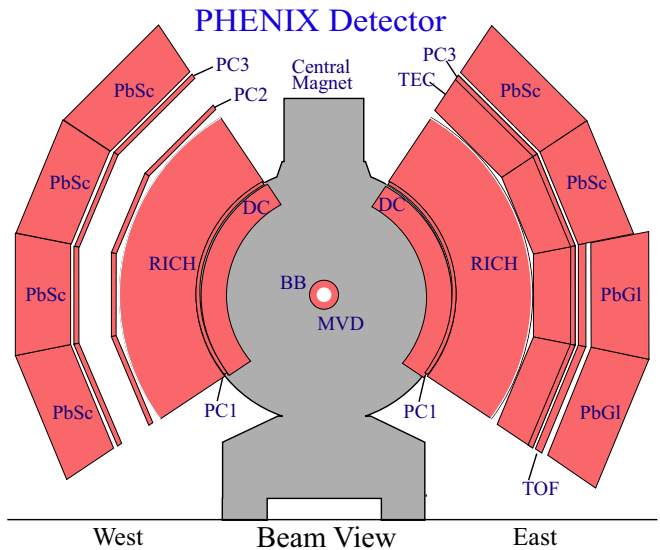}
\caption{(Color online) The two central spectrometer arms of the PHENIX experiment
used to collect the charged hadron and charged and neutral pion
tracks.}\label{figs:PHENIX_detector}
\end{figure}

The PHENIX detector consists of two central spectrometer arms,
two forward muon arms and several global detectors used for triggering,
vertex detection, and centrality selection.  This analysis
utilizes the two central spectrometer arms that each cover a
region of $|\eta| < 0.35$ units of pseudorapidity and $90^{\circ}$
in azimuth. The spectrometer arms are not exactly back-to-back in azimuth,
so while there is
large acceptance for the detection of two particles separated by
$180^{\circ}$ there is also finite
acceptance for two particles
separated
by $90^{\circ}$. Figure~\ref{figs:PHENIX_detector}
shows a beam cross-section view of the PHENIX central spectrometer
arms. A complete overview of the whole PHENIX detector is found in
reference~\cite{Adcox:2003zm}. In this section we will only focus
on those subsystems relevant for the analysis of the dihadron data.

\subsubsection{Global Event Characteristics}
For event characterization the Beam-Beam Counters
(BBCs)~\cite{Allen:2003zt} are utilized.  The BBCs are sets of 64
Cherenkov counters placed symmetrically along the beam line,
covering $3 < |\eta| < 3.9$ units of pseudorapidity and located
144~cm from the center of the interaction region. The BBCs
determine the event vertex and the initial collision time,
$t_{0}$, from the time difference between particles reaching each
BBC.   
For this analysis we
include only events with an offline cut of $|z_{\rm{vertex}}| <$
30~cm.

\begin{figure}[!b]
\includegraphics[width=1.0\linewidth]{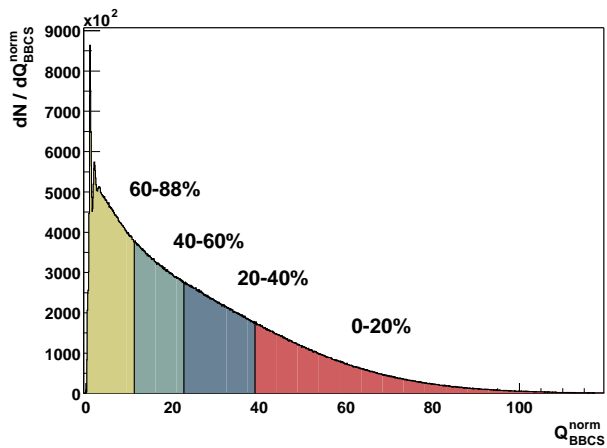}
\caption{(Color online) Total charge
distribution on the Au-going side Beam-Beam Counter (BBC) 
for $d$ + Au~collisions and the
centrality selection (see Table~\ref{table:cent_ncoll}).}
\label{figs:Centrality_def}
\end{figure}

The BBC facing the direction of the Au beam was used to determine the
centrality. Fig.~\ref{figs:Centrality_def} shows the BBC charge
distribution and the centrality classes used in this analysis. The
centrality is defined as
\begin{equation}\label{eq:centrality_def}
\% \textrm{Centrality} = 88.5\%(1-\textrm{frac}(Q_{\rm{BBC}}))
\end{equation}
where frac$(Q_{\rm{BBC}})$ is the fraction of the total BBC charge
distribution integrated from zero to $Q_{\rm{BBC}}$ and $88.5\%$
is the efficiency of the minimum bias trigger. This centrality can
be related to the mean number of Au participants, $\langle
N_{\rm{part}} \rangle$, and mean number of collisions, $\langle
N_{\rm{coll}} \rangle$. To determine the centrality we model the
BBC charge distribution as a negative binomial distribution with a
width and mean proportional to $N_{\rm{part}}$. So, for a given
centrality, there are several negative binomial distributions
(defined by $N_{\rm{part}}$) that contribute to the overall
distribution and as such $N_{\rm{part}}$ is not uniquely defined.
We calculate a weighted average of $N_{\rm{part}}$, where the
weight is given by the negative binomial distribution for a given
$N_{\rm{part}}$ and the probability for having a collision with
$N_{\rm{part}}$.  The latter probabilities were computed using a
Glauber model, with a Hulthen wave function for the deuteron and
an inelastic cross section of 42~mb. Finally, the $\langle
N_{\rm{coll}} \rangle$ was determined for a given $\langle
N_{\rm{part}} \rangle$ from the same Glauber model. The resulting
centrality bins and $\langle N_{\rm{coll}} \rangle$ used in this
analysis are outlined in Table~\ref{table:cent_ncoll}.

\begin{table}[!t]
\caption{\label{table:cent_ncoll}
Table of the mean number of collisions for $d$ + Au~,
$N_{\rm{coll}}$, versus the percentage of the total inelastic
cross section and the nuclear overlap function $T_A(b)$.}
\begin{ruledtabular} \begin{tabular}{ccc}
Percent $\sigma_{\rm{inel}}$ 
& $\langle N_{\rm{coll}} \rangle$ 
& $\langle T_A(b) \rangle (\rm{mb}^{-1})$\\ \hline
0--20\% & $15.4\pm 1.0$ & $0.367 \pm 0.024$ \\
20--40\% & $10.6\pm 0.7$ & $0.252 \pm 0.017$ \\
40--88\% & $4.7\pm 0.3$  & $0.112 \pm 0.007$ \\
\end{tabular} \end{ruledtabular}
\end{table}

The dihadron events were recorded using several different Level-1
triggers. The minimum bias trigger required at least one hit in
each of the BBCs and that the collision vertex (computed online)
satisfies $|z_{\rm{vertex}}|<$ 75~cm. It was sensitive to 88.5\% of
the inelastic $d$ + Au~cross section. PHENIX also employed a
series of Level-1 triggers to select electrons, photons and, with
lower efficiency, high-$p_{T}$ hadrons.  These triggers
are called
the ERT triggers and they utilized
the Ring Imaging Cerenkov (RICH) for electron identification,
together with the Electromagnetic 
Calorimeter (EMC)~\cite{Aphecetche:2003zr}, which consists of 8 sectors, 6 of
which are Lead-Scintillator (PbSc) sampling calorimeters and 2 are
Lead-Glass (PbGl) Cherenkov counters.  The EMC has excellent timing
and energy resolution for electromagnetic showers.
The ERT triggers were produced by summing signals from
tiles, where a tile was 4x5 photo-multipliers (PMTs) in the RICH
and either 2x2 or 4x4 PMTs in the EMC.

The electron trigger was defined by the coincidence between the
minimum bias trigger and the RICH and EMC 2x2 trigger where the
threshold for the RICH tile was 3 photo-electrons and the EMC
threshold varied between 400--800~MeV.  Three different thresholds
were available for the 4x4 photon triggers.  These thresholds
differed between the PbGl and PbSc and varied within and between
the $p+p$~and $d$ + Au~runs.  The lowest threshold setting
(1.4~GeV--2.8~GeV) was most sensitive to hadron showers in the
EMC. The threshold values and rejection factors 
($\rm{rejection}=\rm{N_{minBiasEvents}/N_{triggerEvents}}$) for the ERT
triggers, in coincidence with the minimum bias trigger, are given
in Table~\ref{table:ert_threshold_reject}.  The $h^{\pm}-h^{\pm}$
correlations use only the minimum bias triggered data, while the
$\pi^{0}-h^{\pm}$ correlations use only the ERT photon triggers.
The $\pi^{\pm}-h^{\pm}$ correlations use the minimum bias, ERT
photon and ERT electron triggers. A detailed knowledge of the ERT
trigger efficiency is not necessary, since we present the
conditional yield distributions per trigger, for which this
efficiency cancels out.

\begin{table*}
\caption{\label{table:ert_threshold_reject}
EMC threshold and rejection factors for the electron and
photon ERT triggers in coincidence with the minimum bias trigger
for $p+p$~and $d$ + Au.  The photon triggers are defined by the
energy sum of 4x4 PMTs in the EMC above threshold.  The electron
trigger requires the coincidence of the RICH trigger (threshold of
3 photo-electrons for both $p+p$~and $d$ + Au~runs) and the energy
sum of 2x2 PMTs in the EMC above threshold.}
\begin{ruledtabular} \begin{tabular}{ccccccc}
 & \multicolumn{2}{c}{$p+p$} & \multicolumn{2}{c}{$d$ + Au}
& $p+p$ & $d$ + Au \\  \hline
 & PbSc threshold & PbGl threshold & PbSc threshold & PbGl threshold & rejection & rejection \\ \hline
Gamma1 & 2.1~GeV & 2.1~GeV & 2.8~GeV & 3.5~GeV & 400--1200 & 125--300 \\
 Gamma2 & 2.8~GeV & 2.8~GeV & 2.8--3.5~GeV & 3.5--4.2~GeV &
1500 -- 3100 & 450--900 \\
 Gamma3 & 1.4~GeV & 1.4~GeV & 2.1~GeV & 2.8~GeV & 70--160 & 15--60 \\
 Electron & 0.4--0.8~GeV & 0.4--0.8~GeV & 0.6--0.8~GeV &
0.6--0.8~GeV & 5--1200 & 30--170 \\
\end{tabular} \end{ruledtabular}
\end{table*}

\subsubsection{Tracking and Particle Identification}
In this section we discuss the tracking and identification of the
particles used in the different correlation analyses. There are three
types of particles included: Charged hadrons are used in all
analyses, neutral pions are used as trigger particles for the
$\pi^{0}-h^{\pm}$ correlations and charged pions are used as
trigger particles for the $\pi^{\pm}-h^{\pm}$ correlations.

Charged hadron tracks are measured outside the PHENIX central
magnetic field by the Drift Chamber (DC), located 2.0~m from the
vertex, and two layers of multi-wire proportional chamber (PC1 and
PC3), located 2.5 and 5.0~m, respectively, from the
vertex~\cite{Adcox:2003zp}. The DC determines the momentum and the
azimuthal position of the track, while PC1 determines the polar
angle\cite{Mitchell:2002wu}. 
The momentum resolution is determined to be $0.7\%
\bigoplus 1.1\%p$~(GeV/$c$)\cite{Adler:2003ii}. Tracks are
confirmed by requiring that an associated hit in PC3 lies within a
$2.5\,\sigma$ (for $h^{\pm}-h^{\pm}$) or $3 \sigma$ (for
$\pi-h^{\pm}$) matching window in both the $\phi$ and $z$
directions. This cut reduces the background from particles not
originating in the direction of the vertex. The remaining
background tracks are mainly decays and conversion
particles~\cite{Adler:2003au}. The background level for
single-tracks is less than 5\% below 3 GeV/$c$, increasing to
about 30\% at 5 GeV/$c$. However, the background is smaller for
high-$p_T$ triggered events (see Section~\ref{sec:exptAnalysis}).
The charged particle tracking efficiency for the active region of
the DC, PC1 and PC3 is better than 98\%. Since we perform a pair
analysis, the two track resolution is important. For the DC, the
two track separation is better than 1.5~mm, while at PC1 it is
4~cm and at PC3 it is 8~cm.

Neutral pions are detected by the statistical reconstruction of
their $\gamma\gamma$ decay channel. These decay photons are
detected by the 
EMC
and identified by their
time-of-flight (TOF) and shower shape.  The electromagnetic
shower shape is typically characterized by the $\chi^2$
variable~\cite{Aphecetche:2003zr},
\begin{eqnarray}
\chi^2
=\sum_{i}\frac{(E^{\rm{meas}}_{i}-E^{\rm{pred}}_{i})^2}{\sigma_{i}^2}
\label{eq:chisq}
\end{eqnarray}
where $E^{\rm{meas}}_{i}$ is the energy measured at tower $i$ and
$E^{\rm{pred}}_{i}$ is the predicted energy for an electromagnetic
particle of total energy $\sum_{i}E^{\rm{meas}}_{i}$.  This
$\chi^{2}$ value is useful for the discrimination of
electromagnetic from hadron showers.  The $\chi^2$ and TOF cuts
used give a very clean sample of photons with contamination of
other particles at $\ll 1\%$.

\begin{figure}[!t]
\begin{center}
\includegraphics[width=1.0\linewidth]{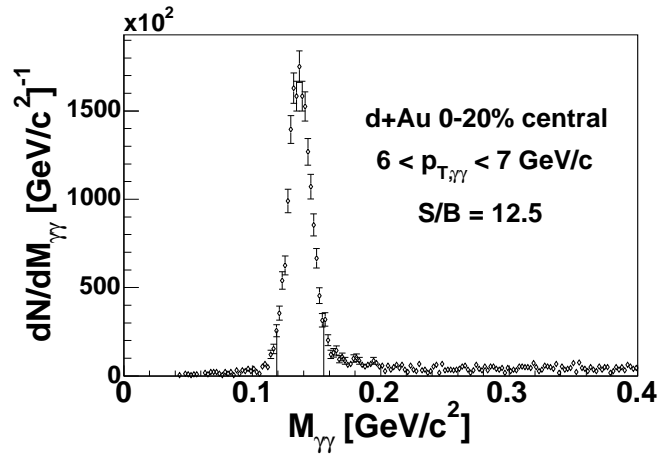}
\end{center}
\caption{Measured $\gamma\gamma$ invariant mass distribution for 6
$<p_T<$ 7 GeV/$c$ in central $d$ + Au~collisions. 
The peak is fitted with 
a gaussian to extract
the centroid mass and $\sigma$. The S/B within 2 $\sigma$ of the
centroid ranges from $\sim$ 6 at 3 GeV/$c$ up to 20 at $\sim$ 8
GeV/$c$. } \label{figs:pi0_invmass}
\end{figure}

Using pairs of photons that pass these EMC cuts, we create the
invariant mass spectra for each photon pair $p_{T}$. A sample
invariant mass distribution with a S/B of approximately twelve is
given in Fig.~\ref{figs:pi0_invmass}. The background distribution
can be reproduced by mixing clusters from different events and
normalizing that distribution to the real event distribution
outside the $\pi^{0}$ mass region. The peak position and width of
the invariant mass distribution were parameterized as a function
of pair $p_{T}$, in order to select $\pi^{0}$ candidates from a
region of invariant mass within $2\sigma$ of the peak position.
The S/B for a $\pi^{0}$ with $p_T>$ 5 GeV is $10-20$, increasing
as a function of $p_T$.  There is a slight dependence on
centrality with the $\pi^{0}$ S/B decreasing with increasing
centrality.

PHENIX identifies high momentum charged pions with the RICH and
EMC detectors. Charged particles with velocities above the
Cherenkov threshold of $\gamma_{\rm{th}} = 35$ (CO$_{2}$ radiator)
emit Cherenkov photons, which are detected by photo-multiplier
tubes (PMTs) in the RICH~\cite{Aizawa:2003zq}. This threshold
corresponds to 18 MeV/$c$ for electrons, 3.5 GeV/$c$ for muons and
4.9 GeV/$c$ for charged pions. In a previous PHENIX
publication~\cite{Adler:2003au}, we have shown that charged
particles with reconstructed $p_T$ above 4.9 GeV/$c$, which have
an associated hit in the RICH, are dominantly charged pions and
background electrons from photon conversions. The efficiency for
detecting charged pions rises quickly past 4.9 GeV/$c$, reaching
an efficiency of $>90\%$ at $p_T>6$ GeV/$c$.

To reject the conversion backgrounds in the pion candidates, the
shower information at the EMC is used. Since most of the
background electrons are genuine low $p_T$ particles that were
mis-reconstructed as high $p_T$ particles, simply requiring a
large deposit of shower energy in the EMC is very effective in
suppressing the electron background. In this analysis a momentum
dependent energy cut at EMC is applied
\begin{eqnarray}
\label{eq:ecut} E>0.3+0.15\,p_T
\end{eqnarray}

\begin{figure}[ht]
\begin{center}
\includegraphics[width=1.0\linewidth]{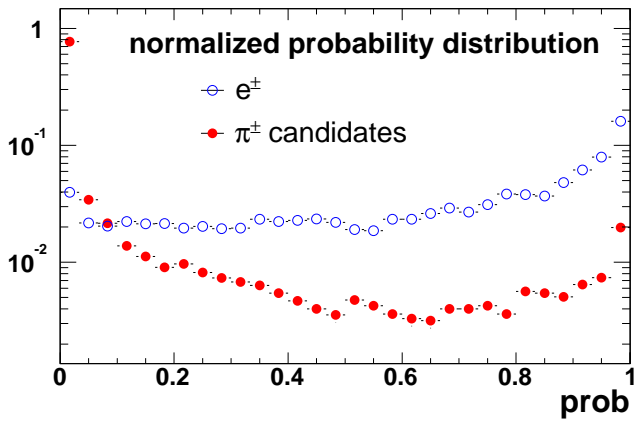}
\caption{\label{fig:prob} (Color online) The probability distribution for charged
pion candidates and electrons. The histogram integral has been
normalized to unity.}
\end{center}
\end{figure}

In addition to this energy cut, the shower shape
information~\cite{Aphecetche:2003zr} is used to further separate
the broad hadronic showers from the narrow electromagnetic showers
and hence reduce the conversion backgrounds.
In this analysis we use the probability (prob) calculated from the
$\chi^2$ value (Eq.~\ref{eq:chisq}) for an EM shower. 
The probability values range
from 0 to 1, with a flat distribution expected for an EM shower
and a peak around 0 for a hadronic shower. Fig.~\ref{fig:prob}
shows the probability distribution for the pion candidates and
electrons, normalized by the integral, where the pions candidates
were required to pass the energy cut and the electrons were
selected using particle ID cuts similar to that used in
Ref.~\cite{Adler:2004ta}. Indeed, the electron distribution is
relatively flat, while the charged pions peak at 0. A cut of
prob~$< 0.2$ selects pions above the energy cut with an efficiency
of $\gtrsim 80$\%. Detailed knowledge of the pion efficiency is
not necessary, since we present in this paper the per trigger pion
conditional-yield distributions, for which this efficiency cancels
out.

\begin{figure}[ht]
\begin{center}
\includegraphics[width=1.0\linewidth]{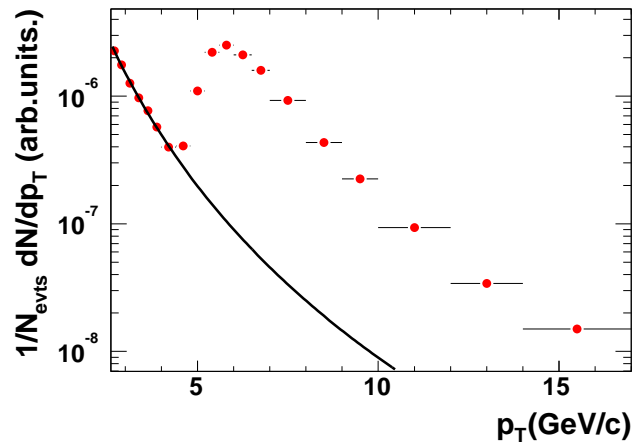}
\caption{\label{fig:pion} (Color online) The raw charged pion transverse momentum
spectrum, with the final cuts applied. The level of the remaining
background is estimated from an extrapolation from low-p$_T$
and is shown as a black line.}
\end{center}
\end{figure}

Since the energy and prob cuts are independent of each other, we
can fix one cut and then vary the second to check the remaining
background level from conversions. The energy cut in
Eq.~\ref{eq:ecut} is chosen such that the raw pion yield is found
to be insensitive to the variation in prob. Fig.~\ref{fig:pion}
shows the raw pion spectra for ERT triggered events as function of
$p_T$, with the above cuts applied. The pion turn on from $4.9-7$
GeV/$c$ is clearly visible. Below $p_T$ of 5 GeV/$c$, the
remaining background comes mainly from the random association of
charged particles with hits in the RICH detector. The background
level is less than 5\% from $5-16$ GeV/$c$, which is the $p_T$
range for the charged pion data presented in this paper.

\subsection{Data Analysis\label{sec:exptAnalysis}}

In this section we outline the method to obtain correlation
functions and distributions.  From these we extract the jet
shapes and yields outlined in Section~\ref{sec:jets}.
For the extraction of the jet-yield from the azimuthal distributions
we discuss how we obtain the absolute normalization of the distribution, while for the jet shape properties, $j_{T_y}$ and 
$\left<sin^{2}\phi_{jj}\right>$, the absolute normalization is not necessary.

\subsubsection{Correlation Functions}
Azimuthal correlations functions are generally defined as
\begin{equation}
C\left(\Delta\phi\right) \propto \frac{N_{\rm{cor}}\left(
\Delta\phi \right)} {N_{\rm{mix}}\left( \Delta\phi \right)}
\end{equation}
Similarly, one can also define the correlation function in
pseudo-rapidity,
\begin{equation}
\label{eq:cfeta} C\left(\Delta\eta\right) \propto
\frac{N_{\rm{cor}}\left( \Delta\eta \right)} {N_{\rm{mix}}\left(
\Delta\eta \right)}
\end{equation}
The same-event pair distribution, $N_{\rm{cor}}(\Delta\phi)$ or
$N_{\rm{cor}}(\Delta\eta)$, is constructed for trigger-associated
particle pairs.  The mixed-event pair distribution,
$N_{\rm{mix}}(\Delta\phi)$ or $N_{\rm{mix}}(\Delta\eta)$, is
determined by combining trigger particles with associated
particles from randomly selected events.

This definition of the correlation function relies on the fact
that detector acceptance and efficiency cancels. It is therefore
important that the pair efficiencies of the average mixed event
background and the average foreground distributions are the same.
For this reason we generate mixed event distributions only for
events with similar centrality and event vertex. More precisely,
mixed events were required to match within $\pm$ 10\% centrality
and the event vertices were also required to be within $\pm$ 3~cm.
For $h^{\pm}-h^{\pm}$ correlations the
real and mixed events are minimum bias data.  For $\pi^0-h^{\pm}$
correlations the real and mixed events are ERT-triggered data. For
$\pi^{\pm}-h^{\pm}$ correlations the real events are ERT-triggered
and minimum bias data while the mixed events mix ERT-triggered
events with minimum bias events.

For $h^{\pm}-h^{\pm}$ and $\pi^{\pm}-h^{\pm}$ correlations, due to
finite two track resolution for charged particles at the DC and
PC, the reconstruction efficiency for same-event charged track pair
drops at small $\Delta\phi$ and $\Delta\eta$. To minimize the
difference of the pair efficiency between
$N_{\rm{cor}}(\Delta\phi)$ and $N_{\rm{mix}}(\Delta\phi)$, the
pairs are required to have a minimal separation of about two times
the resolution at the various tracking detectors. This corresponds
to about 0.28~cm, 8 cm and 15~cm at the DC, PC1 and PC3,
respectively. However, these pair cuts are not required for
$\pi^0-h^{\pm}$ correlations, because different detector
subsystems are used for reconstructing trigger-$\pi^0$ and the
associated charged tracks as outlined earlier.

Given the similarity of the analysis techniques between
$\Delta\phi$ and $\Delta\eta$ correlations, in this paper we focus
on the $\Delta\phi$ correlation. The $\Delta\phi$ correlation
functions are obtained with two different normalizations. For
$h^{\pm}-h^{\pm}$ assorted correlations, the correlation function
is area normalized
\begin{eqnarray}
\label{E:exptCorrFn_e02} C_{\rm{norm}}(\Delta \phi) =
\frac{N_{\rm{cor}}(\Delta \phi)}{N_{\rm{mix}}(\Delta \phi)}\times
\frac{\int d\Delta\phi (N_{\rm{mix}}(\Delta \phi))}{\int
d\Delta\phi (N_{\rm{cor}}(\Delta \phi))}\quad
\end{eqnarray}
The details concerning this normalization are discussed
in Section~\ref{sec:resultsCorrFn}. The second normalization is used
in both the $\pi^{0}-h^{\pm}$ and $\pi^{\pm}-h^{\pm}$
correlations. It was shown in Ref~\cite{jet:method} that the $CY$
can be derived from the measured correlation function with an
appropriate normalization,
\begin{eqnarray}\label{eqn:CondYieldDistDef}
\frac{1}{N^0_{\rm{trig}}}\frac{dN_0}{d\Delta\phi}
=\frac{R_{\Delta\eta}}{N_{\rm{trig}}\epsilon}\frac{N_{\rm{cor}}\left(\Delta\phi\right)}{\frac{2\pi
N_{\rm{mix}}\left(\Delta\phi\right)}{\int d\Delta\phi
N_{\rm{mix}}\left(\Delta\phi\right)}}
\end{eqnarray}
where $N^0_{\rm{trig}}$ and $N_{\rm{trig}}$ are the true and
detected number of triggers respectively, and $\epsilon$ is the
average single particle efficiency for the associated particles in
$2\pi$ in azimuth and $\pm0.35$ in pseudo-rapidity. $R_{\Delta
\eta}$ corrects for the loss of jet pairs outside a given $\Delta
\eta$ acceptance, determined by PHENIX's finite acceptance in
$\eta$. This second normalization is defined so that the integral
of the resulting correlation function should be
$N_{0}/N_{\rm{trig}}^{0}$, the total number of pairs per trigger
particle in a given azimuthal and eta range.

For the normalization in Eq.~\ref{eqn:CondYieldDistDef} there are
two separate efficiencies that must be determined, the
$\Delta\eta$ correction and the single particle efficiency. The
near-side correlation has a well-behaved peak around
$\Delta\eta=0$. As we show in Section~\ref{sec:resultsCorrFn}, the
near-side jet width in $\Delta \phi$ and $\Delta \eta$ are consistent
with each other within errors. So we correct the near-side yield
to the full-jet yield assuming the shape of the jet is Gaussian
and the widths are equal in $\Delta \phi$ and $\Delta \eta$. This
correction, according to Ref.~\cite{jet:method}, is
\begin{eqnarray}
\label{eq:cy1d6} R_{\Delta \eta}  & = & \frac{1}{\int_{-0.7}^{0.7}
d\Delta\eta
\frac{1}{\sqrt{2\pi}\sigma}e^{-\frac{\Delta\eta^2}{2\sigma^2}}
    acc(\Delta\eta)}\quad
\end{eqnarray}
where $acc(\Delta\eta)$ represent the PHENIX pair acceptance
function in $|\Delta\eta|$. It can be obtained by convoluting two
flat distributions in $|\eta|<0.35$, so $acc(\Delta\eta)$ has a
simple triangular shape: $acc(\Delta\eta) =
\left(0.7-|\Delta\eta|\right)/0.7$.  The PHENIX single particle
acceptance is flat in $\eta$ to within 5\%.

\begin{figure}[t]
\begin{center}
\includegraphics[width=1.0\linewidth]{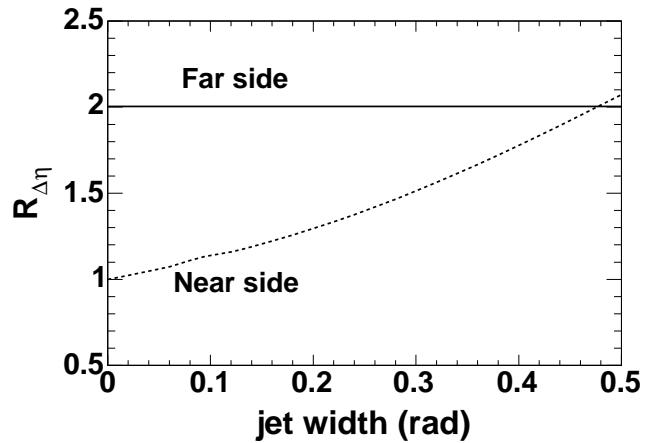}
\caption{\label{fig:jetwidth4} The correction factor
$R_{\Delta\eta}$ plotted as function of jet width for near-side
(dashed line) and far-side (solid line).}
\end{center}
\end{figure}

In the far-side the jet signal is much broader than the PHENIX
acceptance due to the broad range of momentum-fraction $x$ of the
partons that participate in the hard-scattering. In fact, we
studied the far-side jet shape for $\pi^{\pm} - h^{\pm}$
correlation (Fig.~\ref{fig:jetwidth1}b) and found the true jet
correlation strength is almost constant in the PHENIX pair
acceptance $|\Delta\eta|<0.7$. Based on that, we assume that the
far-side jet strength is constant and correct the far-side yield
to the corresponding accessible pair range of $|\Delta \eta |<
0.7$,
\begin{eqnarray}
\label{eq:cy1d7}R_{\Delta \eta}^{far} & = & \frac{2\times
0.7}{\int_{-0.7}^{0.7} d\Delta\eta
    [\frac{0.7-|\Delta\eta|}{0.7}]} = 2
\end{eqnarray}
$R_{\Delta \eta}^{far}$ equals two, because the pair efficiency has a
triangular shape in $|\Delta \eta |< 0.7$, which results in 50\%
average efficiency when the real jet pair distribution is flat in
$|\Delta \eta |< 0.7$. Figure~\ref{fig:jetwidth4} shows the
correction factor $R_{\Delta\eta}$ as a function of jet width. The
typical range of the near-side jet width in all analysis (see
Section~\ref{sec:resultsCorrFn}) is below 0.5 radian. The maximum
correction is about a factor of 2 for the near-side jet.

The single particle efficiency for associated particles,
$\epsilon$, includes detector acceptance and reconstruction
efficiency. It is evaluated in a way similar to previously
published Au + Au~\cite{Adler:2003au} and 
$\dA$~\cite{Adler:2003ii} analyses. However, the jet associated
charged hadron spectrum in $\dA$ is much flatter than the inclusive 
charged hadron spectra~\footnote{For example in Fig.~\ref{fig:mbdaucomppt},
the jet associated yields decreases by a factor of 100 from 0.5 to 5
GeV/$c$. However the typical single inclusive hadron spectra
decrease by a factor of 100000~\cite{Adler:2003ii}.}, so the
corrections due to momentum scale and momentum resolution are much
smaller than that for inclusive charged hadron. For the
same reason, the background contamination at high $p_T$, mainly
coming from decay and photon conversions which are falsely
reconstructed as high $p_T$
tracks~\cite{Adler:2003au,Adler:2003ii}, is also reduced. We
studied both effects using a full GEANT simulation of PYTHIA
events through PHENIX detectors. The jet associated yields were
extracted in the same manner as for the real data analysis. By
comparing it with the input jet associated yield spectra, we can
quantitatively study the effect of momentum smearing and high
$p_T$ background contamination. The corrections due to momentum
scale and resolution are found to be less than 5\% with 3\%
systematic errors. For high-$p_T$ triggered events, the background
contamination to the associated charged hadrons is found to be 5\%
independent of $p_T$ from $1-5$ GeV/$c$.

\subsubsection{Extracting jet properties}
The normalized correlation functions and conditional yield
distributions are both fitted with a sum of two Gaussians to
extract the jet widths and the conditional yield of hadrons in the
near-side ($\Delta \phi \sim 0$) and far-side ($\Delta \phi \sim
\pi$). The fit for the normalized correlation functions is
described in Section \ref{sec:resultsCorrFn}. For the conditional
yield, we fit with the following function
\begin{equation}\label{CY:1}
\frac{1}{N^0_{\rm{trig}}}\frac{dN_0}{d\Delta\phi} =
B+\frac{\rm{Yield}_{N}}{\sqrt{2\pi}\sigma_{N}}
e^{\frac{-\Delta\phi^2}{2\sigma_{N}^2}} + \frac{\rm{Yield}_{F}}
{\sqrt{2\pi}\sigma_{F}}
e^{\frac{-(\Delta\phi-\pi)^2}{2\sigma_{F}^2}},
\end{equation}
where $B$ reflects the combinatoric background level in the real
distribution relative to the mixed distribution and the other two
terms represent the near-side jet and far-side jet signal,
respectively. The resulting widths, $\sigma_N$ and $\sigma_F$, are
then used to calculate the jet shapes via Eqs.~\ref{eq:jt3} and
\ref{eq:LHSRHS}.
For each choice of trigger and
associated particle $p_T$ range, Yield$_{N}$ and Yield$_{F}$
directly reflects the jet associated yield, $dN/dp_{T,\rm{assoc}}$
at the near and far-side, respectively.

Two methods were used to calculate the $dN/dx_E$ distribution. The
first method was used for $\pi^0-h^{\pm}$ correlations.  Since
these correlations are binned in $p_{T}$ there is a distribution
of $x_{E}$ for each trigger-associated $p_{T}$ bin.  This
distribution is approximately Gaussian.  
The fitted peak value is used as the
bin center of the $dN/dx_{E}$ distributions and the fitted
gaussian width is used as the horizontal error bar. To estimate
the bin-width in $x_{E}$, we used the definition
Eq.~\ref{eqn:xEdef} which can be written as (ignoring the sign)
$x_{E} = p_{T,\rm{assoc}}\cos(\Delta\phi)/p_{T,\rm{trig}}$.  We
estimate $\cos(\Delta\phi) \approx 1$ and write the bin-width as
\begin{equation}
\label{eq:xe1} \Delta x_{E} =
\frac{p_{T,\rm{assoc}}^{\rm{max}}-p_{T,\rm{assoc}}^{\rm{min}}}{\langle
p_{T,\rm{trig}} \rangle}
\end{equation}
where we have an associated $p_{T}$ bin from
[$p_{T,\rm{assoc}}^{min}$,$p_{T,\rm{assoc}}^{\rm{max}}$] and a
trigger $p_{T}$ bin with a mean $\langle p_{T,\rm{trig}} \rangle$.

The second method is adopted by $\pi^{\pm}-h^{\pm}$ analysis. It
is statistically based and can be used to calculate the
distribution for any pair variable $p_{T,\rm{trig}}$,
$p_{T,\rm{assoc}}$, $\Delta\phi$, $\Delta\eta$, $x_E$,
$p_{\rm{out}}$ etc. In the following we show two examples: the
$dN/d\Delta\phi$ and $dN/dx_{E}$ distributions. For each pair we
calculate the $\Delta\phi$ and $x_E$ value, then from
Eq.~\ref{eqn:CondYieldDistDef} we calculate the same correction
factor that was used for the $dN/d\Delta\phi$ distribution.
\begin{eqnarray}
\label{eqn:weight} w(\Delta\phi) =
\frac{R_{\Delta\eta}}{N_{\rm{trig}}\epsilon}\frac{1}{\frac{2\pi
N_{\rm{mix}}\left(\Delta\phi\right)}{\int d\Delta\phi
N_{\rm{mix}}\left(\Delta\phi\right)}}
\end{eqnarray}

If this weight is used to fill the $\Delta\phi$ histogram for the
real and mixed distribution, we obtain the $CY$ for the true real
pairs,
\begin{eqnarray}
\frac{1}{N^0_{\rm{trig}}}\frac{dN_0}{d\Delta\phi} = \sum_i
\Delta\phi_{\rm{real}} w(\Delta\phi_{\rm{real}})
\end{eqnarray}
and for the mixed pair the sum should be one by definition,
\begin{eqnarray}
background (\Delta\phi) = \sum_i \Delta\phi_{\rm{mix}}
w(\Delta\phi_{\rm{mix}}) \equiv 1 \quad.
\end{eqnarray}
Thus the jet signal can be extracted as
\begin{eqnarray}
\label{eq:statdndphi}
\frac{1}{N^0_{\rm{trig}}}\frac{dN_0^{\rm{jet}}}{d\Delta\phi} &=&
\sum_i \Delta\phi_{\rm{real}} w(\Delta\phi_{\rm{real}})
-\\\nonumber &&B\sum_i \Delta\phi_{\rm{mix}}
w(\Delta\phi_{\rm{mix}})\quad,
\end{eqnarray}
consistent with Eq~\ref{CY:1}.

When this weight is used to fill the $x_E$ histogram for both real
and mixed distributions, we obtain the $dN/dx_E$ by subtracting
the mixed $x_E$ distribution from the real $x_E$ distribution,
\begin{eqnarray}
\label{eq:statdndxe}
\frac{1}{N^0_{\rm{trig}}}\frac{dN_0^{\rm{jet}}}{dx_E} &=& \sum
x^{\rm{real}}_E w(\Delta\phi_{\rm{real}}) - \\\nonumber &&B \times
\sum_i x^{\rm{mix}}_E w(\Delta\phi_{\rm{mix}})\quad.
\end{eqnarray}

Equation~\ref{eq:statdndphi} is rather trivial, because the weighting
procedure is equivalent to Eq.~\ref{eqn:CondYieldDistDef}, for
which we know the shape of the distribution (Eq.~\ref{CY:1}). But the
advantage of the weighting procedure is that it allows for the
determination of the absolute background pair distribution in any
pair variables. 

Similarly, the statistical method is used to extract the
$p_{T,\rm{assoc}}$ and $p_{\rm{out}}$ spectra as
\begin{eqnarray}
\frac{1}{N^0_{\rm{trig}}}\frac{dN_0^{\rm{jet}}}{dp_{T,\rm{assoc}}}
&=& \sum p^{\rm{real}}_{T,\rm{assoc}}
w(\Delta\phi_{\rm{real}}) - \\
\nonumber && B \times \sum_i p^{\rm{mix}}_{T,\rm{assoc}}
w(\Delta\phi_{\rm{mix}})\\
\frac{1}{N^0_{\rm{trig}}}\frac{dN_0^{\rm{jet}}}{dp_{\rm{out}}} &=&
\sum p^{\rm{real}}_{\rm{out}}
w(\Delta\phi_{\rm{real}}) - \\
\nonumber && B \times \sum_i p^{\rm{mix}}_{\rm{out}}
w(\Delta\phi_{\rm{mix}})
\end{eqnarray}
By construction, the integral of the jet yield should be conserved
independent of the pair variable used, {\it i.e.}:

\begin{eqnarray}
&&\int d\Delta\phi\frac{dN_0}{d\Delta\phi}= \int
dx_E\frac{dN_0^{\rm{jet}}}{dx_E}=\\\nonumber &&\int
dp_{T,\rm{assoc}} \frac{dN_0^{\rm{jet}}}{dp_{T,\rm{assoc}}} = \int
dp_{\rm{out}}\frac{dN_0^{\rm{jet}}}{dp_{\rm{out}}}
\end{eqnarray}

\subsection{Systematic Uncertainties\label{sec:exptSyst}}

The correlation analyses presented here consist of several steps
ranging from the generation of correlation functions to the
extraction of the final physics variables ($j_T$, $\sin(\phi_{jj})$, per
trigger yields, etc.) from these correlation functions. Systematic
error estimations for each of these steps have been evaluated and
combined to determine the overall error quoted for each
measurement. All errors quoted are maximum extent.

Systematic errors associated with the generation of correlation
functions can result from shape distortions in either the
foreground or the background distributions. These distortions can
arise if the requisite quality cuts (see section
\ref{sec:exptCollect}) are not stable. In order to minimize such
errors, the track-pair and quality cuts were assigned such that
the correlation functions were essentially insensitive to
reasonable cut variations. Systematic errors associated with such
cut variations are estimated to be less than 4\%. A further source
of systematic errors is related to the efficiency of the
background rejection when requiring a confirmation hit in the
outer pad chamber. The yields have been corrected for remaining
background. The systematic error on the background estimate is
$\approx$ 3\% for tracks with a transverse momentum ($p_T$) $<$
4 GeV/$c$ and $\approx$ 7\% for particles with 4 $< p_{T} < $ 6
GeV/$c$. For the calculation of the conditional yields, the
systematic error is dominated by the uncertainties associated with
the determination of the efficiency corrected single particle
yields. These systematic errors have been estimated to be
$\approx$ 10\% as obtained from Ref~\cite{Adler:2003ii}. This
error has two parts; the normalization error includes the error on
PC3 matching and active area. The momentum smearing error includes
contributions from momentum resolution and momentum scale.

A separate error is estimated for $\pi^{0}-h^{\pm}$ correlations
due to the background contamination of the $\pi^{0}$s within the
mass bin.  To estimate the width and yield contribution of the
background $\gamma\gamma$ pairs, we created correlations of
$\gamma\gamma$ outside the $\pi^{0}$ mass with hadrons.  From
these we extrapolated the background contribution at the $\pi^{0}$
mass.  These systematic errors are $p_{T}$-dependent.  For the
near and far angle width the variation is $1-3$\%, the near yield
variation is 1\%, the far yield variation is $1-5$\% and increases
with increasing $p_{T}$.

\begin{figure}[ht]
\begin{center}
\includegraphics[width=1.0\linewidth]{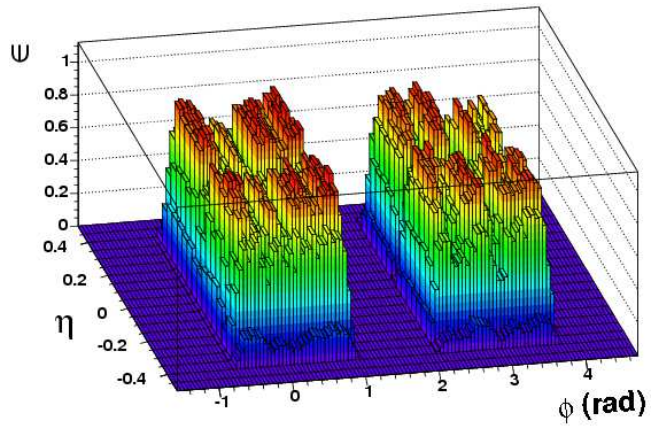}
\caption{\label{fig:cycheck1} (Color online) A typical PHENIX single particle
acceptance for charged hadrons.}
\end{center}
\end{figure}

The event mixing technique has been used to correct for the
limited detector acceptance and inefficiency. In addition, the
$CY$ has been corrected for limited $\Delta\eta$ coverage. To
cross check these procedures we have run a detailed simulation
using the PYTHIA event generator~\cite{Sjostrand:2001yu} coupled to
a single particle acceptance filter that randomly
accepts charged particles according to the detector efficiency. In
the following, we shall use $\pi^{\pm}-h^{\pm}$ as an example for
this cross check. Figure~\ref{fig:cycheck1} shows a typical PHENIX
two dimensional single particle acceptance used in this analysis.

\begin{figure}[th]
\includegraphics[width=1.0\linewidth]{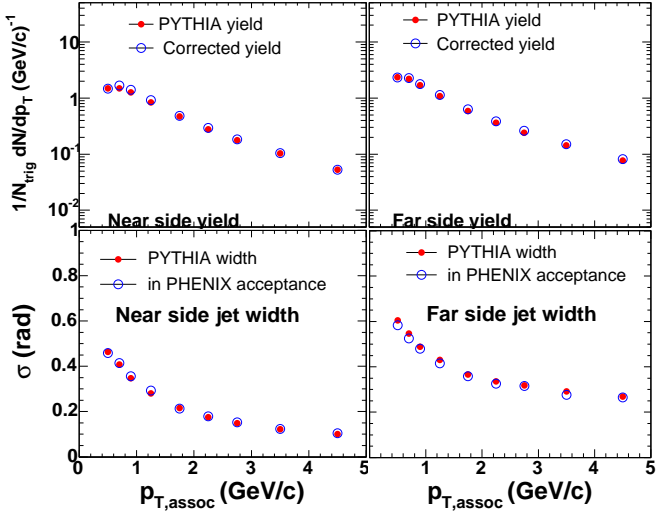}
\caption{\label{fig:cycheck2} (Color online) The comparison of near-side yield
(top left panel), near-side width (bottom left panel), far-side
yield (top right panel) and far-side width (bottom right panel) as
function of $p_T$ of charged hadrons. These are obtained for
$\pi^{\pm}-h^{\pm}$ correlation from PYTHIA, with a trigger pion
of $6-10$ GeV/$c$. The filled circles represent the quantities
calculated with PHENIX acceptance filter.}
\end{figure}

We generated 1 million PYTHIA events, each required to have at
least one $>6$ GeV/$c$ charged pion. To speed up the event
generation, a $Q^2$ cut of 100 GeV$^2$ on the underlying
parton-parton scattering is required. These events were filtered
through the single particle acceptance filter. As an
approximation, we ignore the $p_T$ dependence of acceptance. The
same event and mixed pair $\Delta\phi$ distributions were then
built by combining the accepted $\pi^{\pm}$ and charged hadrons.
The jet width and raw yield were extracted by fitting the
$\frac{dN_{\rm{fg}}}{d\Delta\phi}/\frac{dN_{\rm{mix}}}{d\Delta\phi}$
with a constant plus double gaussian function. The raw yields were
then corrected via Eq.~\ref{eq:cy1d6} to full jet yield for the
near-side and the yield in $|\Delta\eta|<0.7$~for the far-side. We
also extracted the true $CY$ and jet width without the acceptance
requirement. The comparison of the $CY$ and jet width with and
without the acceptance requirement are shown in
Fig.\ref{fig:cycheck2}. The trigger particles are $\pi^{\pm}$ with
$6<p_{T,\rm{trig}}<10$ GeV/$c$, the associated particles are
$h^{\pm}$. In the near-side, the corrected yield (top left panel)
and width (bottom left panel) are compared with those extracted
without acceptance filter. In the far-side, the yield corrected
back to $|\Delta\eta|<0.7$~(top right panel) and the width (bottom
right panel) are compared with those extracted without the
acceptance filter. The data requiring the acceptance filter are
always indicated by the filled circles, while the expected yield
or width are indicated with open circles.

\begin{figure}[ht]
\includegraphics[width=1.0\linewidth]{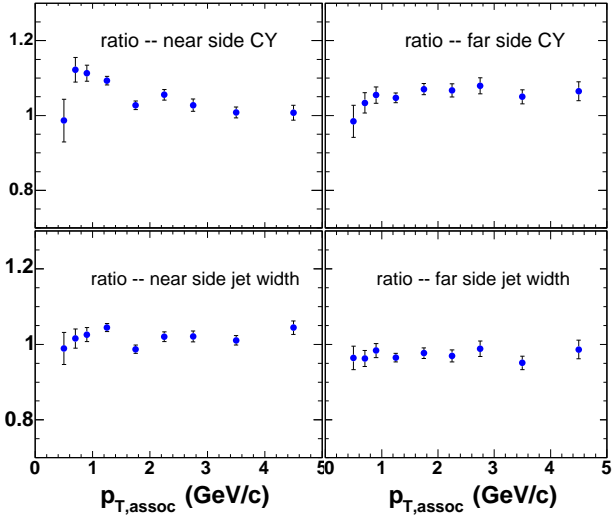}
\caption{\label{fig:cycheck3} (Color online) The ratio of the jet width or
corrected yield obtained from the event mixing method to those without
the acceptance filter.}
\end{figure}

The agreement between the two data sets can be better seen by
plotting the ratios, which are shown in Fig.~\ref{fig:cycheck3}.
The yields agree within 10\% and the widths agree within 5\%.
Since $\sqrt{\left<j_{T}^{2}\right>}$, $\mean{\sin^2(\phi_{jj})}$
are derived from the jet widths, the agreement in width naturally
leads to the agreement in the $\sqrt{\left<j_{T}^{2}\right>}$ and
$\mean{\sin^2(\phi_{jj})}$. One notices that there are some
systematic differences in the comparison of the yield at low
$p_{T,\rm{assoc}}$. This might indicate that the Gaussian
assumption is not good enough when the jet width is wide and the
extrapolation for $|\Delta\eta|>0.7$ become sizeable. (At
$p_{T,\rm{assoc}}=0.5$ GeV/$c$, the jet width $\sigma_{N} =
0.5$~(rad) and the extrapolation is about 20\%.).

The approximations in the formulas used to extract $j_{T_y}$ and
$\sin^2(\phi_{jj})$ are used to estimate the systematic error on
these quantities. We estimate the systematic uncertainty in the
formulation at the level of 5\% for the $\sqrt{\langle j_T^2
\rangle}$ and 3-4\% for the $\sqrt{\langle \sin^2(\phi_{jj})
\rangle}$.

\begin{table*}[th]
\caption{\label{tab:widtherror} Summary of the systematic errors
on the widths and $j_T$, $\langle \sin^2(\phi_{jj})\rangle$.}
\begin{ruledtabular} \begin{tabular}{cc}
Error source &  \\  \hline
tracking cuts, pair cuts & $<4\%$ \\ 
assumptions used in formula& $<5\%$ \\
 S/B correction ($\pi^{0}$ only) & $1-3\%$ \\
\end{tabular} \end{ruledtabular}

\caption{\label{tab:hadroncyerror} Summary of the systematic
errors on the Conditional Yields for $h^{\pm}-h^{\pm}$ analysis.}
\begin{ruledtabular} \begin{tabular}{ccc}
Error source & $<$ 4 GeV/$c$ & $4-6$ GeV/$c$\\ \hline
 quality cuts  & $<$ 4\% & $<$ 4\% \\ 
background correction &3\% & 30\% \\  
error on single particle yields& 10\% & 10\% \\ 
\end{tabular} \end{ruledtabular}

\caption{\label{tab:pizerocyerror} Summary of the systematic
errors on the Conditional Yields for $\pi^0-h^{\pm}$ analysis.}
\begin{ruledtabular} \begin{tabular}{ccccc}
 & Pair cuts & \multicolumn{3}{c}{3\%} \\
\rabo{Single Particle} & normalization & \multicolumn{3}{c}{6.5\%} \\ 
\rabo{$\epsilon_{\rm{single}}$} & p smearing (reso+scale) & \multicolumn{3}{c}{3\%}
\\  & near-side yield & \multicolumn{3}{c}{1\%} \\ \hline
S/B & $p_{T,\rm{assoc}}$ (GeV/$c$) & $<$2 & $2-3$ & $>$3 \\
 & far-side yield & 5\% & 2\% & 1\% \\
\end{tabular} \end{ruledtabular}

\caption{\label{tab:pioncyerror} Summary of the systematic errors
on the Conditional Yields for $\pi^{\pm}-h^{\pm}$ analysis.}
\begin{ruledtabular} \begin{tabular}{ccccccc}
 & normalization & \multicolumn{5}{c}{6.5\%} \\  
\rabo{Single Particle}
 & p smearing (reso+scale) & \multicolumn{5}{c}{3\%} \\ 
\rabo{$\epsilon_{\rm{single}}$}  & trigger
pion background & \multicolumn{5}{c}{5\%} \\ 
  & centrality dependent part & \multicolumn{5}{c}{5\%} \\ \hline
  & $p_{T,\rm{assoc}}$ (GeV/$c$) & $<$ 1 & $1-2$ & $2-3$ & $3-4$ & $4-5$ \\ 
\cline{2-7} & pair cuts & 1\% & 1\% & 2\% & 3\% & 4\% \\
 Yield Extraction & near-side yield & 20\% & 10\% & 6\% & 6\% & 6\% \\
 & far-side yield & \multicolumn{5}{c}{6\%} \\
 & error on the fit & $10-20$\% & 6\% & 4\% & 4\% & 4\% \\
\end{tabular} \end{ruledtabular}

\caption{\label{tab:pioncyerror1} Summary of the systematic errors
on the $p_{\rm{out}}$ distribution for $\pi^{\pm}-h^{\pm}$ analysis.}
\begin{ruledtabular} \begin{tabular}{ccccc}
$p_{\rm{out}}$ (GeV/$c$) & $<$ 0.5 & $0.5-1$ &$1-2$&$2-2.5$\\ \hline
yield extraction (near) & 8\% &15\% & 20\%& 20\%\\ 
yield extraction (far)  & 8\% &15\% & 20\%& 30\%\\ 
other errors & 10.6\% &10.6\% & 10.6\%& 10.6\%\\
\end{tabular} \end{ruledtabular}
\end{table*}

Table~\ref{tab:widtherror} summarizes the systematic errors for
the extracted widths, $\sqrt{\langle j_T^2 \rangle}$ and
$\sqrt{\langle \sin^2(\phi_{jj}) \rangle}$, while
Tables~\ref{tab:hadroncyerror}, \ref{tab:pizerocyerror}, and
\ref{tab:pioncyerror} summarize the list of systematic errors on
the $CY$ for the hadron-hadron, neutral pion-hadron, and charged
pion-hadron correlations, respectively.
Table~\ref{tab:pioncyerror1} outlines the systematic errors on the
$p_{\rm{out}}$ extraction from pion-hadron correlations.


\section{Results\label{sec:results}}

We present the minimum bias and centrality dependent results on
extracted jet widths and yields in Section
\ref{sec:resultsCorrFn}, which are used in Section
\ref{sec:resultsMinBias} to calculate quantities describing the
jet-structures: the values of $\sqrt{\langle j_T^2 \rangle}$, 
$\langle \sin^2(\phi_{jj}) \rangle$,  
and jet fragmentation conditional yields $dN/dp_T$
and $dN/dx_E$. The minimum-bias $d$ + Au~results are compared with
results from $p+p$ in Section \ref{sec:resultsDAuvspp} to
establish the extent of effects due to medium modification in $d$
+ Au with as much statistical precision as possible.  The $d$ + Au
centrality dependence of the derived quantities is presented in
Section \ref{sec:resultsCentrality}. This provides a larger
lever-arm in nuclear thickness function, at the cost of dividing
the available minimum-bias data into different centrality bins.

\subsection{Correlation Functions, Widths and Yields\label{sec:resultsCorrFn}}

\begin{figure}[ht]
\includegraphics[width=1.0\linewidth]{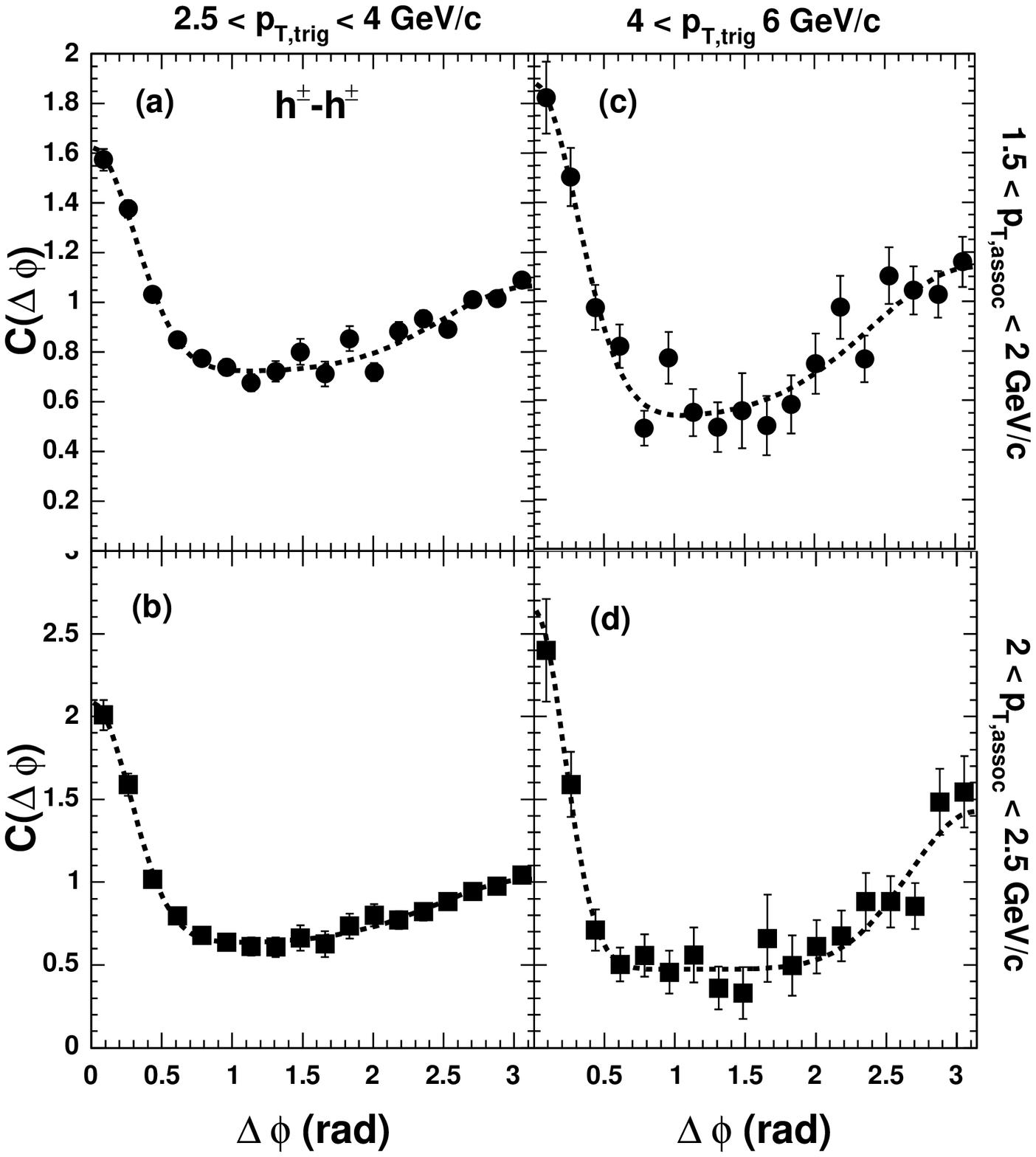}
\caption{ Assorted $h^{\pm}-h^{\pm}$ correlation functions from
$d$ + Au~collisions for
centrality 0-80\% and several \pt~cuts. The trigger \pt~range is
$2.5<p_{T,\rm{trig}}<4.0$ GeV/$c$ (a -- b) and
$4.0<p_{T,\rm{trig}}<6.0$ (c -- d) respectively, while the
associated hadron is in the range $1.5<p_T<2.0$ GeV/$c$ or
$2.0<p_T<2.5$ GeV/$c$. The correlations are for the centrality
class $0-80$\%. The dashed line represents a fit to the
correlation function using Eq.~\ref{E:exptCorrFn_e04}.}
\label{Fi:exptCorrFn_fi01}
\end{figure}

\begin{figure}[ht]
\includegraphics[width=1.0\linewidth]{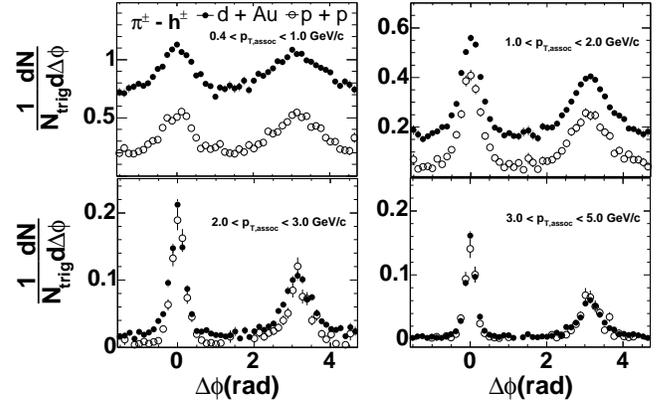}
\caption{ Fully corrected assorted charged pion-hadron conditional
pair distributions for $d$ + Au~collisions
centrality $0-88$\% and $p+p$~collisions. The trigger
$\pi^{\pm}$'s are within 5 $<p_{T,\rm{trig}}<$ 10 GeV/$c$ and are
correlated with hadrons with $p_{T,\rm{assoc}}$ $0.4-1.0$ GeV/$c$,
$1.0-2.0$ GeV/$c$, $2.0-3.0$ GeV/$c$ and $3.0-5.0$ GeV/$c$ (from
top to bottom and left to right). }
\label{Fi:exptCorrFn_distro_fi01}
\end{figure}

\begin{figure}[ht]
\includegraphics[width=1.0\linewidth]{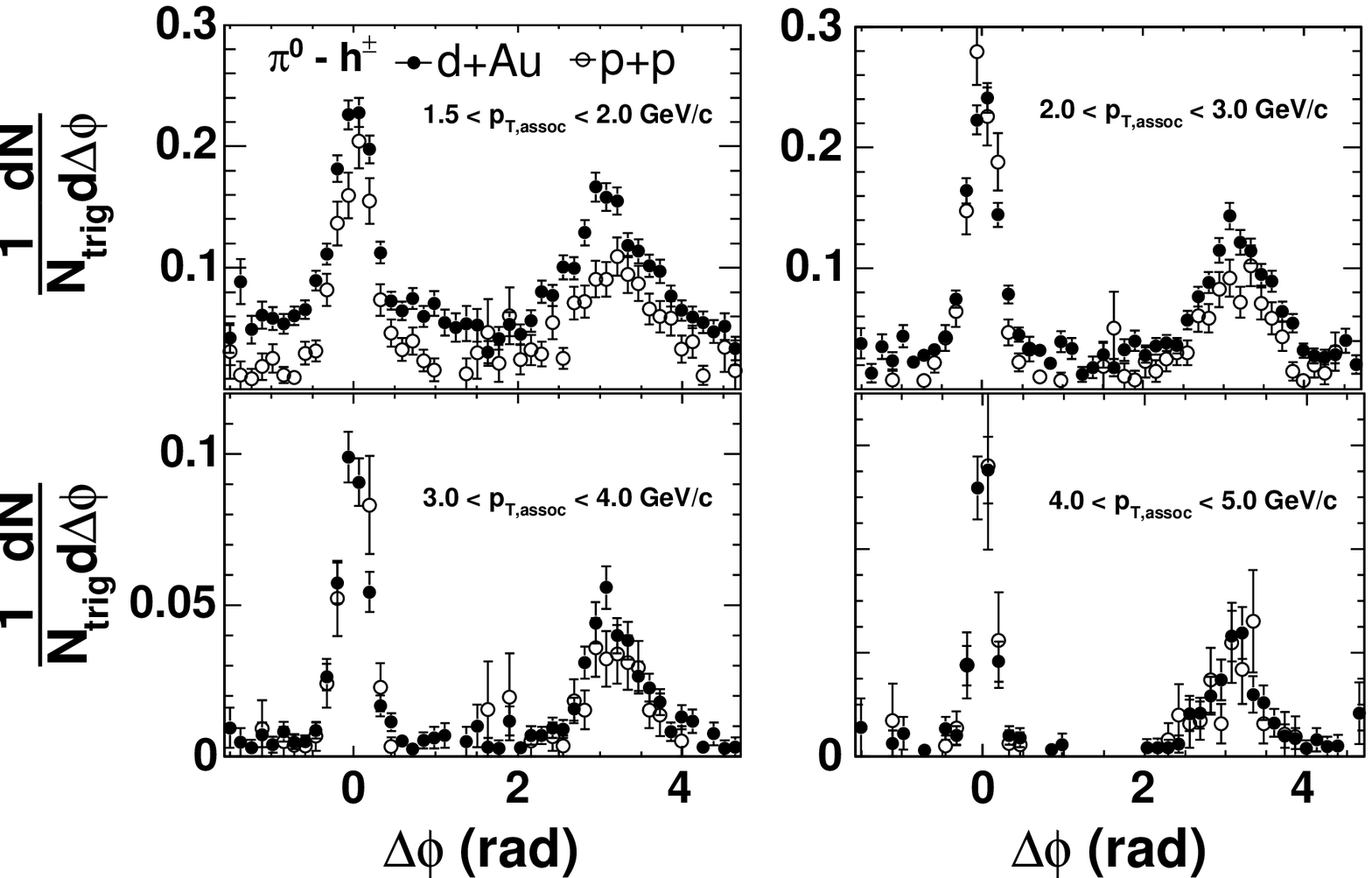}
\caption{ Fully corrected assorted $\pi^{0}$-hadron conditional
pair distributions for $d$ + Au~collisions centrality $0-88$\%
and $p+p$~collisions. The
trigger
$\pi^{0}$'s are within 5 $< p_{T,\rm{trig}} <$ 10 GeV/$c$ and are
correlated with hadrons with $p_{T,\rm{assoc}}$ of (upper-left)
$1.5-2$ GeV/$c$, (upper-right) $2-3$ GeV/$c$, (lower left) $3-4$
GeV/$c$, and (lower right) $4-5$ GeV/$c$.}
\label{Fi:exptCorrFn_distro_fi02}
\end{figure}

The baseline data from which jet structures are extracted are the
correlation functions and conditional pair distributions that were
defined in Section~\ref{sec:jets}. Figure~\ref{Fi:exptCorrFn_fi01}
shows representative correlation functions between two charged
hadrons, while Figs.~\ref{Fi:exptCorrFn_distro_fi01} and
\ref{Fi:exptCorrFn_distro_fi02} show representative conditional
yield distributions triggered on neutral pions ($\pi^0$) and
charged pions respectively. All three correlation sets
(Fig.~\ref{Fi:exptCorrFn_fi01} to \ref{Fi:exptCorrFn_distro_fi02})
show relatively narrow peaks centered at $\Delta\phi = 0$ and
$\pi$ radians. The widths of these structures decrease with larger
$p_T$, which is consistent with narrowing of the jet cone for
increasing $p_T$. The fractional area under the jet peak
relative to the flat underlying background also increases
significantly as function of associated particle $p_T$, indicating
increasing (di)jet contributions to the correlation
function. In particular, Fig.~\ref{Fi:exptCorrFn_distro_fi01}
shows that for events where there is a high $p_T$ trigger, a large
fraction of the low $p_T$ (as low as $0.4-1$ GeV/$c$) particles are
coming from the dijet fragmentation, and the jet contribution
dominates at $p_T>2$ GeV/$c$. Events tagged with a high
$p_T$ jet are much harder than a typical minimum bias event.

We characterize the jet correlations shown in
Figs.~\ref{Fi:exptCorrFn_fi01} to \ref{Fi:exptCorrFn_distro_fi02}
by assuming that there are only
two contributions to the correlation function
 --~(di)jet correlations and an isotropic
underlying event.
This scenario can then be expressed as:

\begin{equation}\label{E:assumpt}
C(\Delta \phi) = A_{o}( 1 + J(\Delta \phi) )
\end{equation}

where $A_{o}$ denotes the isotropic background
and $J(\Delta \phi)$ is the jet-function.
Approximating the jet-function
as the sum of two Gaussians, we fit the correlations with;
\begin{equation}\label{E:exptCorrFn_e04}
C(\Delta \phi) = A_{o}(1+\frac{\lambda_{N}}{\sqrt{2\pi}\sigma_{N}}
e^{\frac{-\Delta\phi^2}{2\sigma_{N}^2}} + \frac{\lambda_{F}}
{\sqrt{2\pi}\sigma_{F}} e^{\frac{-(\Delta\phi-\pi)^2}{2\sigma_{F}^2}}).
\end{equation}
Here, $\lambda_{N,F}$ are the normalized Gaussian areas and
$\sigma_{N,F}$ are the Gaussian widths for the near and far-side
jets respectively. For the pair distribution functions we fit with
the same shaped function, but with a different normalization (Eq.
\ref{CY:1}) as outlined in section \ref{sec:exptAnalysis}.

\begin{figure}[ht]
\includegraphics[width=1.0\linewidth]{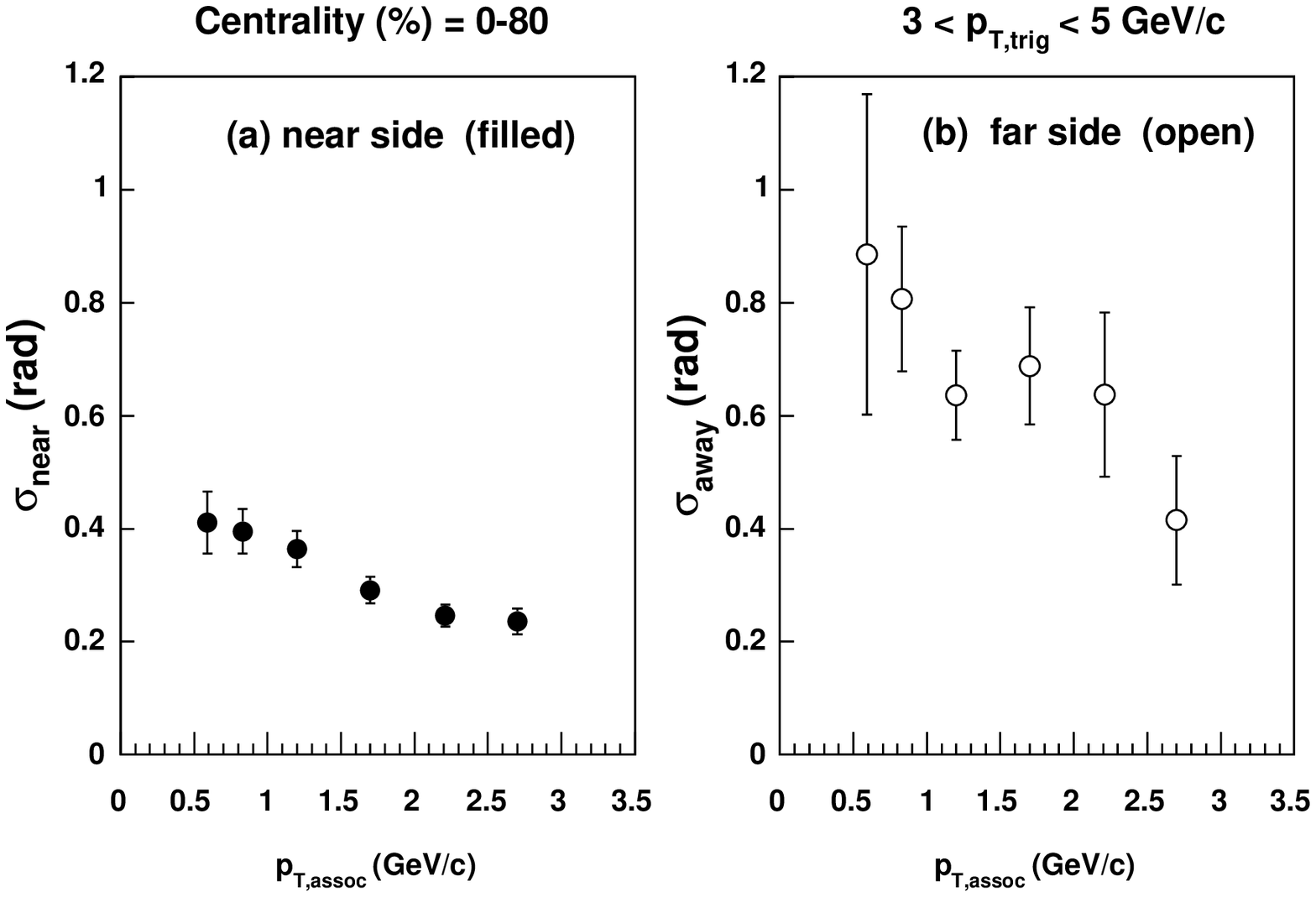}
\caption{Near (a) and far-side (b) widths as a function of
$p_{T,\rm{assoc}}$ for charged hadron correlations from
minimum-bias $d$ + Au~collisions (see text). Bars are statistical errors.}
\label{Fi:exptCorrFn_fi05}
\end{figure}

Figure \ref{Fi:exptCorrFn_fi05} shows the associated-$p_T$~dependence of the
extracted widths for both the near- and far-side peaks 
\footnote{The results here
are not sensitive to the slightly different range in centrality used.}
from the
charged-hadron correlation functions with the trigger range 
for the charged hadron being $3-5$ GeV/$c$.
The data are tabulated in Table~\ref{Ta:hh_widths_3_5}.

\begin{table} [ht]
\caption{\label{Ta:hh_widths_3_5}
Near and far-side widths as a function of
$p_{T,\rm{assoc}}$ for charged hadron triggers ($3-5$ GeV/$c$)
and associated charged hadrons from $d$ + Au~collisions.}
\begin{ruledtabular} \begin{tabular}{ccc}
$\mean{p_{T,\rm{assoc}}}$ & $\sigma_{near}$ & $\sigma_{far}$ \\
(GeV/$c$) & (rad) & (rad) \\ \hline
0.59 & $ 0.411 \pm 0.055$ & $ 0.89 \pm 0.28$ \\ 
0.83 & $ 0.395 \pm 0.039$ & $ 0.807\pm 0.128$ \\ 
1.2 & $ 0.364 \pm 0.032$ & $ 0.636 \pm 0.079$ \\ 
1.7 &  $ 0.291\pm 0.023$ & $ 0.688 \pm 0.103$ \\ 
2.2 & $ 0.246 \pm 0.019$ & $ 0.637 \pm 0.146$ \\ 
2.7 & $ 0.236\pm 0.023$ & $ 0.415 \pm 0.114$ \\ 
\end{tabular} \end{ruledtabular}
\end{table}

\begin{figure}[ht]
\includegraphics[width=1.0\linewidth]{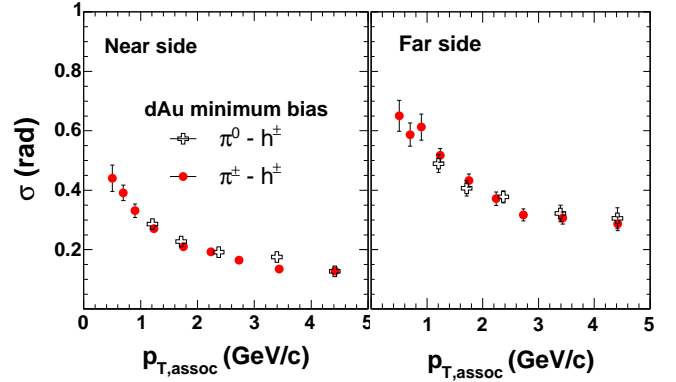}
\caption{\label{fig:mbdawidth} (Color online) a) near-side width,
b) far-side width as a function of $p_{T,\rm{assoc}}$ for a
charged pion (closed symbols) and neutral pion 
(open symbols)
triggers from the $p_{T,\rm{trig}}$ range of $5-10$ GeV/$c$
in minimum-bias $d$ + Au~collisions (see text). Bars are statistical errors.}
\end{figure}

In Fig.
\ref{fig:mbdawidth} we present the same quantities from the
high-$p_T$ identified pion correlations, where there is excellent
agreement between the $\pi^0$ and charged-pion data sets. For both
types of identified pions 
the trigger $p_T$ range is $5-10$ GeV/$c$ and
these data are tabulated in
Tables~\ref{Ta:pippmh_widths_5_10} and ~\ref{Ta:pi0h_widths_5_10}.

\begin{table} [ht]
\caption{\label{Ta:pippmh_widths_5_10}
Near and far-side widths as a function of
$p_{T,\rm{assoc}}$ for charged pion triggers ($5-10$ GeV/$c$)
and associated charged hadrons from minimum-bias $d$ + Au~collisions.}
\begin{ruledtabular} \begin{tabular}{ccc}
$\mean{p_{T,\rm{assoc}}}$ & $\sigma_{near}$ & $\sigma_{far}$ \\
(GeV/$c$) & (rad) & (rad) \\ \hline
0.50 & $ 0.440 \pm 0.044 $ & $ 0.651 \pm 0.052 $ \\ 
0.70 & $ 0.391 \pm 0.026 $ & $ 0.587 \pm 0.039 $ \\ 
0.90 & $ 0.331 \pm 0.023 $ & $ 0.613 \pm 0.044 $ \\ 
1.23 & $ 0.271 \pm 0.010 $ & $ 0.517 \pm 0.024 $ \\ 
1.75 & $ 0.210 \pm 0.008 $ & $ 0.433 \pm 0.022 $ \\ 
2.24 & $ 0.193 \pm 0.009 $ & $ 0.372 \pm 0.023 $ \\ 
2.73 & $ 0.165 \pm 0.007 $ & $ 0.317 \pm 0.020 $ \\ 
3.44 & $ 0.135 \pm 0.006 $ & $ 0.307 \pm 0.020 $ \\ 
4.42 & $ 0.128 \pm 0.008 $ & $ 0.287 \pm 0.023 $ \\ 
\end{tabular} \end{ruledtabular}
\end{table}

\begin{table} [ht]
\caption{\label{Ta:pi0h_widths_5_10}
Near and far-side widths as a function of
$p_{T,\rm{assoc}}$ for neutral pion triggers ($5-10$ GeV/$c$)
and associated charged hadrons
from minimum-bias $d$ + Au~collisions.}
\begin{ruledtabular} \begin{tabular}{ccc}
$\mean{p_{T,\rm{assoc}}}$ & $\sigma_{near}$ & $\sigma_{far}$ \\
(GeV/$c$) & (rad) & (rad) \\ \hline
1.21 & $ 0.284 \pm 0.011  $ & $ 0.494 \pm 0.022  $ \\ 
1.71 & $ 0.227 \pm 0.007  $ & $ 0.410 \pm 0.019  $ \\ 
2.37 & $ 0.193 \pm 0.005  $ & $ 0.380 \pm 0.015  $ \\ 
3.39 & $ 0.177 \pm 0.006  $ & $ 0.322 \pm 0.020  $ \\ 
4.41 & $ 0.130 \pm 0.007  $ & $ 0.315 \pm 0.026  $ \\ 
\end{tabular} \end{ruledtabular}
\end{table}

The far-side widths shown in Figs.~\ref{Fi:exptCorrFn_fi05} and
\ref{fig:mbdawidth} are larger than the near-side widths, as
expected, since the far-side structure is a convolution of two
jet fragmentations as well as any $k_T$ of the scattered partons.
The widths of the correlation functions also steadily decrease as
a function of $p_{T,\rm{assoc}}$ as expected from (di)jet
fragmentation.
For completeness we also tabulate the near- and far-side widths
extracted as a function of $p_{T,\rm{trig}}$ for identified pions.
These data are tabulated in
Tables~\ref{Ta:pippmh_widths_pttrig} and ~\ref{Ta:pi0h_widths_pttrig}.

\begin{table} [ht]
\caption{\label{Ta:pippmh_widths_pttrig}
Near and far-side widths as a function of
$p_{T,\rm{trig}}$ for charged pion triggers and associated charged
hadrons ($2-4.5$ GeV/$c$) from minimum-bias $d$ + Au~collisions.}
\begin{ruledtabular} \begin{tabular}{ccc}
$\mean{p_{T,\rm{trig}}}$ & $\sigma_{near}$ & $\sigma_{far}$ \\
(GeV/$c$) & (rad) & (rad) \\ \hline
 5.44 & $ 0.176 \pm 0.008 $ & $ 0.393 \pm 0.030 $ \\
        6.31 & $ 0.165 \pm 0.007 $ & $ 0.342 \pm 0.020 $ \\
        7.27 & $ 0.162 \pm 0.007 $ & $ 0.322 \pm 0.022 $ \\
        8.60 & $ 0.157 \pm 0.008 $ & $ 0.301 \pm 0.019 $ \\
        10.6 & $ 0.149 \pm 0.020 $ & $ 0.231 \pm 0.039 $ \\
        13.2 & $ 0.177 \pm 0.019 $ & $ 0.329 \pm 0.042 $ \\
\end{tabular} \end{ruledtabular}
\end{table}

\begin{table} [ht]
\caption{\label{Ta:pi0h_widths_pttrig}
Near and far-side widths as a function of
$p_{T,\rm{trig}}$ for neutral pion triggers and associated charged
hadrons ($2.5 - 5$ GeV/$c$) from minimum-bias $d$ +
Au~collisions.} 
\begin{ruledtabular} \begin{tabular}{ccc}
$\mean{p_{T,\rm{trig}}}$ & $\sigma_{near}$ & $\sigma_{far}$ \\
(GeV/$c$) & (rad) & (rad) \\ \hline
 5.39 & $ 0.207 \pm 0.007 $ & $ 0.409 \pm 0.025 $ \\
        6.40 & $ 0.151 \pm 0.008 $ & $ 0.304 \pm 0.034 $ \\
        7.66 & $ 0.144 \pm 0.012 $ & $ 0.295 \pm 0.035 $ \\
\end{tabular} \end{ruledtabular}
\end{table}

\begin{figure}[ht]
\includegraphics[width=1.0\linewidth]{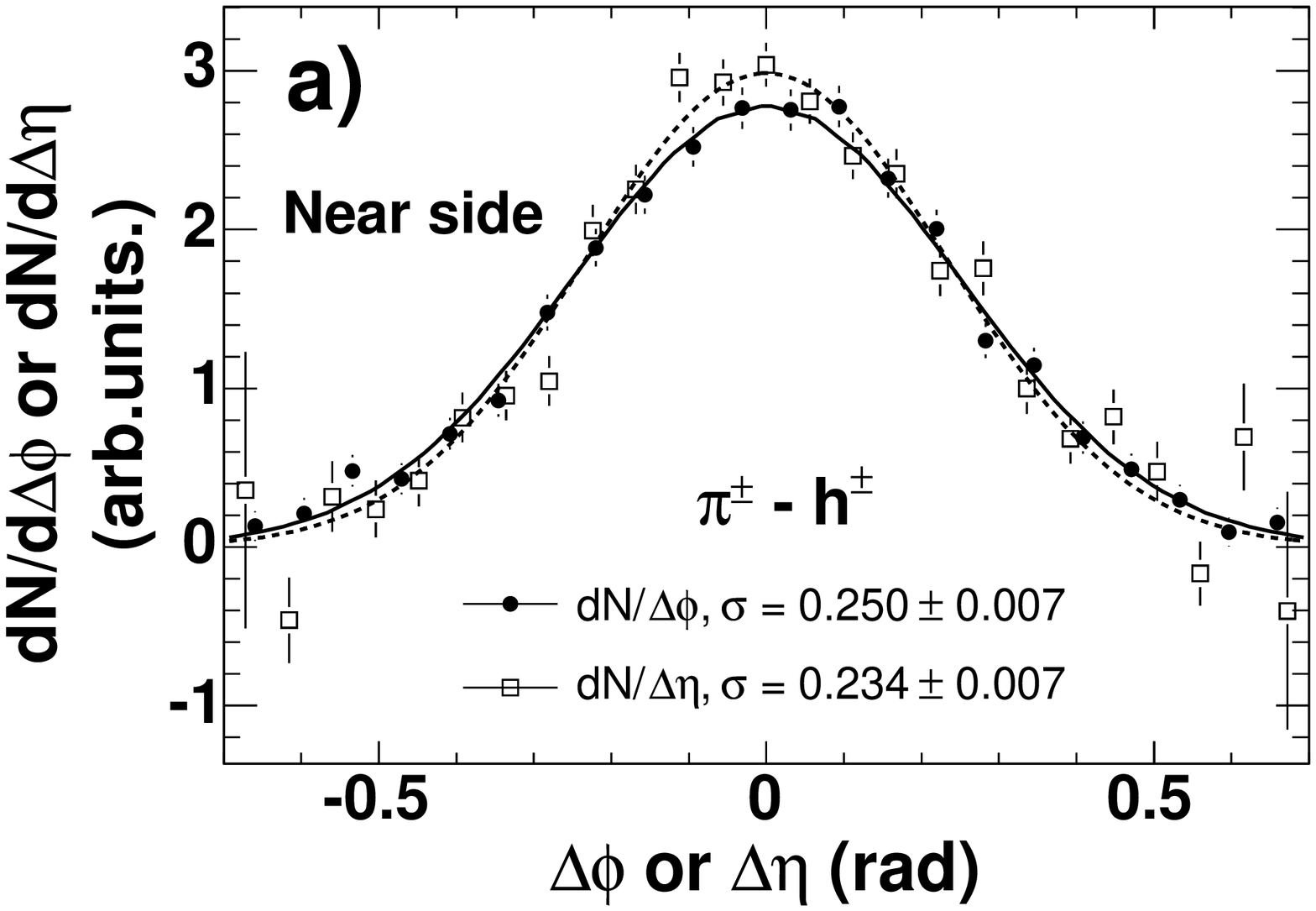}
\includegraphics[width=1.0\linewidth]{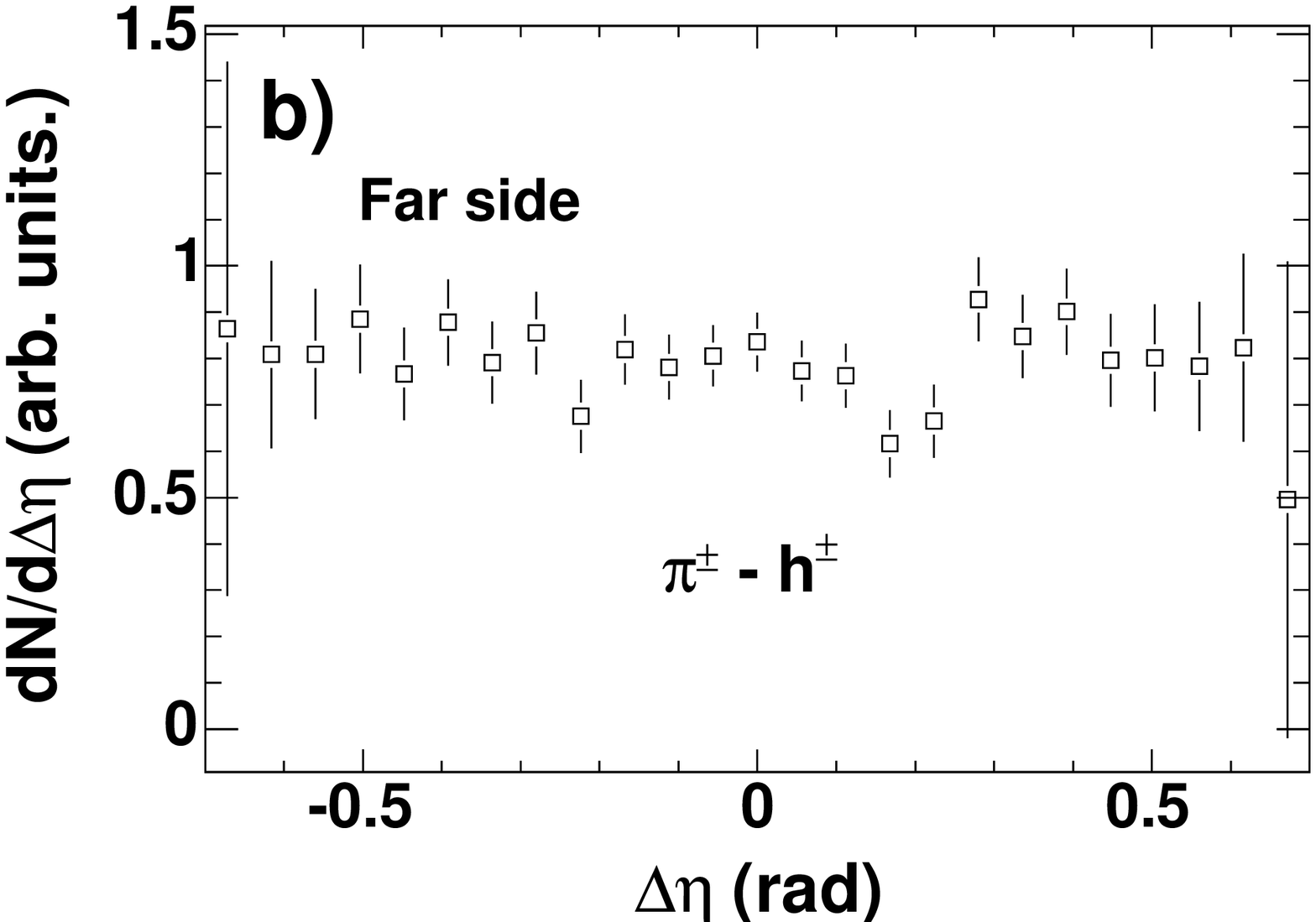}
\caption{\label{fig:jetwidth1} The jet shapes in $\Delta\phi$ and
$\Delta\eta$ from $\pi^{\pm}-h^{\pm}$ correlation with
$5<p_{T,\rm{trig}}<10$ GeV/$c$ and $1.0<p_{T,\rm{assoc}}<2.0$
GeV/$c$
from minimum-bias $d$ + Au~collisions;
a) The near-side jet shape in
$\Delta\phi$ (solid circles)
and $\Delta\eta$ (open boxes). b) The far-side jet shape in
$\Delta\eta$.}
\end{figure}

Although PHENIX single particle acceptance is limited to
$|\eta|<0.35$, it can sample jet pairs in twice as large of a
window in $\Delta\eta$ ($|\Delta\eta|<0.7$) with varying pair
efficiency. Similar to azimuthal correlation, the pair efficiency
in $\Delta\eta$ can be estimated via mixed-events and can
subsequently be divided out (Eq.~\ref{eq:cfeta}). Assuming that
the underlying event is flat\footnote{$\eta$ dependence of the
single particle yield is very weak in
$0<|\eta|<1$~\cite{Adams:2004da}. Thus the underlying pair
distribution in $|\Delta\eta|<0.7$ is almost flat. } in
$|\Delta\eta|<0.7$, we fix the background level to be equal to
that in azimuthal correlation function Eq.~\ref{E:assumpt} and
subsequently extract the jet distribution as a function of
$\Delta\eta$. In Fig.~\ref{fig:jetwidth1}a, we compare the
near-side jet shape in $\Delta\phi$ and $\Delta\eta$ in the
angular range of $|\Delta\phi,\Delta\eta|<0.7$ for the
$\pi^{\pm}-h^{\pm}$ correlation with $1.0<p_{T,\rm{assoc}}<2.0$
GeV/$c$. There is no significant difference in jet shape between
$\Delta\eta$ and $\Delta\phi$ and the widths are consistent in
both directions. Figure~\ref{fig:jetwidth1}b shows the far-side
jet shape in $\Delta\eta$, the associated pair distribution is
flat within $\pm10\%$.

\begin{figure}[ht]
\includegraphics[width=1.0\linewidth]{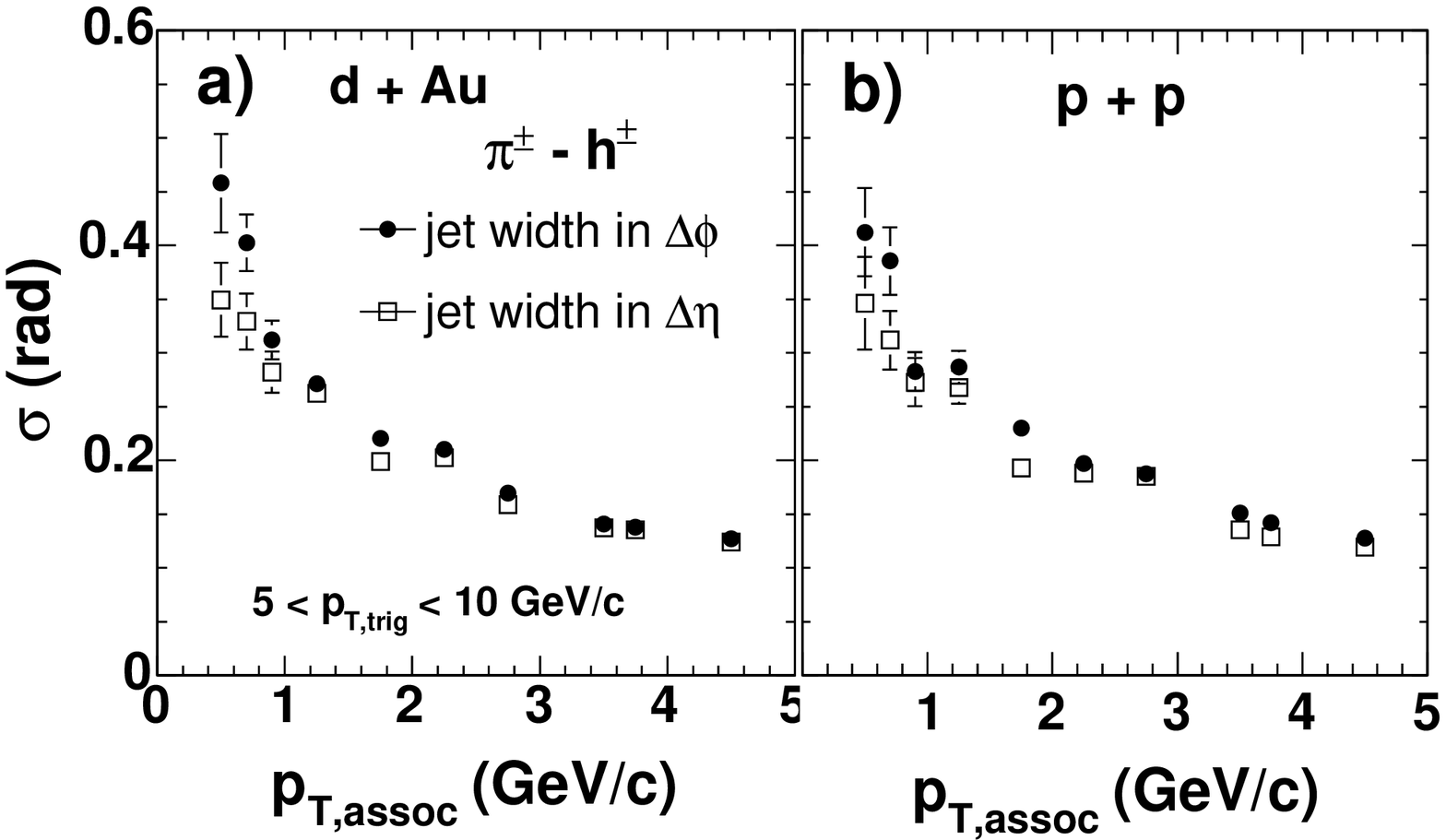}
\caption{\label{fig:jetwidth3} The comparison of jet width as
function of $p_{T,\rm{assoc}}$ in $\Delta\phi$ (solid circles) and
$\Delta\eta$ (open boxes) from $\pi^{\pm}-h^{\pm}$ correlation
with $5<p_{T,\rm{trig}}<10$ GeV/$c$. a) results for $d$ + Au. b)
results for $\pp$. Bars are statistical errors.}
\end{figure}

Figure~\ref{fig:jetwidth3}a shows the comparison of the near-side jet
width in $\Delta\phi$ and $\Delta\eta$ from $d$ + Au. There is
overall very good agreement between the two data sets. However,
the width in $\Delta\eta$ is systematically lower than that in
$\Delta\phi$ at small $p_{T,\rm{assoc}}$. This is due to the fact
that the the underlying background is not completely flat in
$\Delta\eta$, but varies by up to 10\% in $|\Delta\eta|<0.7$. Thus
the procedure of dividing by the mixed-event distribution 
(Eq.~\ref{eq:cfeta})
introduces some distortion of the jet shape at large $\Delta\eta$,
and consequently leads to a slightly different value for the jet
width. In fact for $\pp$ collisions, Figure~\ref{fig:jetwidth3}b
indicates a similar discrepancy between $\Delta\phi$ and
$\Delta\eta$ at small $p_{T,\rm{assoc}}$ for $p+p$ collisions.
Thus this deviation is not likely due to the cold medium effect in
$d$ + Au.

We extract not only the widths of the jet-structures but also the
conditional yields of how many hadrons are in the near-side and
far-side structures for each high-$p_T$ trigger. The conditional
yield defined in Eq.~\ref{eq:cond6} can be obtained from either a
correlation function or conditional pair distribution, both of
which produce identical results. For the conditional pair
distributions the conditional yield is directly extracted from the
fit parameters (Eq.~\ref{CY:1}), while for correlation functions
several normalization factors need to be applied to obtain the per
trigger yield~\cite{Stankus_RP,Ajit_Methods}, as described below.

For correlation functions, it is convenient to define the fraction
of jet-correlated particle pairs per event, $\frac{n_{\rm{jet\,
pair}}}{n_{\rm{total\,pair}}}$. Following the basic ansatz outlined in
Eq.~\ref{E:assumpt}, the fraction of jet-correlated particle pairs
is obtained by summing the jet function over all bins in $\Delta
\phi$ and dividing by the total sum of the correlation function

\begin{equation}\label{E:assumpt2}
\frac{n_{\rm{jet\,pair}}}{n_{\rm{total\,pair}}} = \frac{\sum A_{o}
J(\Delta \phi)}{\sum C(\Delta \phi)}.
\end{equation}

\begin{figure}[ht]
\includegraphics[width=1.0\linewidth]{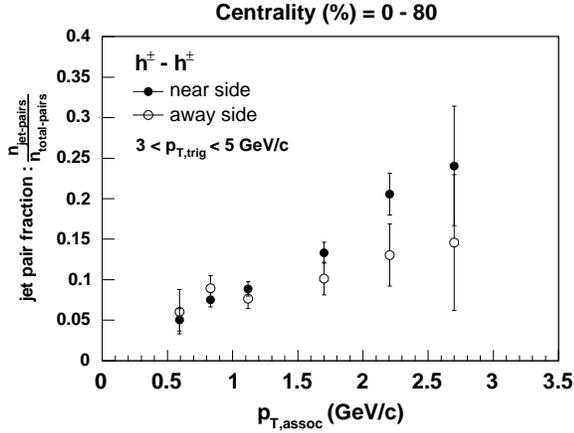}
\caption{Average jet pair fraction per event as a function of
$p_{T,\rm{assoc}}$. Results are shown for the trigger hadron
selection $3.0 < p_T < 5.0$ GeV/$c$
from $d$ + Au~collisions and a centrality of $0-80$\%.
Bars are statistical errors.} \label{Fi:exptCorrFn_fi03}
\end{figure}

Such pair fractions are shown as a function of $p_{T, \rm{assoc}}$
for a trigger hadron of $3.0 < p_T < 5.0$ GeV/$c$ and a centrality
selection of $0-80$\% in Fig.~\ref{Fi:exptCorrFn_fi03}. The
results, shown for both the near and far-side jets, indicate an
increase in the average fraction of jet-correlated particle pairs
with $p_T$~as might be expected if jet fragmentation becomes the
dominant particle production mechanism as $p_T$~is increased.

The pair-fraction is multiplied by the ratio
$\frac{n_{\rm{pairs}}}{n_{\rm{trig}} n_{\rm{assoc}}}$
\begin{equation}\label{E:exptCorrFn_e05}
\frac{n_{\rm{jet\,pair}}}{n_{\rm{trig}} n_{\rm{assoc}}} =
\frac{n_{\rm{jet\,pair}}} {n_{\rm{total\,pair}}} \times
\frac{n_{\rm{pairs}}}{n_{\rm{trig}} n_{\rm{assoc}}}
\end{equation}
where $n_{\rm{pairs}}$ denotes the average number of detected
particle pairs per event and $n_{\rm{trig}}$, and $n_{\rm{assoc}}$
are the detected single particle yields per event for trigger and
associated particles respectively. This gives the average number
of jet-correlated pairs per event over the combinatoric background
$\frac{n_{\rm{jet\,pair}}}{n_{\rm{trig}} n_{\rm{assoc}}}$. The
conditional per-trigger yield, $\frac{n_{\rm{jet
pair}}}{n_{\rm{trig}}}$, is obtained via multiplication by the
efficiency corrected single particle yield
($n_{\rm{assoc}}^{\rm{eff-corr.}}$) for the selected associated
$p_T$~bin of interest;
\begin{equation}\label{E:exptCorrFn_e06}
\frac{n_{\rm{jet\,pair}}}{n_{\rm{trig}}} = \frac{n_{\rm{jet
pair}}}{n_{\rm{trig}} n_{\rm{assoc}}} \times
n_{\rm{assoc}}^{\rm{eff-corr.}}.
\end{equation}
The per trigger yields for hadron triggers (found using 
Eq.~\ref{E:exptCorrFn_e06})
are corrected for the azimuthal acceptance and tracking efficiency 
but are reported within the PHENIX $\eta$ acceptance for the central arms,
{\it i.e.} no $R(\Delta \eta)$ corretions is applied to the hadron-triggered
conditional yields.

\begin{figure}[ht]
\includegraphics[width=1.0\linewidth]{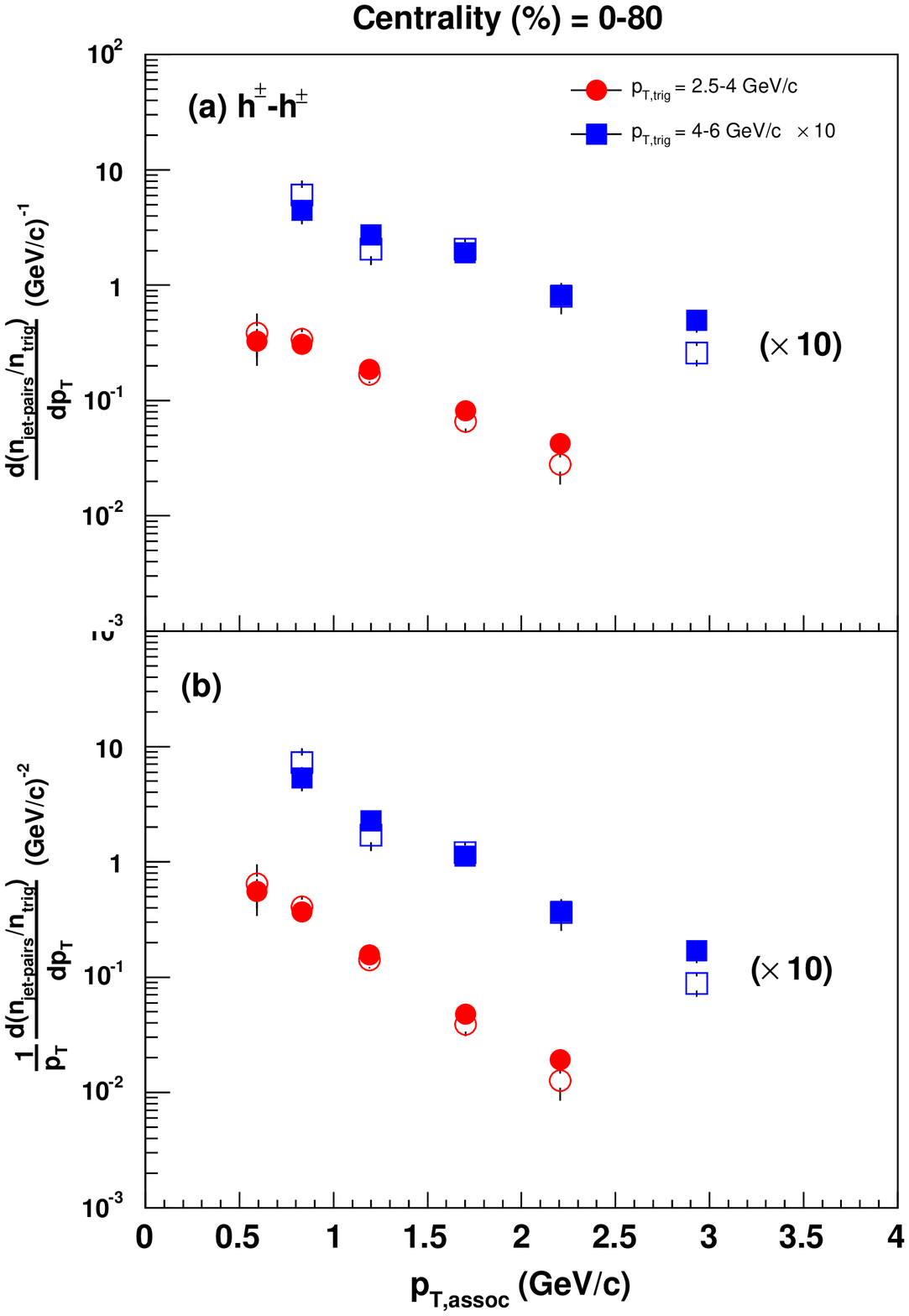}
\caption{(Color online) Per trigger yield (a) and invariant conditional yield (b)
as a function of $p_{T, \rm{assoc}}$ for  trigger hadron ranges of
$2.5 < p_T < 4.0$ GeV/$c$ and $4.0 < p_T < 6.0$ GeV/$c$,
respectively. The closed points are the near-side yields and the
open points are the far-side yields. The centrality range is
$0-80$\% in $d$ + Au~collisions. The yields are corrected for
efficiency and are reported in the PHENIX eta acceptance. Bars are statistical errors.}
\label{Fi:exptCorrFn_fi04}
\end{figure}

Figure~\ref{Fi:exptCorrFn_fi04} plots the near- and far-side
invariant conditional yields extracted via
Eq.~\ref{E:exptCorrFn_e06} for different trigger $p_T$~selections
as indicated. An approximate exponential decrease with $p_T$ is
observed, {\it i.e.} there are more low-$p_T$ particles associated
with each high-$p_T$ trigger hadron.
The data are tabulated in Tables~\ref{Ta:hh_yields_2.5-4} and
\ref{Ta:hh_yields_4-6}.

\begin{table} [ht]
\caption{\label{Ta:hh_yields_2.5-4}
Near and far-side conditional yields as a function of
$p_{T,\rm{assoc}}$ for charged hadron triggers ($2.5-4$ GeV/$c$)
and associated charged hadrons from $d$ + Au~collisions.}
\begin{ruledtabular} \begin{tabular}{ccc}
$\mean{p_{T,\rm{assoc}}}$ & $dN/dp_{T} near$  & $dN/dp_{T} far$ \\
(GeV/$c$) & & \\ \hline
  0.592 & $ 0.327 \pm 0.092 $ & $ 0.383 \pm  0.182 $ \\ 
  0.831 & $ 0.307 \pm  0.045 $ & $ 0.339 \pm  0.079 $ \\ 
  1.190 & $ 0.174 \pm 0.018 $ & $ 0.158 \pm  0.024 $  \\ 
  1.702 & $ 0.081 \pm 0.009 $ & $ 0.066 \pm  0.014 $ \\ 
  2.205 & $ 0.042 \pm 0.006 $ & $ 0.028 \pm  0.009 $ \\ 
\end{tabular} \end{ruledtabular}
\end{table}

\begin{table} [ht]
\caption{\label{Ta:hh_yields_4-6}
Near and far-side conditional yields as a function of
$p_{T,\rm{assoc}}$ for charged hadron triggers ($4-6$ GeV/$c$)
and associated charged hadrons from $d$ + Au~collisions.}
\begin{ruledtabular} \begin{tabular}{ccc}
$\mean{p_{T,\rm{assoc}}}$ & $dN/dp_{T} near$  & $dN/dp_{T} far$ \\
(GeV/$c$) & & \\ \hline
  0.831 & $ 4.437 \pm 1.040 $ & $ 6.031 \pm  2.010 $\\ 
  1.200 & $ 2.725 \pm 0.506 $ & $ 2.051 \pm  0.562 $\\ 
  1.700 & $ 1.907 \pm 0.278 $ & $ 2.046 \pm  0.447 $\\ 
  2.210 & $ 0.819 \pm 0.152 $ & $ 0.804 \pm  0.244 $\\ 
  2.931 & $ 0.497 \pm 0.107 $ & $ 0.258 \pm  0.061 $\\ 
\end{tabular} \end{ruledtabular}
\end{table}

\begin{figure}[ht]
\begin{center}
\includegraphics[width=1.0\linewidth]{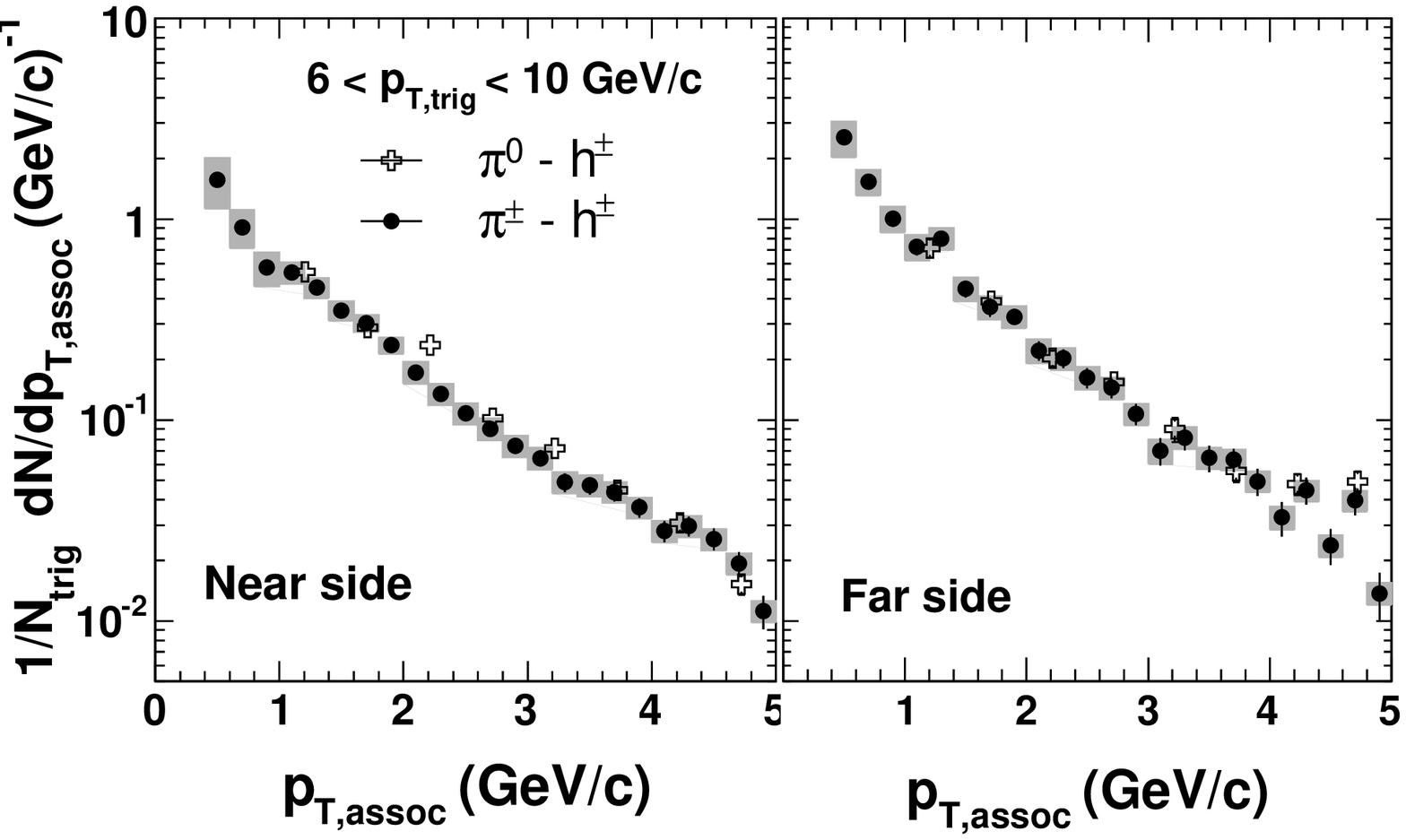}
\caption{\label{fig:mbdaucomppt} Conditional yield as function of
$p_{T,\rm{assoc}}$ for near-side (left panel) and far-side (right
panel) correlation from minimum-bias $d$ + Au~collisions. Bars are statistical errors. The shaded boxes represent the total systematic errors on each point.}
\end{center}
\end{figure}

In Fig.~\ref{fig:mbdaucomppt} the conditional yields for
identified pion triggers are plotted as function of
$p_{T,\rm{assoc}}$ for both near-side correlation and far-side
correlation. For this high-$p_T$ data, the conditional yields are
extracted from the fits to the data in
Figs.~\ref{Fi:exptCorrFn_distro_fi01} and
\ref{Fi:exptCorrFn_distro_fi02} using Eq.~\ref{CY:1}, then corrected
for pair efficiency in $\Delta\eta$ using Eq.~\ref{eq:cy1d6} and
Eq.~\ref{eq:cy1d7}. The conditional yields are tabulated in
Tables~\ref{Ta:jj_yields} and \ref{Ta:nathan_yields}.

\begin{table} [ht]
\caption{\label{Ta:jj_yields}
Near and far-side conditional yields as a function of
$p_{T,\rm{assoc}}$ for charged pion triggers ($5-10$ GeV/$c$)
and associated charged hadrons from minimum-bias $d$ + Au~collisions.}
\begin{ruledtabular} \begin{tabular}{ccc}
$\mean{p_{T,\rm{assoc}}}$ & $dN/dp_{T} near$  & $dN/dp_{T} far$ \\
(GeV/$c$) & & \\  \hline
0.5 & $ 1.57 \pm 0.083 $&$ 2.54 \pm 0.15 $\\ 
0.7 & $ 0.911 \pm 0.049 $&$ 1.53 \pm 0.13 $\\ 
0.9 & $ 0.574 \pm 0.031 $&$ 1.00 \pm 0.087 $\\ 
1.1 & $ 0.542 \pm 0.032 $&$ 0.727 \pm 0.068 $\\ 
1.3 & $ 0.456 \pm 0.026 $&$ 0.801 \pm 0.058 $\\ 
1.5 & $ 0.351 \pm 0.022 $&$ 0.451 \pm 0.044 $\\ 
1.7 & $ 0.303 \pm 0.018 $&$ 0.365 \pm 0.038 $\\ 
1.9 & $ 0.235 \pm 0.015 $&$ 0.327 \pm 0.031 $\\ 
2.1 & $ 0.172 \pm 0.012  $&$ 0.222 \pm 0.025 $\\ 
2.3 & $ 0.135 \pm 0.010 $&$ 0.203 \pm 0.022 $\\ 
2.5 & $ 0.108 \pm 0.008 $&$ 0.162 \pm 0.018 $\\ 
2.7 & $ 0.0905 \pm 0.0075 $&$ 0.145 \pm 0.017 $\\ 
2.9 & $ 0.0742 \pm 0.0064 $&$ 0.107 \pm 0.013 $\\ 
3.1 & $ 0.0645 \pm 0.0059 $&$ 0.070 \pm 0.011 $\\ 
3.3 & $ 0.0490 \pm 0.0052  $&$ 0.0819 \pm 0.0113 $\\ 
3.5 & $ 0.0473 \pm 0.0047 $&$ 0.0647 \pm 0.0097 $\\ 
3.7 & $ 0.0439 \pm 0.0045 $&$ 0.0636 \pm 0.0084 $\\ 
3.9 & $ 0.0367 \pm 0.0040 $&$ 0.0495 \pm 0.0075 $\\ 
4.1 & $ 0.0281 \pm 0.0034 $&$ 0.0327 \pm 0.0064 $\\ 
4.3 & $ 0.0297 \pm 0.0034  $&$ 0.0446 \pm 0.0068 $\\ 
4.5 & $ 0.0256 \pm 0.0031 $&$ 0.0238 \pm 0.0048 $\\ 
4.7 & $ 0.0192 \pm 0.0027 $&$ 0.0397 \pm 0.0061 $\\ 
4.9 & $ 0.0112 \pm 0.0021 $&$ 0.0137 \pm 0.0036 $\\ 
\end{tabular} \end{ruledtabular}
\end{table}

\begin{table} [ht]

\caption{\label{Ta:nathan_yields}
Near and far-side conditional yields as a function of
$p_{T,\rm{assoc}}$ for neutral pion triggers ($5-10$ GeV/$c$)
and associated charged hadrons from minimum-bias $d$ + Au~collisions.}
\begin{ruledtabular} \begin{tabular}{ccc}
$\mean{p_{T,\rm{assoc}}}$& $dN/dp_{T} near$  & $dN/dp_{T} far$ \\
(GeV/$c$) & & \\ \hline
1.21 & $ 0.545 \pm 0.035 $ & $ 0.718\pm 0.076 $ \\ 
1.71 & $ 0.289 \pm 0.014 $ & $ 0.389 \pm 0.030 $\\ 
2.21 & $ 0.236 \pm 0.009 $ & $ 0.203 \pm 0.018 $ \\ 
2.72 & $ 0.102 \pm 0.006 $ & $ 0.155 \pm 0.012 $ \\ 
3.22 & $ 0.0724 \pm 0.0049 $ & $ 0.090 \pm 0.012 $ \\ 
3.73 & $ 0.0448 \pm 0.0039 $ & $ 0.0560 \pm 0.0071 $ \\ 
4.23 & $ 0.0308 \pm 0.0029 $ & $ 0.0481 \pm 0.0058 $ \\ 
4.72 & $ 0.0152 \pm 0.0017 $ & $ 0.0495 \pm 0.0060 $ \\ 
\end{tabular} \end{ruledtabular}
\end{table}

The agreement between the two pion-triggered
data sets is good, which indicates that the jet fragmentation
function is independent of whether a neutral pion or a charged
pion trigger is used. The difference in the magnitudes of far-side
and near-side yield reflect the fact that the far-side
correlations measures a hadron triggered effective fragmentation
while the near-side correlation measures dihadron fragmentation.

The conditional yields presented in FigOAs.~\ref{Fi:exptCorrFn_fi04}
and  \ref{fig:mbdaucomppt} can be considered as the basic
information, while the near
and far $dN/dx_E$ distributions in
Section~\ref{sec:resultsMinBias} have a closer relationship
to parton fragmentation functions as was described in
Section~\ref{sec:jets}.

\begin{figure}[ht]
\includegraphics[width=1.0\linewidth]{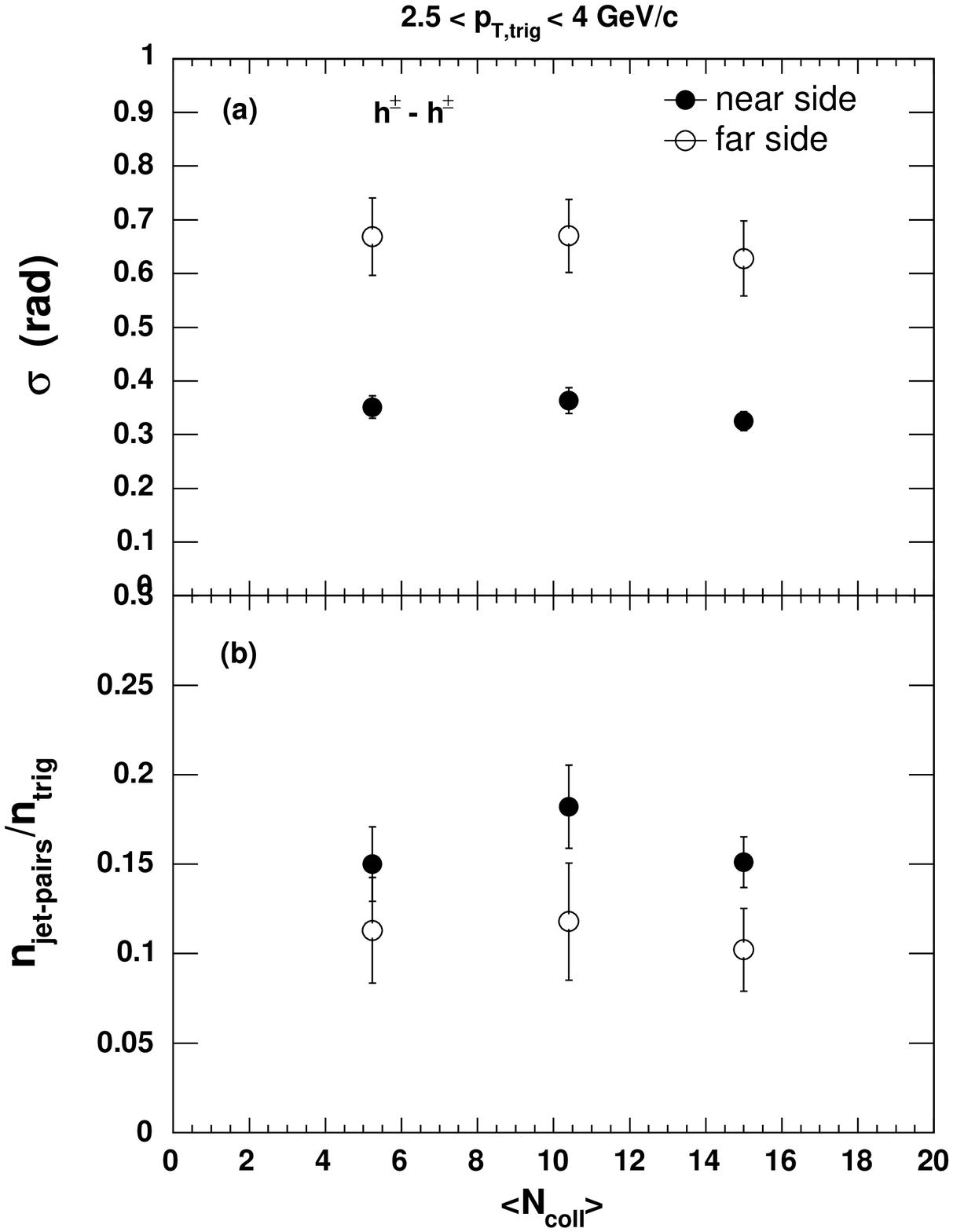}
\caption{Near- and far-side widths as well as conditional yields
in the PHENIX pseudorapidity acceptance for several centrality selections
from $d$ + Au~collisions. Results are shown for the trigger
hadron selection $2.5
< p_T < 4.0$ GeV/$c$ and the associated $p_T$ range of $1 < p_{T}
< 2.5$ GeV/$c$. Bars are statistical errors.} \label{Fi:CentRaw}
\end{figure}
\begin{figure}[ht]
\includegraphics[width=1.0\linewidth]{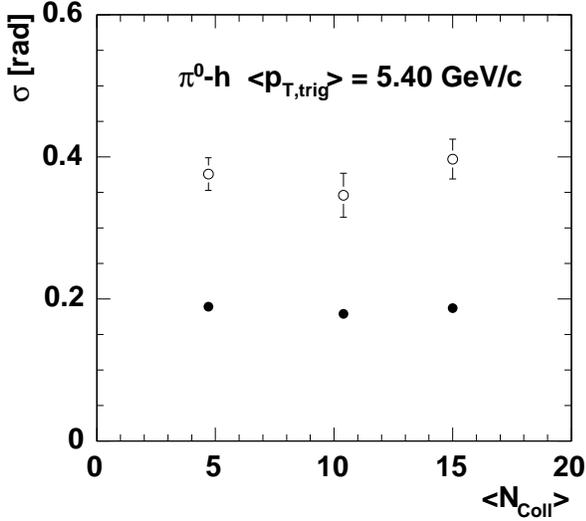}
\caption{Near- and far-side widths for several centrality selections
from $d$ + Au~collisions.
Results are shown for $\pi^0 - h$ correlations
with the trigger $\pi^0$ $5 < p_T < 10$ GeV/$c$ and the
associated $p_T$ range of $2 < p_{T} < 3$ GeV/$c$. Bars are statistical errors.} \label{Fi:CentPi0}
\end{figure}

Since multiple scattering should increase with centrality, we examine
whether these jet characteristics exhibit any centrality
dependence. Figure~\ref{Fi:CentRaw} reports on widths and
conditional yields in the PHENIX $\eta$ acceptance for
dihadron correlations with a trigger hadron in the range
$2.5<p_{T,\rm{trig}}<4$ GeV/$c$. The data are tabulated in
Table~\ref{Ta:hh_widths_cent} and \ref{Ta:hh_yields_cent}.
Centrality dependent widths for
the $\pi^0 - h^{\pm}$ correlations are shown in Fig. \ref{Fi:CentPi0}
for a mean $\pi^0$\,$p_T$ of approximately 5.4 GeV/$c$.
The data for the widths are tabulated in Table~\ref{Ta:pi0h_widths_cent}
while the yields can be found in Table~\ref{Ta:pi0h_yields_cent}.

\begin{table} [ht]
\caption{\label{Ta:hh_widths_cent}
Near and far-side widths as a function of N$_{coll}$
for charged hadron triggers ($2.5-4$ GeV/$c$)
and associated charged hadrons (1-2.5 GeV/$c$) from $d$ + Au~collisions.}
\begin{ruledtabular} \begin{tabular}{ccc}
Centrality & $\sigma_{near} (rad) $ & $\sigma_{far} (rad)$ \\ \hline
0-20 \% & $0.351 \pm 0.021$ & $0.669 \pm 0.072$ \\  
20-40 \% & $0.364 \pm 0.024$ & $0.670 \pm 0.068$ \\  
40-80 \% & $0.325 \pm 0.018$ & $0.628 \pm 0.070$ \\  
\end{tabular} \end{ruledtabular}
\end{table}

\begin{table} [ht]
\caption{\label{Ta:hh_yields_cent}
Near and far-side conditional yields as a function of N$_{coll}$
for charged hadron triggers ($2.5-4$ GeV/$c$)
and associated charged hadrons (1-2.5 GeV/$c$) from $d$ + Au~collisions.}
\begin{ruledtabular} \begin{tabular}{ccc}
Centrality & $dN/dp_{T} near$  & $dN/dp_{T} far$ \\ \hline
0-20 \% & $ 0.150 \pm 0.021 $ & $0.113  \pm  0.030  $ \\  
20-40 \% & $ 0.182 \pm 0.023 $ & $ 0.118  \pm  0.033 $ \\  
40-80 \% & $ 0.151 \pm 0.014  $ & $ 0.102  \pm  0.023 $ \\  
\end{tabular} \end{ruledtabular}
\end{table}

\begin{table} [ht]
\caption{\label{Ta:pi0h_widths_cent}
Near and far-side widths as a function of centrality
for neutral pion triggers ($5-10$ GeV/$c$)
and associated charged hadrons(2-3 GeV/$c$)
from $d$ + Au~collisions.}
\begin{ruledtabular} \begin{tabular}{ccc}
Centrality & $\sigma_{near} (rad)$ & $\sigma_{far} (rad)$ \\ \hline
0-20 \% &  $0.199 \pm 0.009$ & $0.387 \pm 0.024 $ \\  
20-40 \% & $0.195 \pm 0.009$ & $0.401 \pm 0.031$ \\  
40-88 \% & $0.190 \pm 0.008$ & $0.376 \pm 0.024 $ \\  
\end{tabular} \end{ruledtabular}
\end{table}

\begin{table} [ht]
\caption{\label{Ta:pi0h_yields_cent}
Near and far-side conditional yields as a function of centrality
from $d$ + Au~collisions
for neutral pion triggers ($5-10$ GeV/$c$)
and associated charged hadrons(2-3 GeV/$c$).}
\begin{ruledtabular} \begin{tabular}{ccc}
Centrality & $dN/dp_{T} near$  & $dN/dp_{T} far$ \\ \hline
0-20 \% &  $0.0816 \pm 0.0037$ & $0.116 \pm 0.008 $ \\  
20-40 \% & $0.0947 \pm 0.0045$ & $0.141 \pm 0.011$ \\  
40-88 \% & $0.0967 \pm 0.0043$ & $0.144 \pm 0.009 $ \\  
\end{tabular} \end{ruledtabular}
\end{table}

Neither the
widths nor the per-trigger yields change significantly with
centrality, indicating that the influence of multiple scattering
on jet properties is small in this region.
This work is extended in Section~\ref{sec:resultsCentrality} where we present
the centrality dependence of various jet-structure observables.

\subsection{Jet Properties in Minimum Bias $\dA$ Collisions\label{sec:resultsMinBias}}

From the angular widths and yields in the previous section, we
calculate the following quantities that characterize the jet
structures: transverse momentum of hadrons with respect to the
jet ($j_T$), the dijet acoplanarity $\langle \sin^2(\phi_{jj})\rangle$,
and the
$dN/dx_{E}$ distributions. These quantities are first presented
for minimum-bias $d$ + Au~collisions, which have the highest
statistical precision, and are then compared with results from
$p+p$ in Section~\ref{sec:resultsDAuvspp}.

\begin{figure}[t]
\begin{center}
\includegraphics[width=1.0\linewidth]{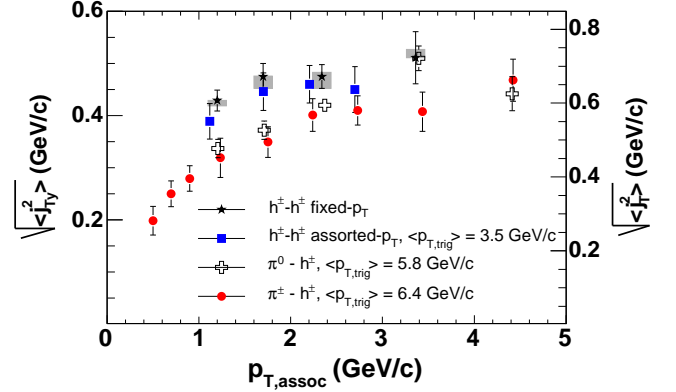}
\caption{\label{fig:mbdaujt} (Color online) the extracted
$\sqrt{\langle j_T^2 \rangle}$ for minimum bias $d$ + Au
collisions
from all four dihadron correlations. The trigger
$p_T$ ranges are $3-5$ GeV/$c$, $5-10$ GeV/$c$ and $5-10$ GeV/$c$
for $h^{\pm}-h^{\pm}$ (filled boxes), $\pi^{0}-h^{\pm}$ (open
crosses) and $\pi^{\pm}-h^{\pm}$ (filled circles) assorted-$p_T$
correlations. The $h^{\pm}-h^{\pm}$ fixed-$p_T$ correlation is
shown by filled stars. The statistical and systematic errors are
combined for the assorted-$p_T$ correlations, while the
fixed-$p_T$ correlation shows the systematic error separately as
shaded boxes.}
\end{center}
\end{figure}

Figure~\ref{fig:mbdaujt} shows the compilation of
$\sqrt{\langle j_T^2 \rangle}$ values extracted from $\pi^{0}-h$
correlations (open crosses), $\pi^{\pm}-h$ correlations (filled
circles), and assorted $p_T$ and fixed $p_T$ correlation results
for $h-h$ at low $p_T$ shown in solid boxes and solid stars,
respectively. The $\sqrt{\langle j_T^2 \rangle}$ values were calculated
using Eq.~\ref{eq:jt3}. The systematic errors are mainly due to
the uncertainties of Eq.~\ref{eq:jt3} and the fitting procedure,
which are about 5\% (independent of $p_T$) and are approximately the
same for all four analysis. The $\sqrt{\langle j_T^2 \rangle}$ values for
$\pi^{\pm}-h$ and $\pi^{0}-h$ indicate a steady increase at
$p_T$ below 2~GeV/$c$ followed by a saturation around $560-640$ MeV/$c$
at $p_T>2$ GeV/$c$. The initial increase is due to the `seagull
effect'~\cite{Apeldoorn:1975,seagulleffect:1994}. The approximately constant
behavior of $\sqrt{\langle j_T^2 \rangle}$ above 2 GeV/$c$ is consistent
with the scaling behavior of the fragmentation functions. The
$\sqrt{\langle j_T^2 \rangle}$ results from the $h-h$ correlation analyses
have a similar increase and saturation behavior. They seem to
reach a slightly higher plateau value at a lower $p_T$ (around 1.5
GeV/$c$). 

A combined fit based on $\sqrt{\langle j_T^2 \rangle}$ data
points at $p_T>2$ GeV/$c$ gives a plateau value of
$\sqrt{\langle j_T^2 \rangle}=
0.64\pm0.02\textrm{(stat)}\pm0.04\textrm{(sys)}$ GeV/$c$ for
minimum bias $d$ + Au
collisions.

\begin{figure}[t]
\begin{center}
\includegraphics[width=1.0\linewidth]{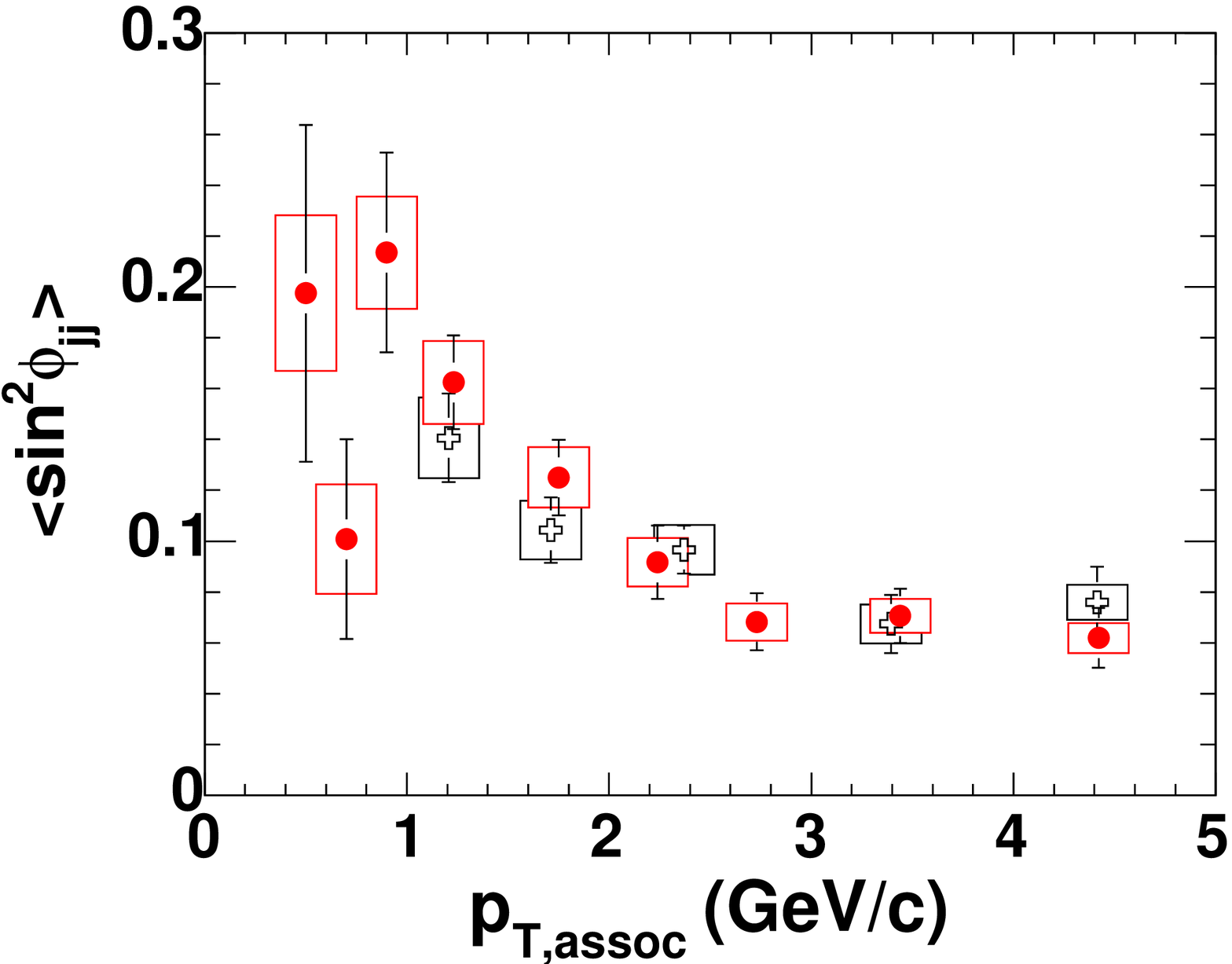}
\caption{\label{fig:sinsqvsPtassoc} (Color online)
$\left<\sin^{2}\phi_{jj}\right>$ for minimum bias $d$ + Au
collisions as function of
associated particle $p_T$
for $\pi^{\pm}-h^{\pm}$ (filled circles) and $\pi^{0}-h^{\pm}$
(open crosses) correlations where
the trigger particles have a $p_T$ between $5-10$ GeV/$c$.
The data points are plotted at the mean
$p_{T,\rm{assoc}}$.  Bars are statistical errors. The boxes represent the total systematic errors on each point.}
\end{center}
\end{figure}

\begin{figure}[t]
\begin{center}
\includegraphics[width=1.0\linewidth]{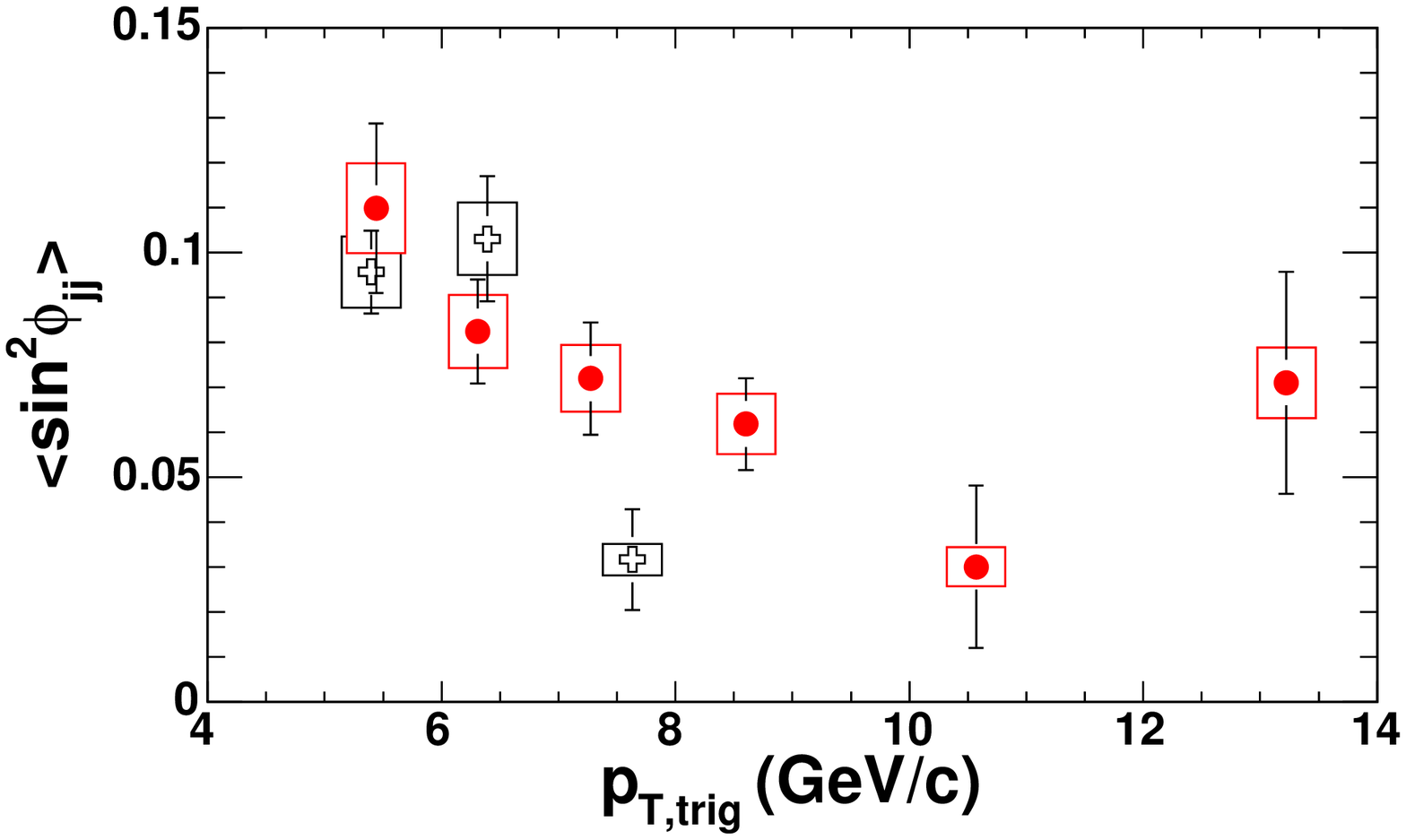}
\caption{\label{fig:sinsqvsPttrig} (Color online)
$\left<\sin^{2}\phi_{jj}\right>$ for minimum bias $d$ + Au
collisions as function of
trigger particle $p_T$
for $\pi^{\pm}-h^{\pm}$ (filled circles) and $\pi^{0}-h^{\pm}$
(open crosses) correlations. The data points are plotted at the mean
$p_{T,\rm{trig}}$. Bars are statistical errors. The boxes represent the total systematic errors on each point.}
\end{center}
\end{figure}

A key quantity that provides information on multiple scattering in
the cold nuclear medium is ${\langle \sin^2(\phi_{jj})\rangle}$,
where $\phi_{jj}$ is the azimuthal angle between the jet axes.
As described in Section~\ref{sec:jets},
we calculate $\left<\sin^{2}\phi_{jj}\right>$
from the
experimental values of the near- and far-side widths (Eq.~\ref{eq:LHSRHS}).
Figure~\ref{fig:sinsqvsPtassoc} shows
$\left<\sin^{2}\phi_{jj}\right>$ as function of
$p_{T,\rm{assoc}}$ for high-$p_T$ pion triggers.
We observe that the RMS of the sine of the 
angle between the jet axes, $\left<\sin^{2}\phi_{jj}\right>$,
decreases as higher-$p_T$ associated particles are selected.
There is good
agreement between the two data sets.
Figure \ref{fig:sinsqvsPttrig} plots $\left<\sin^{2}\phi_{jj}\right>$
as a function of $p_{T,trig}$, where a similar decrease with $p_T$ is observed.

In the next section
we will calculate
the quadrature difference $\left<\sin^{2}\phi_{jj}\right>$
between $d$+Au~and $p+p$ collisions.
and use that to quantify the affect of additional
final-state scattering in $d$+Au~collisions.

\begin{figure}[t]
\begin{center}
\includegraphics[width=1.0\linewidth]{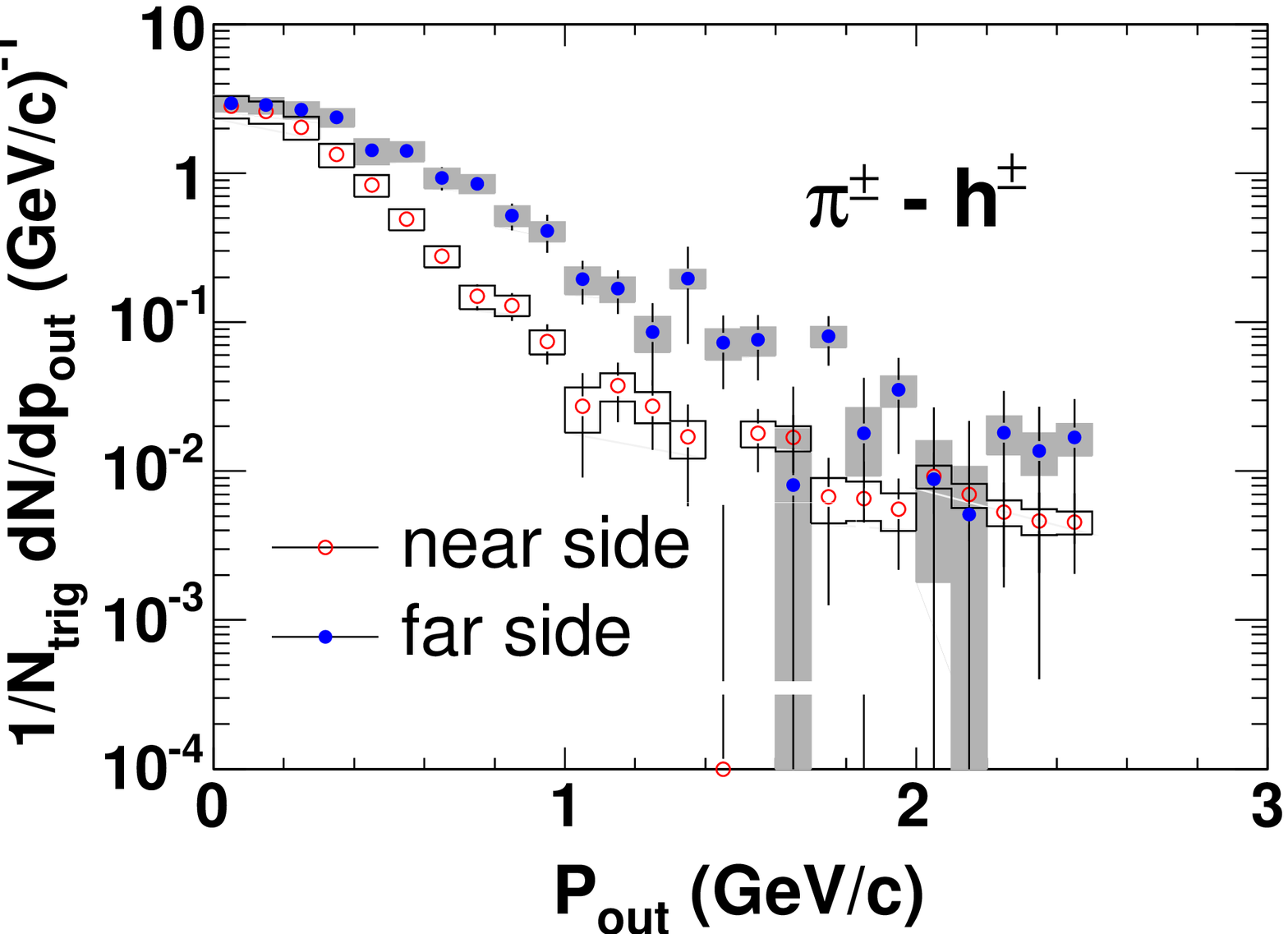}
\caption{\label{fig:mbdaupout} (Color online) Near-side and
far-side $p_{\rm{out}}$ distributions for minimum bias $d$ + Au
collisions obtained from $\pi^{\pm}-h^{\pm}$ correlation. The
trigger is $5<p_{T,\rm{trig}}<10$ GeV/$c$, the associated particle
is $0.5<p_{T,\rm{assoc}}<5.0$ GeV/$c$. Bars are statistical errors. The boxes represent the total systematic errors on each point.}
\end{center}
\end{figure}

In Section~\ref{sec:jets} we defined
the near-side and
far-side $p_{\rm{out}}$.
With this observable, it is possible to move beyond calculating means or
RMS values and hence the $p_{\rm{out}}$ distribution potentially
carries more information about the dijet acoplanarity. The
measured $p_{\rm{out}}$ distributions for $\pi^{\pm}-h^{\pm}$ are
shown in Fig.~\ref{fig:mbdaupout} for the near-side and far-side.
The far-side $p_{\rm{out}}$ has a broader distribution than the
near-side $p_{\rm{out}}$, reflecting the fact that $k_T$ is larger
than $j_T$. The $p_{\rm{out}}$ distributions have a power law
tail, possibly due to strong radiative processes driving large
values of $p_{\rm{out}}$.

\begin{figure}[t]
\begin{center}
\includegraphics[width=1.0\linewidth]{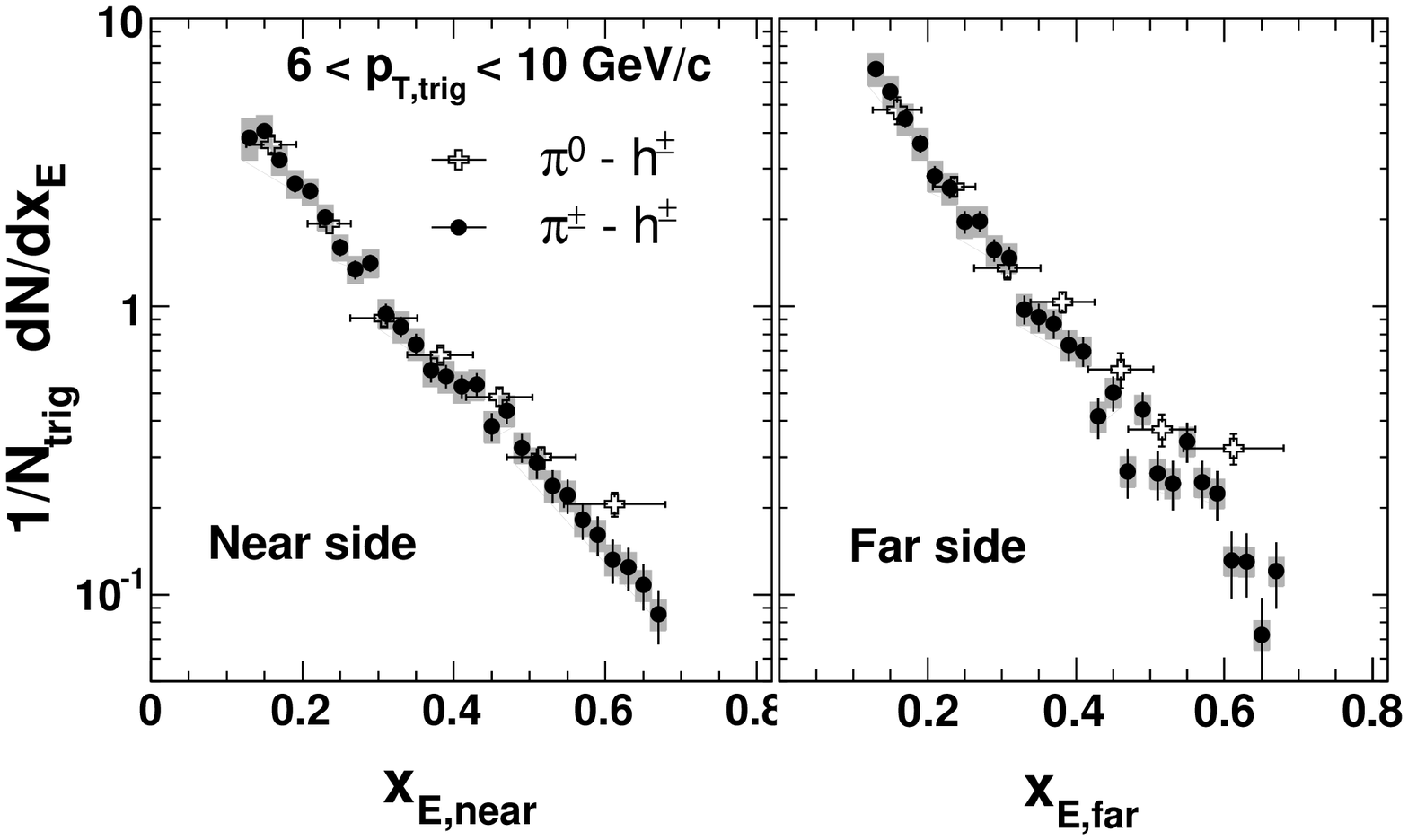}
\caption{\label{fig:mbdauxe} Conditional yield as a function of
$x_E$ for near-side (left panel) and far-side (right panel) for
$\pi^{\pm}-h^{\pm}$ (filled circles) and $\pi^{0}-h^{\pm}$ (open
crosses) from minimum-bias $d$ + Au~collisions. Bars are statistical errors. The boxes represent the total systematic errors on each point.}
\end{center}
\end{figure}

In Section \ref{sec:resultsCorrFn} we reported the yields of
associated hadrons per trigger particle, or the conditional yield. A
more comprehensive way of quantifying the fragmentation function is to plot
the conditional yields as a function of $x_E$. This is shown in
Fig.~\ref{fig:mbdauxe} for $\pi^{\pm}-h$ (solid circles) and
$\pi^0-h$ (open crosses).

\begin{figure}[t]
\begin{center}
\includegraphics[width=1.0\linewidth]{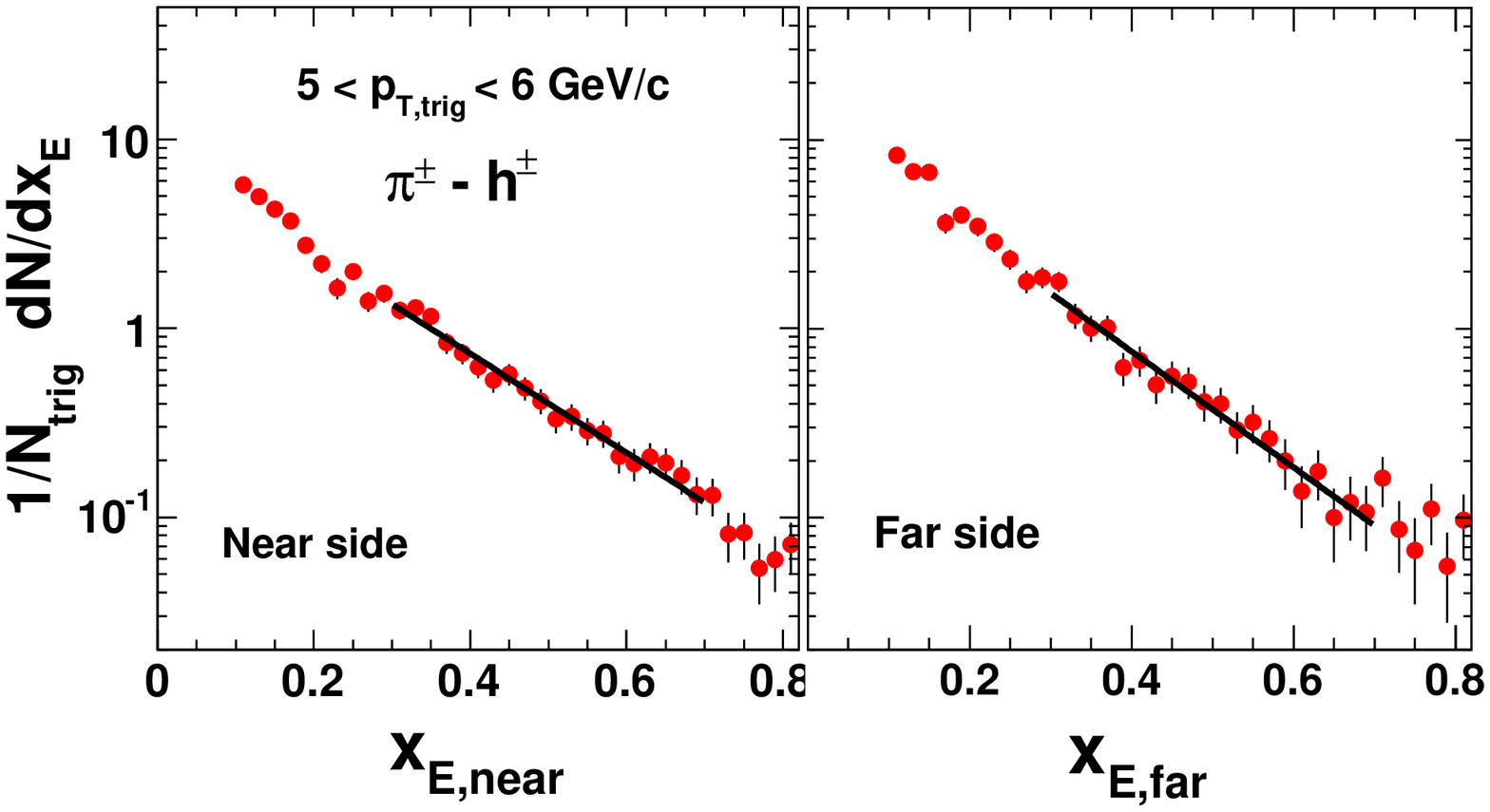}
\caption{\label{fig:mbdauxe2} (Color online) Conditional yield as
a function of $x_E$ for near-side (left panel) and far-side (right
panel) correlations for $\pi^{\pm}-h^{\pm}$ correlations from
minimum-bias $d$ + Au collisions. The trigger pions are $5 <
p_{T,\rm{trig}} < 6$ GeV/$c$, and the black lines are fits to
an exponential function for $0.3 < x_E < 0.7$. Bars are statistical errors.}
\end{center}
\end{figure}

Previously, in ISR experiments, the slope of the $x_E$
distribution has been determined to be around 5.3
(GeV/$c$)$^{-1}$~\cite{jet:ccor2}. In Fig.~\ref{fig:mbdauxe2}, the
conditional yields as a function of $x_E$ are plotted for trigger
$p_T$ range of $5-6$ GeV/$c$. In order to compare data with the
previous ISR results, we have determined the exponential inverse
slope $0.3<x_E<0.7$ and obtain the inverse slope parameter of
$6.0\pm0.3$ (GeV/$c$)$^{-1}$ in the near-side and $7.1\pm0.5$
(GeV/$c$)$^{-1}$ in the far-side. The near-side $x_E$ inverse
slope is smaller than that for the far-side, reflecting the
difference between dihadron fragmentation and single hadron
fragmentation. By requiring a trigger particle on the near-side,
one reduces the jet energy available for fragmenting to the second
hadrons and consequently a smaller inverse slope for the
near-side. Note, however, that the $x_E$ distributions do not have
pure exponential shape, and the fitted inverse slope is sensitive
to the fitting ranges.

\begin{figure}[t]
\begin{center}
\includegraphics[width=1.0\linewidth]{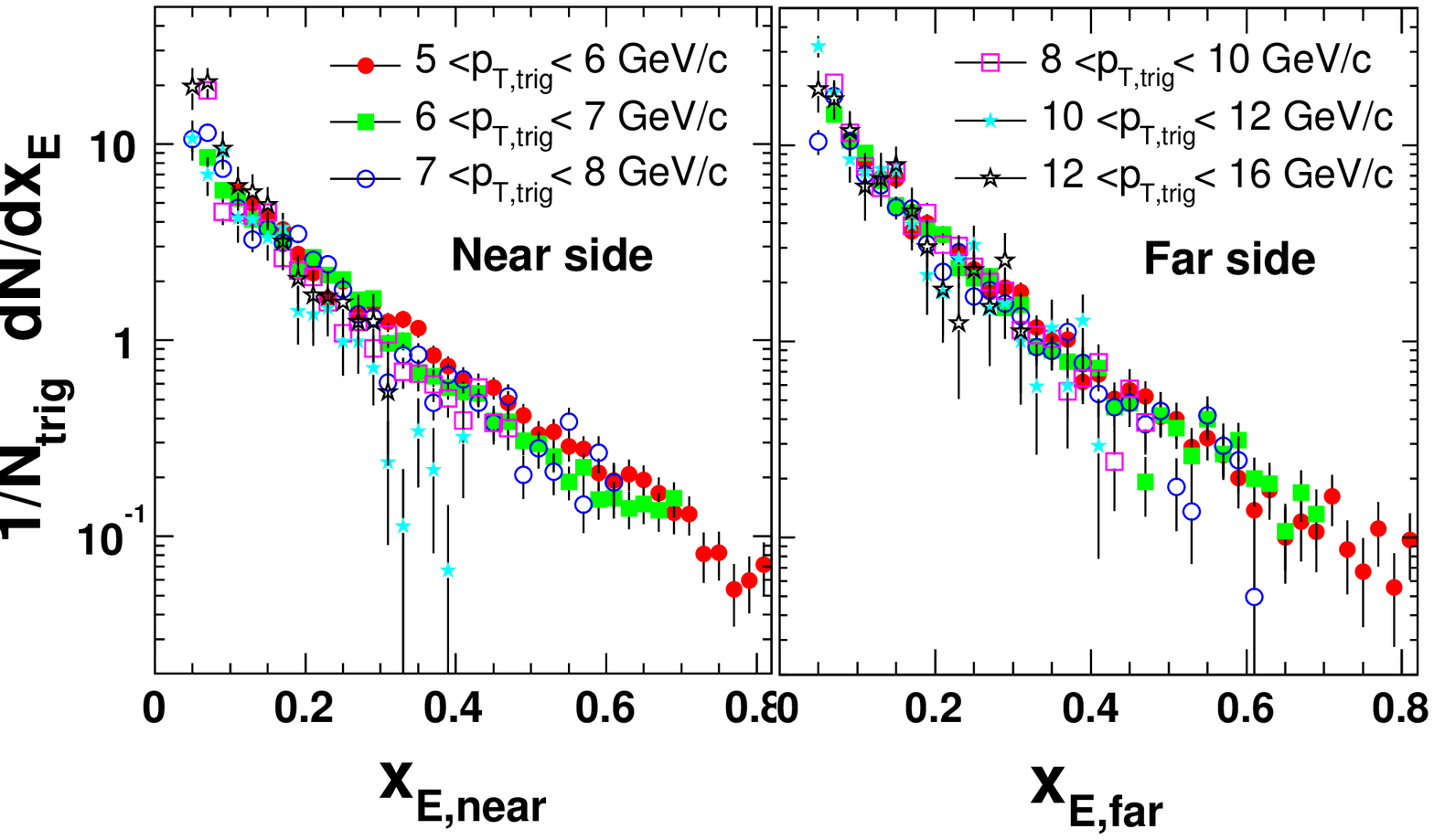}
\caption{\label{fig:mbdauxe1} (Color online) Conditional yield as
a function of $x_E$ for near-side (left panel) and far-side (right
panel) correlation for $\pi^{\pm}-h^{\pm}$ correlation for several
different trigger $p_T$'s for minimum-bias $d$ + Au~collisions. Bars are statistical errors.}
\end{center}
\end{figure}

It is well known that fragmentation functions $D(z)$ approximately
scale in $e^+e^-$ or $p+p$ collisions, {\it i.e.} are independent
of jet energy. To check whether this is still true in $d$ + Au
collisions, we plot in Fig.~\ref{fig:mbdauxe1} the conditional
yields as a function of $x_E$ for different ranges of trigger $p_T$
from $\pi^{\pm}-h^{\pm}$ correlations. All curves fall on top of
each other, indicating a universal behavior of the jet
fragmentation function.

\begin{figure}[t]
\begin{center}
\includegraphics[width=1.0\linewidth]{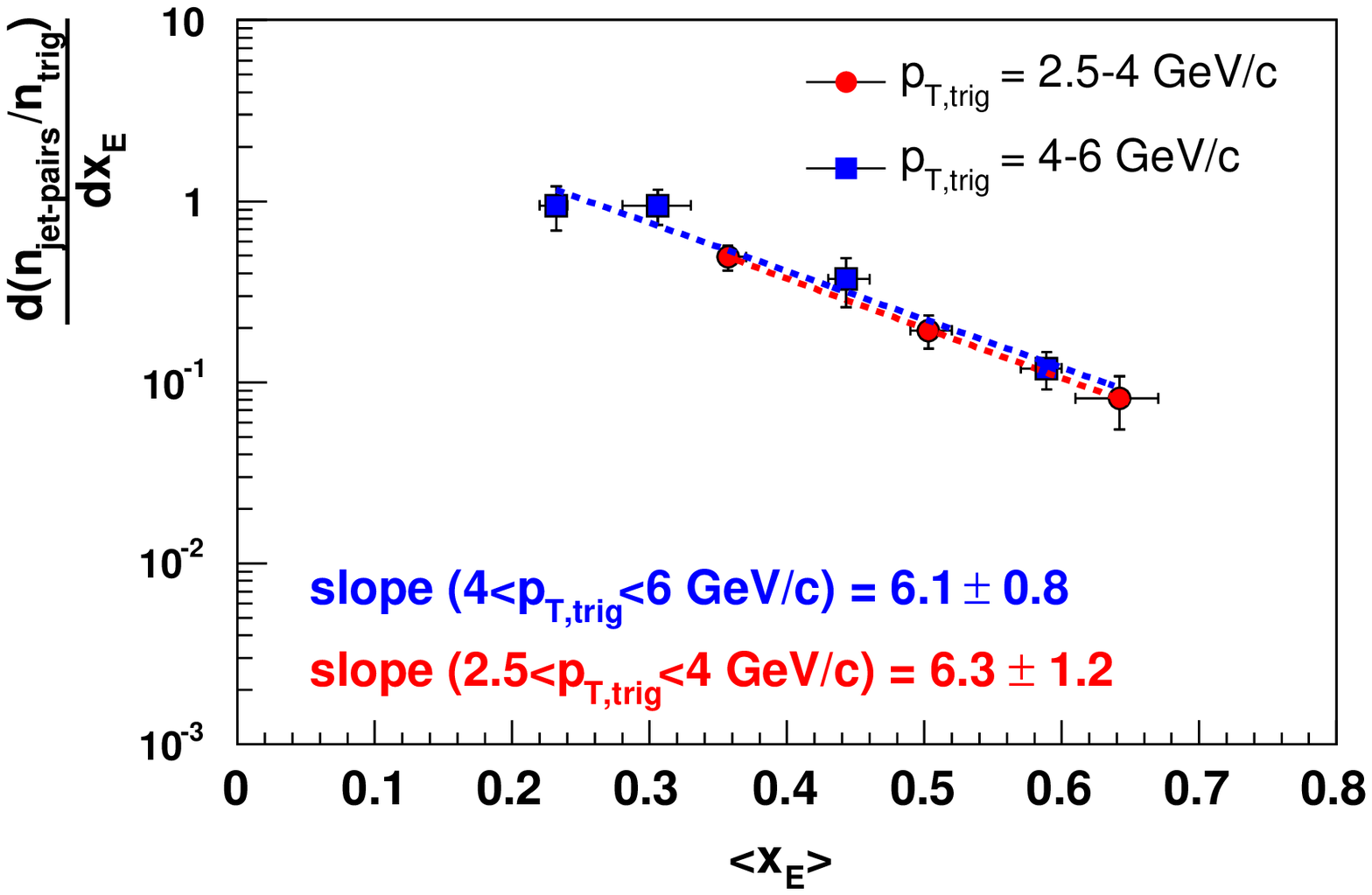}
\caption{\label{Fi:xEhh} (Color online) Conditional yield as a
a function of $\left<x_E\right>$ for the far-side $h-h$
correlations. Data is shown for two trigger $p_T$ ranges,
$2.5<p_{T,\rm{trig}}<4$ GeV/$c$ (open circles) and
$4<p_{T,\rm{trig}}<6$ GeV/$c$ (open boxes), respectively. The
dashed lines are exponential fits to the data. Bars are statistical errors.}
\end{center}
\end{figure}

At lower $p_T$ we have $x_E$ distributions from the h-h
correlations. In Fig.~\ref{Fi:xEhh}, far-side conditional yields as
obtained from charged hadron correlation functions are plotted
versus $\left<x_{E}\right>$. Here, $\left<x_{E}\right>$ has been
calculated from the $\left<p_{T,\rm{trig}}\right>$,
$\left<p_{T,\rm{assoc}}\right>$ and extracted angular widths.
Results are shown for two trigger $p_T$ ranges,
$2.5<p_{T,\rm{trig}}<4$ GeV/$c$ and $4<p_{T,\rm{trig}}<6$ GeV/$c$,
respectively. The dashed lines represent exponential fits to the
data. The slopes extracted are $6.3\pm1.2$ for the lower trigger
$p_T$ range and $6.1\pm0.8$ for the higher trigger $p_T$ window,
respectively. Within the statistics of the charged hadron dataset
we do not observe a strong sensitivity of the slope of the $x_E$
distributions to the trigger $p_T$.

\begin{figure}[ht]
\begin{center}
\includegraphics[width=1.0\linewidth]{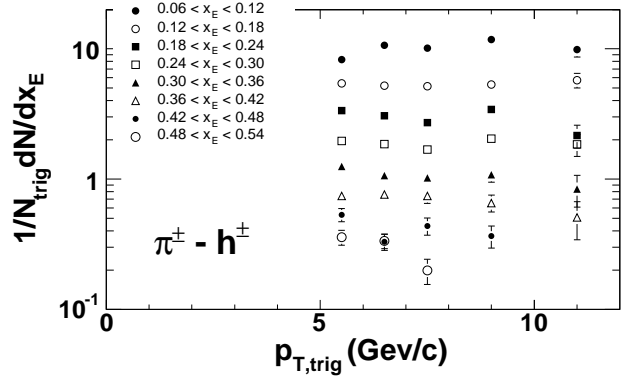}
\caption{\label{fig:mbdauxevspt} Far-side conditional yield as
a function of $p_{T,\rm{trig}}$ for different ranges of $x_E$ for
$\pi^{\pm}-h^{\pm}$ correlation for minimum-bias $d$ + Au
collisions. Bars are statistical errors.}
\end{center}
\end{figure}

A direct way of quantifying the scale dependence of the $x_E$
distribution is to plot the far-side conditional yields versus
$p_{T,\rm{trig}}$ for a fixed range of $x_E$. This is shown in
Fig.~\ref{fig:mbdauxevspt} where the conditional yields are found
to be independent of $p_{T,\rm{trig}}$, {\it i.e.} there is no
significant deviation from scaling. We will quantify any scaling
violation in this data when we compare these $x_E$ distributions from
$d$ + Au to distributions from
$p+p$ collisions in Section~\ref{sec:resultsDAuvspp}.

\subsection{Comparison between $\dA$ and $\pp$ \label{sec:resultsDAuvspp}}

As discussed in Section~\ref{sec:intro}, multiple scattering in
the cold nuclear-medium may broaden the far-side correlation and
possibly modify the fragmentation properties. In the previous
section we presented the measured jet structures from minimum-bias
$d$ + Au collisions. In this section we compare that data to
results from $p+p$ collisions. The goal is to establish the extent
to which the nuclear-medium modifies the properties of jets.

\begin{figure}
\begin{center}
\includegraphics[width=1.0\linewidth]{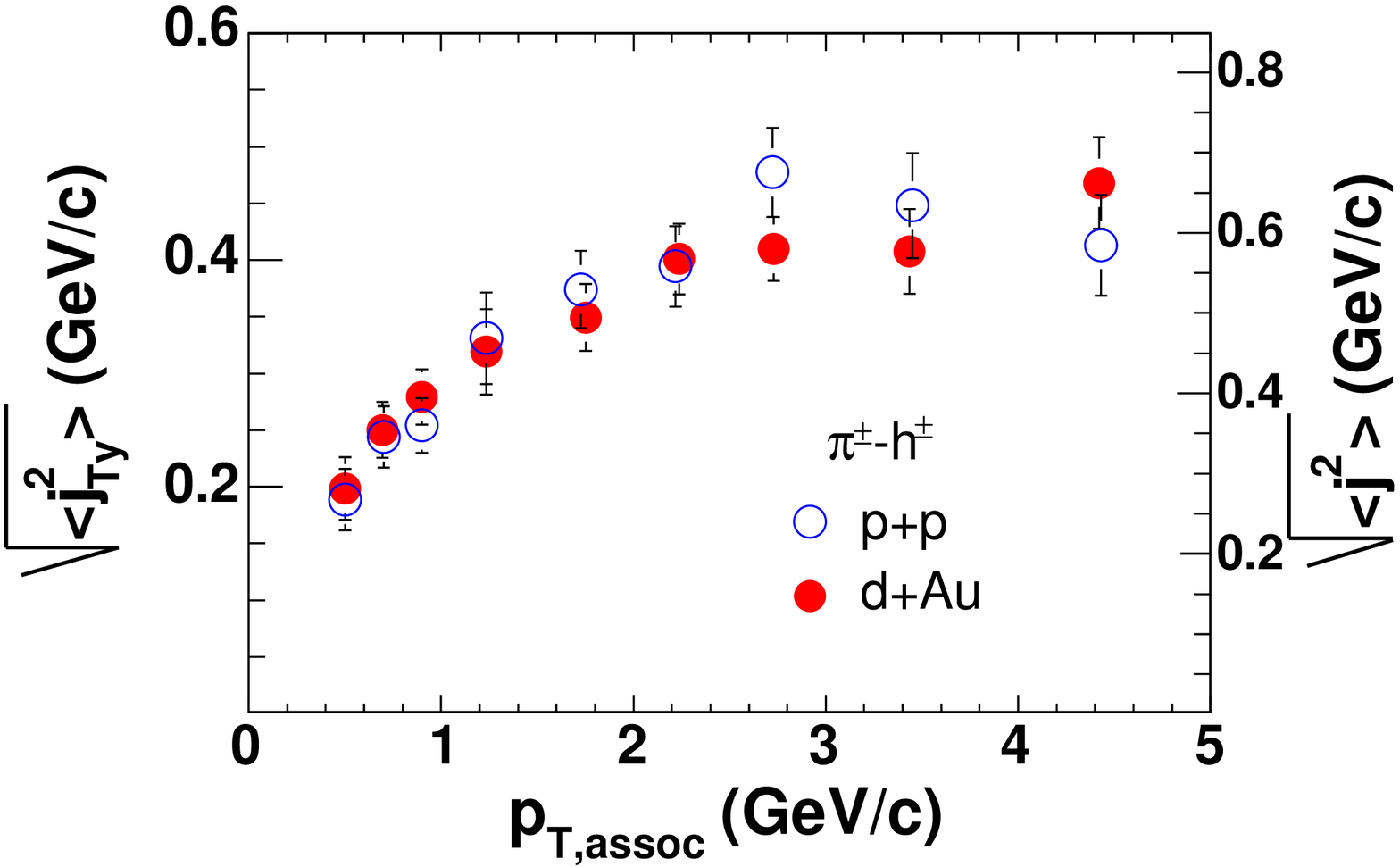}
\includegraphics[width=1.0\linewidth]{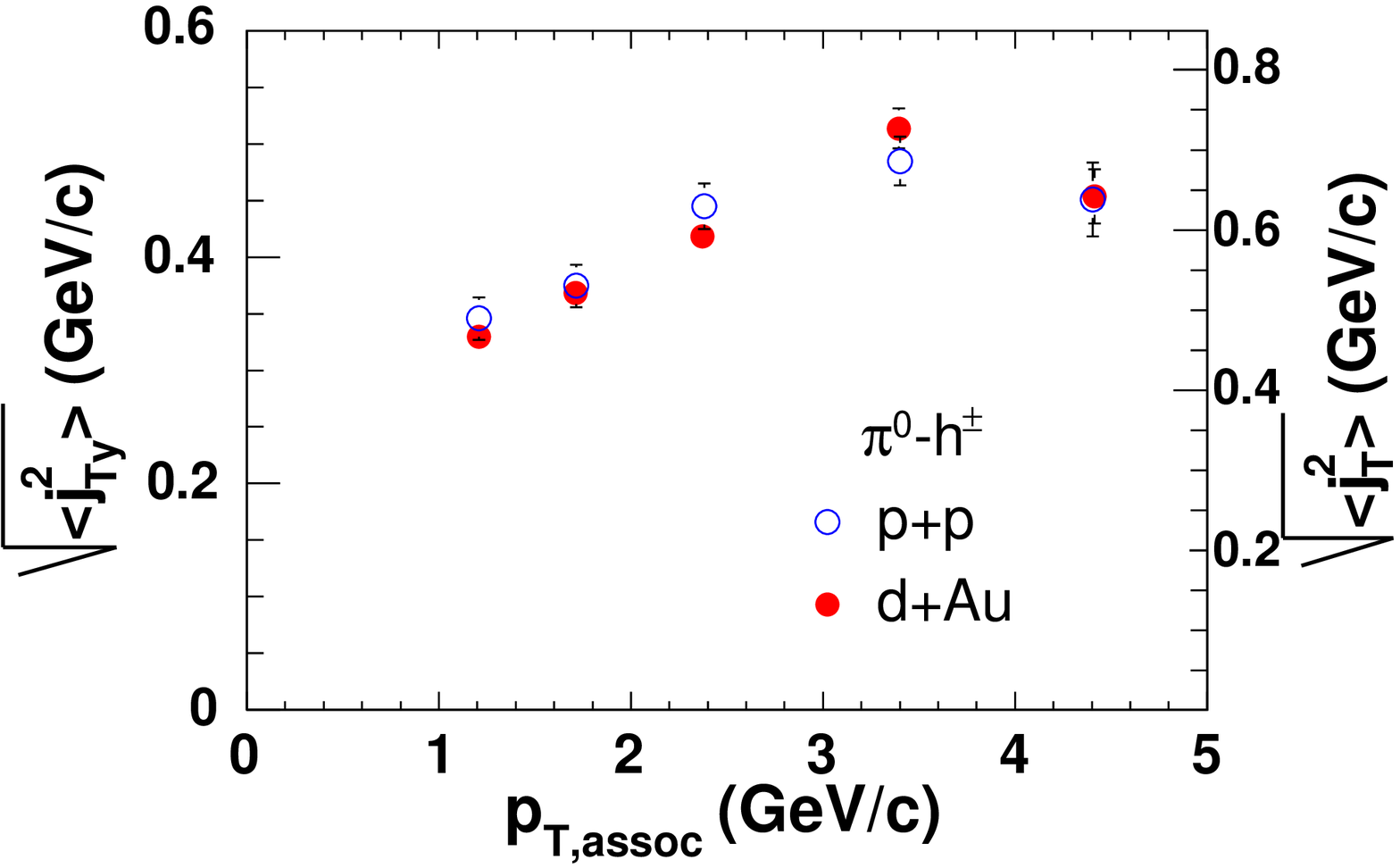}
\caption{\label{fig:dAupp} (Color online) The comparison of the
$\sqrt{\langle j_T^2 \rangle}$ values
between $d$ + Au
(filled circles) and $p+p$ (open circles) for $\pi^{\pm}-h^{\pm}$
correlations (top panel) and $\pi^{0}-h^{\pm}$ correlations (lower
panel). The trigger pion range is 5-10
GeV/$c$.  Bars are statistical errors.
The systematic errors are given in Table~\ref{tab:widtherror}.}
\end{center}
\end{figure}

Figure~\ref{fig:dAupp} shows the comparison of the extracted
$\sqrt{\langle j_T^2 \rangle}$ values as function of
$p_{T,\rm{assoc}}$ from $d$ + Au and $p+p$~ collisions. The
$\sqrt{\langle j_T^2 \rangle}$ values show no change from $p+p$ to
$d$ + Au collisions.

\begin{figure}
\begin{center}
\includegraphics[width=1.0\linewidth]{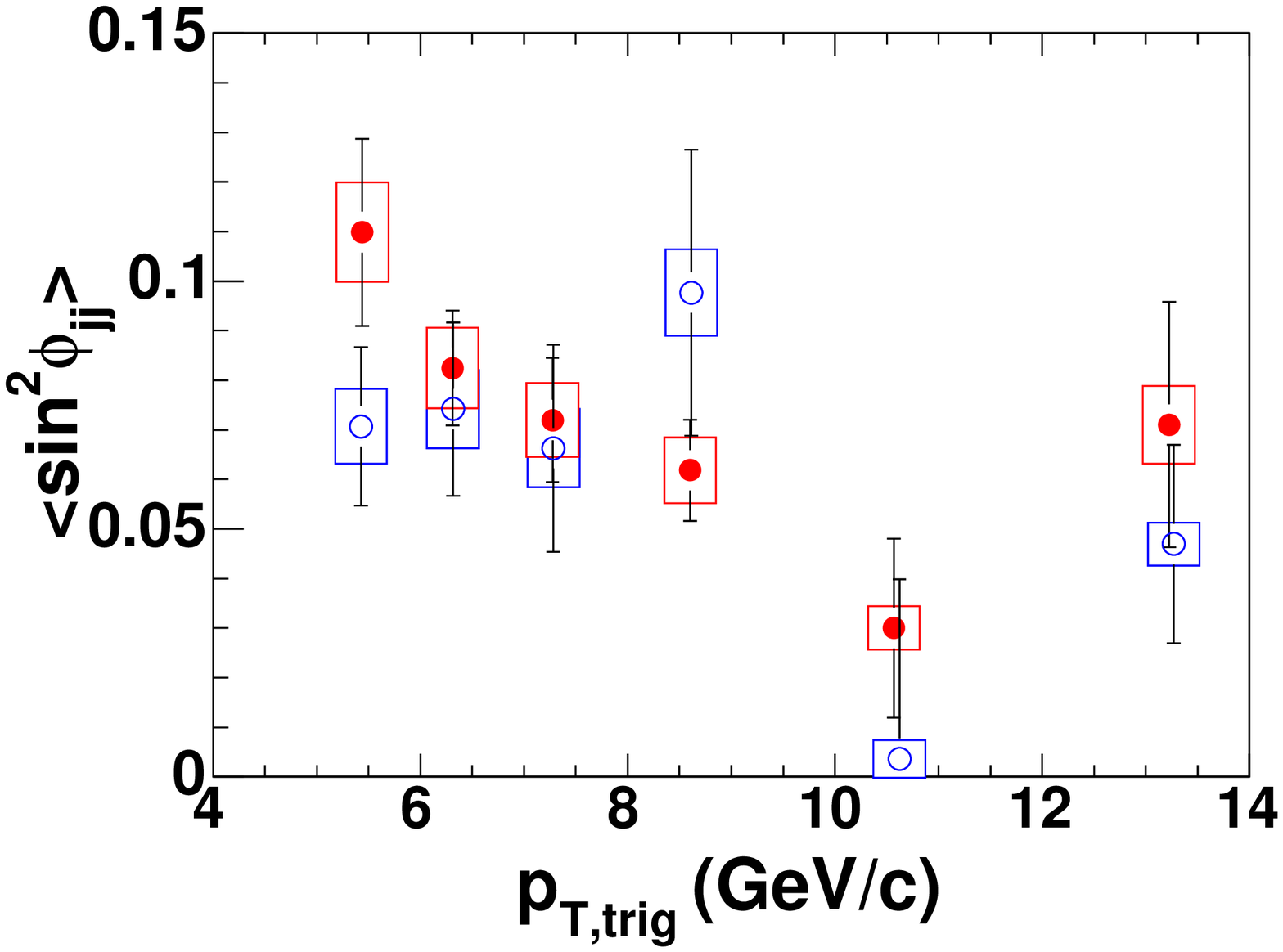}
\includegraphics[width=1.0\linewidth]{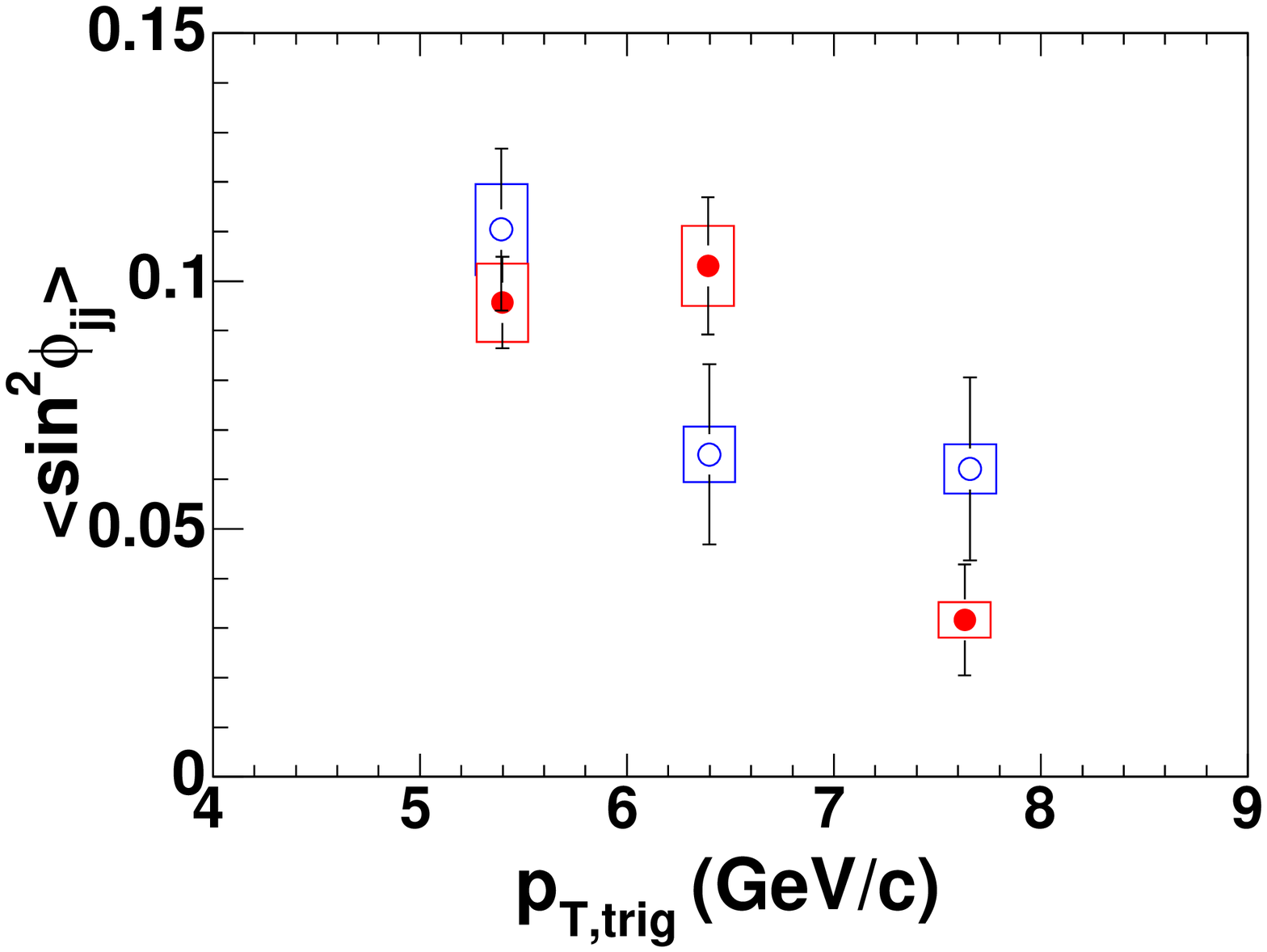}
\caption{\label{fig:dAuppII} (Color online) The comparison of the $\langle
\sin^2(\phi_{jj}) \rangle$ values between $d$ + Au (filled
circles) and $p+p$ (open circles) for $\pi^{\pm}-h^{\pm}$
correlations (top panel) and $\pi^{0}-h^{\pm}$ correlations (lower
panel). The associated hadron range is $2 < p_{T,\rm{assoc}} <
4.5$ GeV/$c$ for the charged pion triggers and $2.5 <
p_{T,\rm{assoc}} < 5$ GeV/$c$ for the neutral pions. Bars are statistical errors. The boxes represent the total systematic errors on each point.}
\end{center}
\end{figure}

Similarly, the $\langle \sin^2(\phi_{jj}) \rangle$
values shown in Fig.~\ref{fig:dAuppII} for
$d$ + Au are comparable to those from $p+p$ within errors,
although the values from $d$ + Au collisions are systematically
higher for $\pi^{\pm}-h^{\pm}$.
Since there is no strong difference between the $d$+Au~and $p+p$~results,
there is little indication for increased multiple-scattering in the
$d$+Au~final state.

\begin{figure}
\begin{center}
\includegraphics[width=1.0\linewidth]{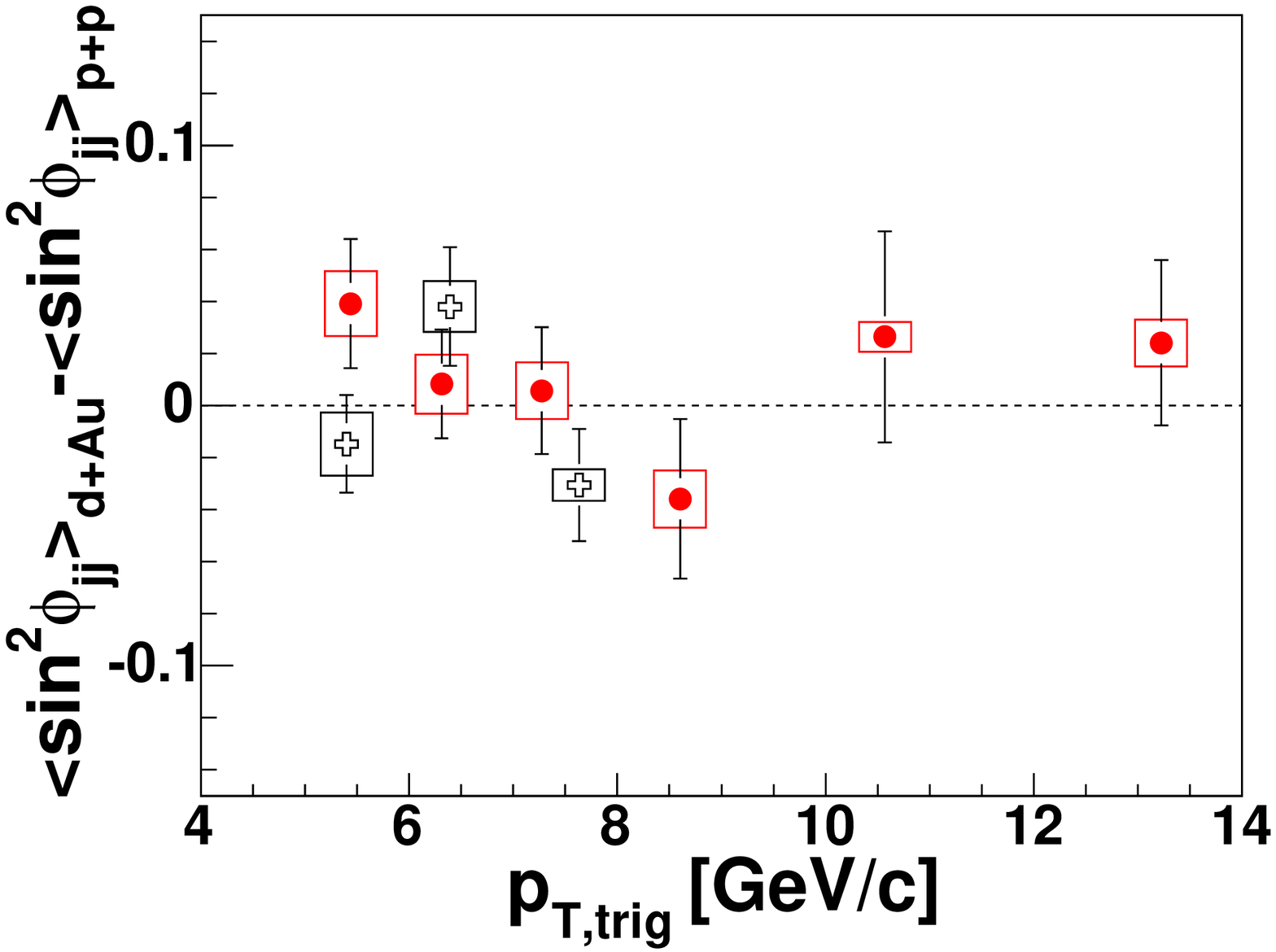}
\caption{(Color online) Quadrature difference between minimum-bias $d$ + Au and
$p+p$ $\left<\sin^{2}\phi_{jj}\right>$ values.  Closed circles are
$\pi^{\pm}-h$ values and the open circles are $\pi^{0}-h$ values.
Bars are statistical errors. The boxes represent the total systematic errors on each point.} \label{fig:deltaSin2qjj}
\end{center}
\end{figure}

Any additional radiation can be quantified by calculating the
point-by-point quadrature difference in
$\left<\sin^{2}\phi_{jj}\right>$ between
$d$ + Au and $p+p$~collisions.
This is shown in Fig.~\ref{fig:deltaSin2qjj} and this
difference is consistent with zero.
The average value for $\pi^{0}-h$ is $\Delta
\left<\sin^{2}\phi_{jj}\right> =
-0.005\pm0.012$(stat)$\pm$0.003(sys); for $\pi^{\pm}-h$
$\Delta\left<\sin^{2}\phi_{jj}\right> =
0.011\pm0.011$(stat)$\pm$0.010(sys). Combining the two data sets,
we find $\Delta \left<\sin^{2}\phi_{jj}\right> =
+0.004\pm0.008$(stat)$\pm$0.009(sys).

\begin{figure}
\begin{center}
\includegraphics[width=1.0\linewidth]{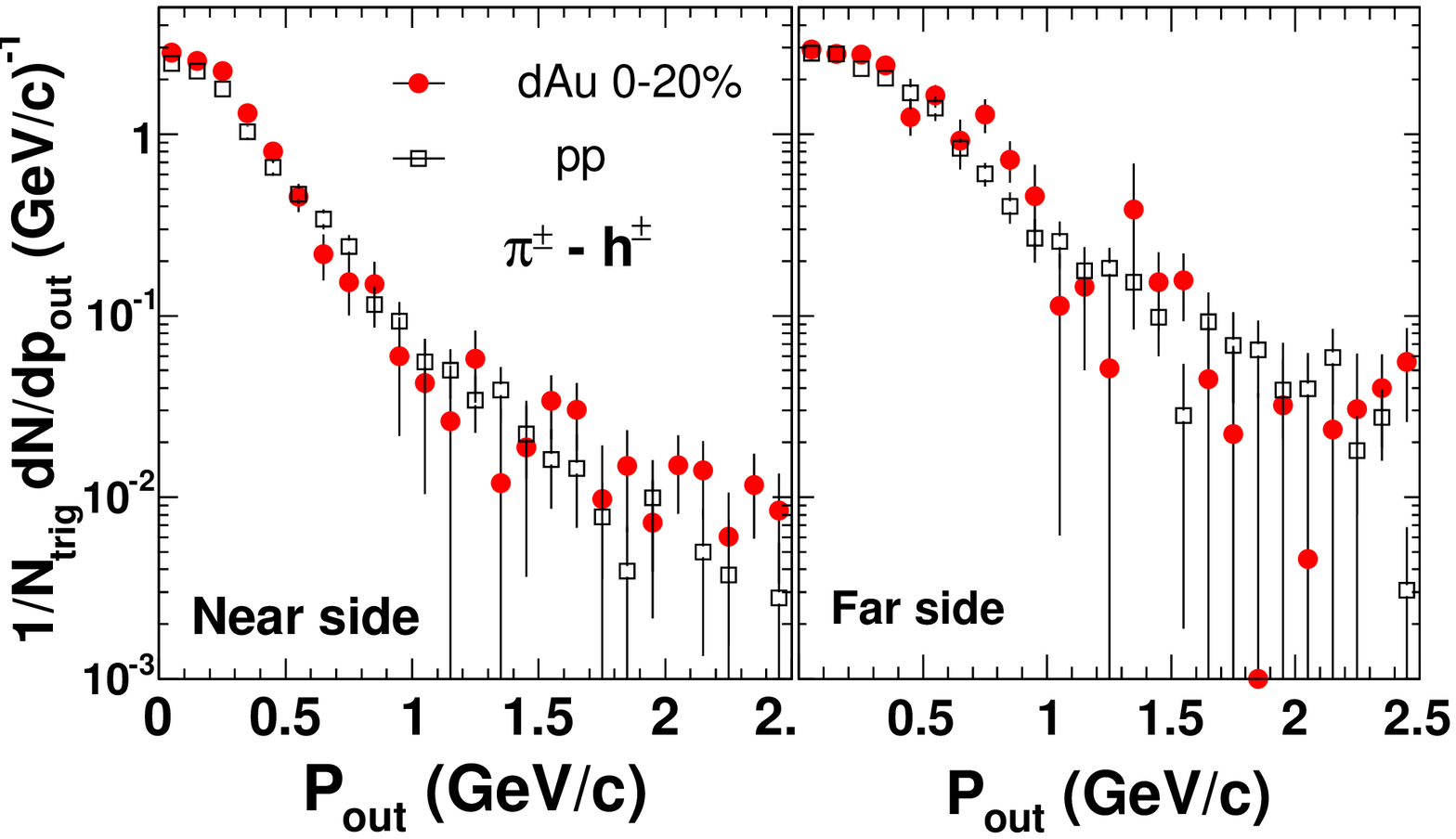}
\caption{\label{fig:dAupppout} (Color online) The comparison of the $p_{out}$
distribution at the near-side (left panel) and far-side (right
panel) between central $d$ + Au collisions and $p+p$ collisions.
Results are obtained for $\pi^{\pm}-h^{\pm}$ correlations
with the associated hadron range $0.5 < p_{T,\rm{assoc}} <
5$ GeV/$c$ and trigger pion range of $5 < p_{T,\rm{trig}} <
10$ GeV/$c$. Bars are statistical errors.}
\end{center}
\end{figure}

Figure~\ref{fig:dAupppout} shows the comparison of the $p_{out}$
distribution between central $d$ + Au and $p+p$ from
$\pi^{\pm}-h^{\pm}$ correlation, no apparent differences are
observed in both the near and far-side. This is consistent with
the observations that both $\sqrt{\langle j_T^2 \rangle}$ and
$\langle \sin^2(\phi_{jj})  \rangle$ are similar between $d$ + Au and $p+p$.

\begin{figure}
\begin{center}
\includegraphics[width=1.0\linewidth]{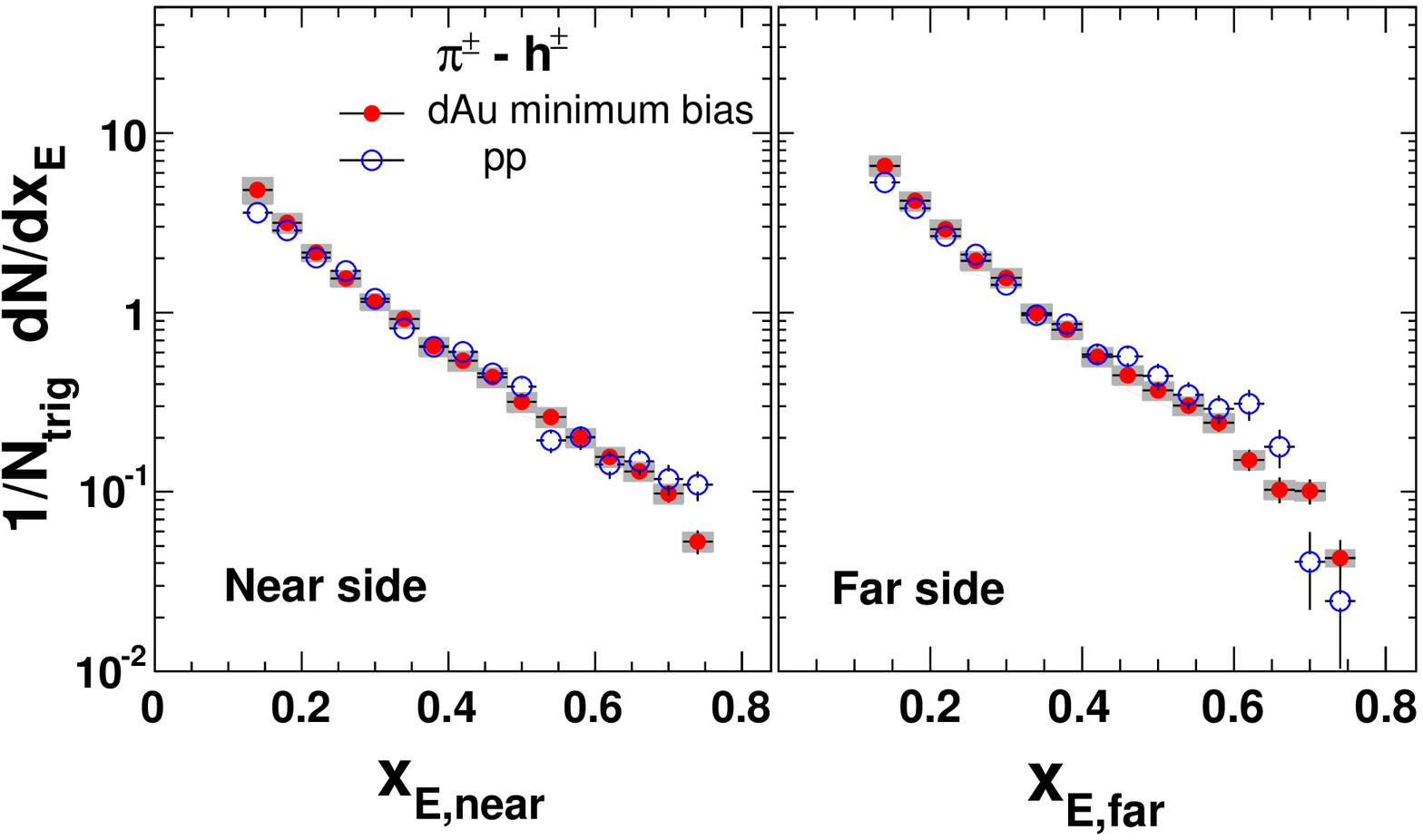}
\caption{\label{fig:dAuppxe} (Color online) The comparison of the $x_E$
distribution from $\pi^{\pm}-h^{\pm}$ correlation at the near-side
(left panel) and far-side (right panel) between minimum bias $d$ +
Au collisions (filled circles) and $p+p$ collisions (open
circles). The trigger $\pi^{\pm}$ are from $5-10$ GeV/$c$. Bars are statistical errors. The boxes represent the total systematic errors on each point.}
\end{center}
\end{figure}

A second set of comparisons between $d$ + Au and $p+p$
collisions is the number of hadrons in the near- and far-angle jet
structures associated with a high-$p_T$ trigger.
Figure~\ref{fig:dAuppxe} shows the comparison of the conditional
yield as function of $x_E$ and no apparent difference between $d$
+ Au and $p+p$ collisions is observed for either the near- or
far-side.

\begin{figure}[ht]
\begin{center}
\includegraphics[width=1.0\linewidth]{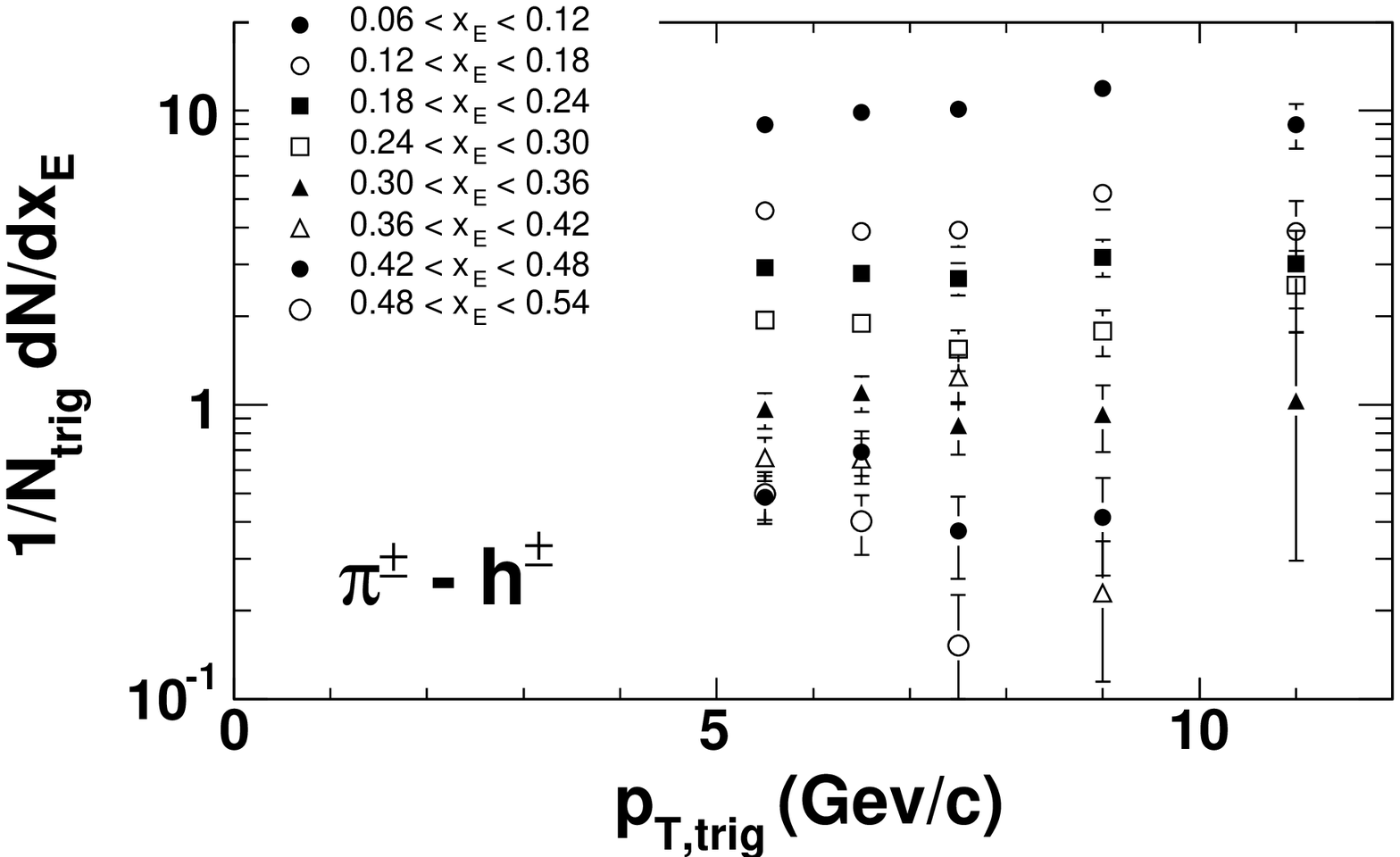}
\caption{Far-side conditional yield as function of
$p_{T,\rm{trig}}$ for different ranges of $x_E$ from $p+p$
collisions, triggers are $\pi^{\pm}$ from $5-10$ GeV/$c$. Bars are statistical errors.}
\label{fig:ppxevspt}
\end{center}
\end{figure}

\begin{figure}[ht]
\begin{center}
\includegraphics[width=1.0\linewidth]{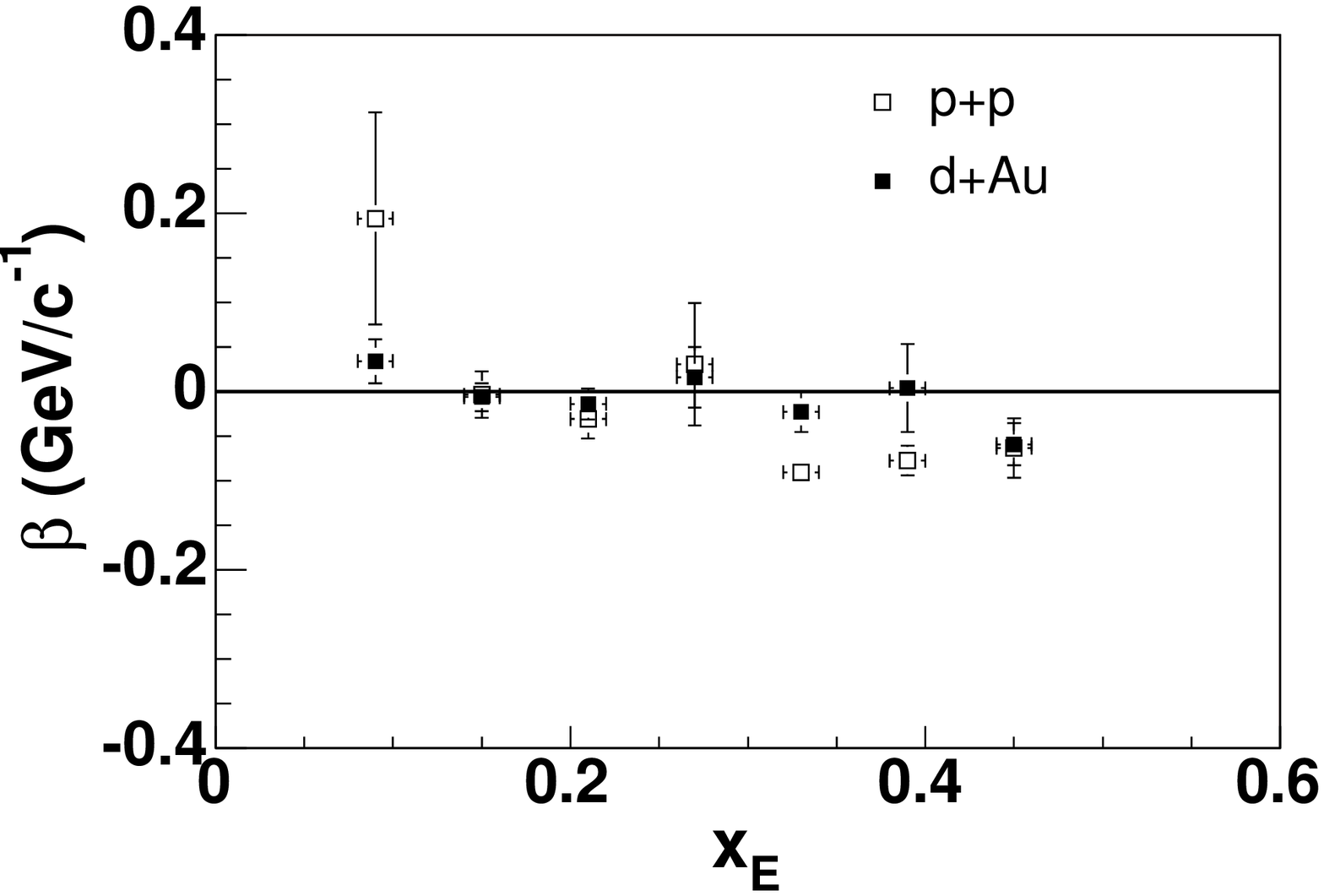}
\caption{The fitted fractional change in $dN/dx_E$ per unit
$p_{T,\rm{trig}}$ ($\beta$ in Eq.~\ref{eq:violation}) of the
far-side conditional yield for different ranges of $x_E$. The
$p+p$ data are shown as open squares, while the $d$ + Au data are
shown as filled squares. Bars are statistical errors.} \label{fig:violation}
\end{center}
\end{figure}

In the previous section we examined the level of scaling
violations in the far-side $dN/dx_E$ distribution for $d$ + Au
collisions by plotting different $x_E$ ranges as a function of
$p_{T,\rm{trig}}$ (Fig.~\ref{fig:mbdauxevspt}). The comparable
plot for $p+p$ collisions is shown in Fig.~\ref{fig:ppxevspt}.
For both $d$ + Au and $p+p$ collisions the amount of scaling
violations, {\it i.e.} the dependence of $dN/dx_E$ on
$p_{T,\rm{trig}}$, can be quantified by fitting the data in each
$x_E$ range with a straight line as a function of
$p_{T,\rm{trig}}$.
\begin{equation}
\frac{dN}{dx_E} = \frac{dN}{dx_E}_0 (1 + \beta p_T)
\label{eq:violation}
\end{equation}
The fitted slopes ($\beta$) represent the fractional change in
$dN/dx_E$ per GeV/$c$ and are shown in Fig.~\ref{fig:violation}.
For the $d$ + Au data $\beta$ is consistent with zero, {\it i.e.}
there is no significant scaling violation across the whole $x_E$
range, while there may be a slight scaling violation at high $x_E$
for the $p+p$ data. On a point-by-point basis there is no
systematic difference between the $d$ + Au and $p+p$ data.

Taken as a whole, all the results presented in this section
indicate that there is no significant change in jet fragmentation
between $d$ + Au and $p+p$ collisions due to the presence of the
cold nuclear-medium. In addition there is no strong evidence for
an increase in $\langle \sin^2(\phi_{jj}) \rangle$ due to multiple scattering
in the Au nucleus. Using the minimum-bias $d$ + Au data has the
advantage of the highest statistical precision. In the next
section we examine whether there is any change in jet-structures as
a function of collision centrality in $d$ + Au, {\it i.e.} we
split the statistics into a few centrality classes to increase the
lever arm of the nuclear-thickness function.

\subsection{Centrality Dependence\label{sec:resultsCentrality}}

As discussed in Section \ref{sec:intro}, $\mean{\sin^2(\phi_{jj})}$ is
expected to increase as $d$ + Au collisions become more central
due to increased multiple scattering. Models of
multiple scattering \cite{Accardi:2002ik} predict that the
increase in $\mean{k_{T}^2}$ is proportional to $T_{A}(b)$, the
nuclear thickness function.
We are not aware of predictions of how
$\mean{\sin^2(\phi_{jj})}$ will change with centrality, but
note that any increase in $\mean{k_T^2}$
will also increase $\mean{\sin^2(\phi_{jj})}$.

\begin{figure}
\begin{center}
\includegraphics[width=1.0\linewidth]{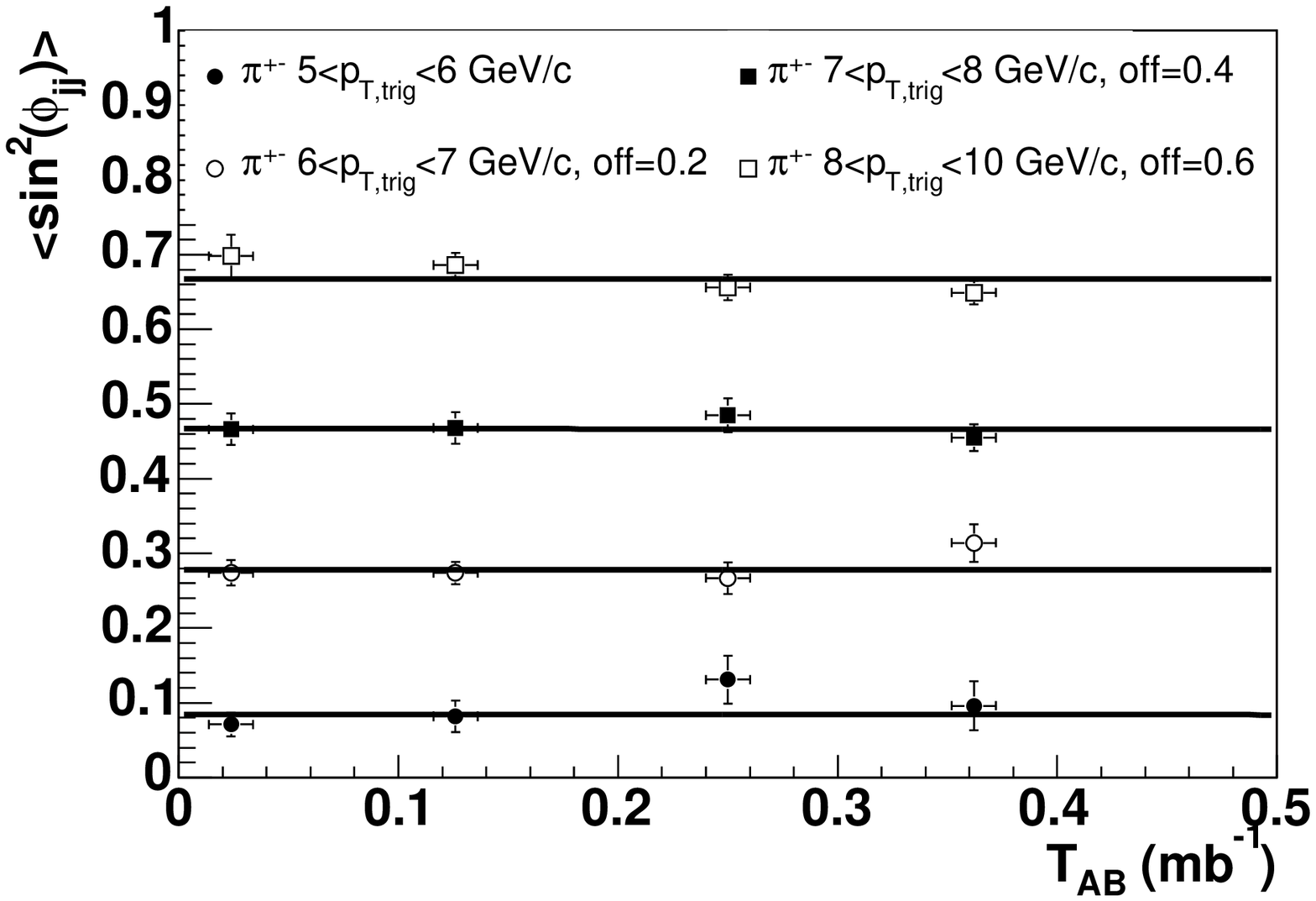}
\includegraphics[width=1.0\linewidth]{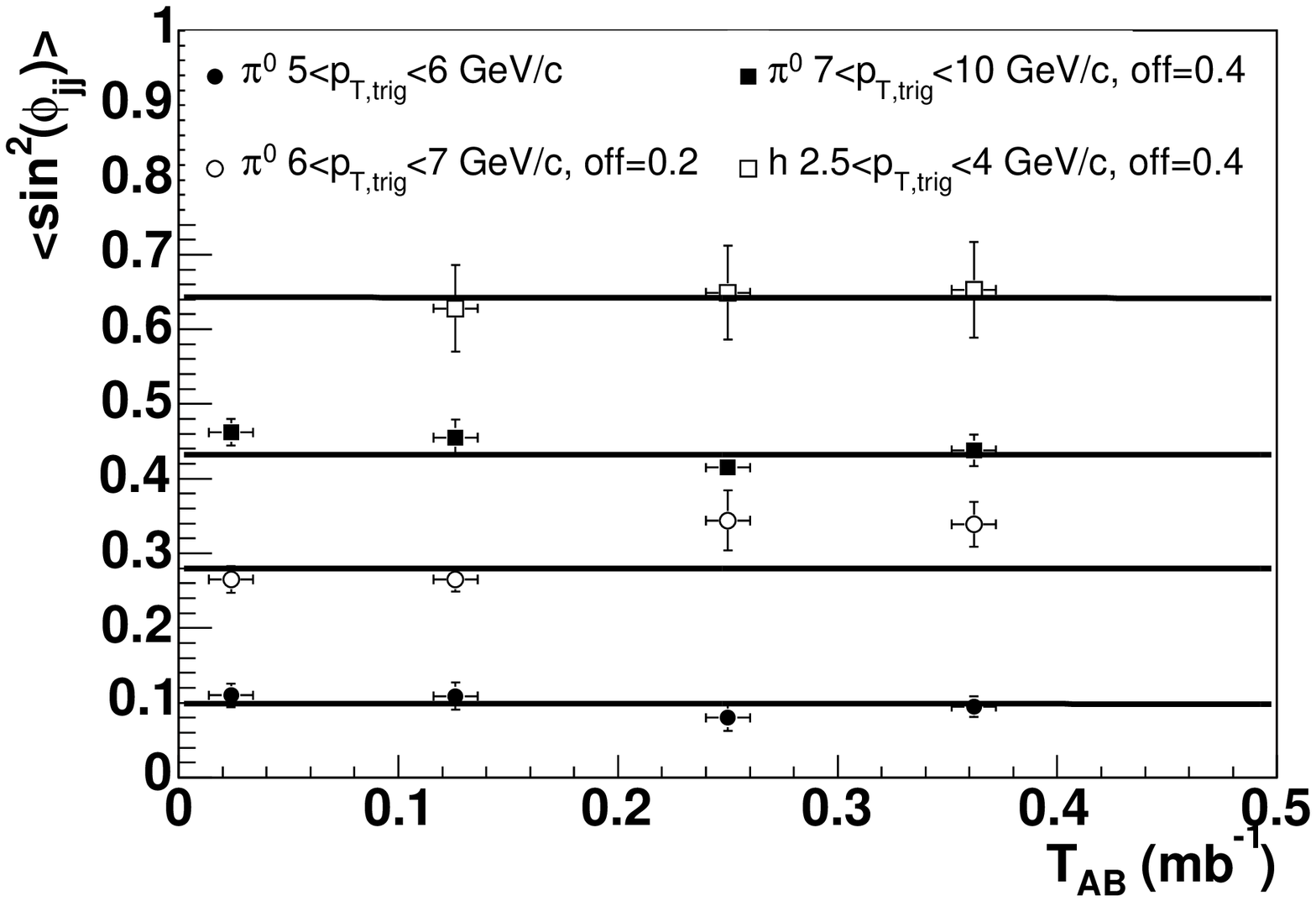}
\caption{The measured values of $\langle \sin^2(\phi_{jj}) \rangle$ are
shown as a function of the nuclear thickness function $T_{A}(b)$. The
left-most points are from $p+p$ collisions and the other points are
from the three centrality classes in $d$ + Au collisions. The data
sets are offset as indicated for clarity, On the top plot is the
data for $\pi^\pm$ and the bottom-plot is the data for $\pi^0$ and
hadron triggered data. The lines shown are a simultaneous fit to
all the data with Eq.~\ref{eq:centphiTAB} to extract a common fractional
increase in
$\mean{\sin^2(\phi_{jj})}$ with $T_{A}(b)$. Bars are statistical errors.}
\label{fig:centsinphi}
\end{center}
\end{figure}

To probe this physics, we have
measured angular correlations in three centrality bins for $d$ +
Au collisions (0-20\%, 20-40\%, and 40-88\%) to extract
angular-widths of the jet-structures and hence
$\sqrt{\langle j_T^2 \rangle}$ and $\mean{\sin^2(\phi_{jj})}$.
Figure~\ref{fig:centsinphi} shows the independent data sets of
$\mean{\sin^2(\phi_{jj})}$ 
including results from $p+p$
collisions as well as the three centrality classes from $d$ + Au
collisions.

All the $\mean{\sin^2(\phi_{jj})}$ data in Fig.~\ref{fig:centsinphi}
have been simultaneously fit with the following linear equation in $T_{A}(b)$
\begin{equation}
\label{eq:centphiTAB}
\langle \sin^2(\phi_{jj})\rangle  =
\langle \sin^2(\phi_{jj})\rangle _0 (1 + a_{frac} T_{A}(b))
\end{equation}
The slope parameter, $a_{frac}$, is assumed to be common to all data
sets, while the pre-factors, $\langle \sin^2(\phi_{jj})\rangle_0$, depend on
the $p_T$ of the trigger and associated particles. The extracted
slope  $a_{frac} =  -0.01 \pm 0.40$~mb with chi-squared per
degree-of-freedom, $\chi^2/\nu=27/22$. This slope is consistent
with zero, {\it i.e.} we do not observe any significant increase
in $\langle \sin^2(\phi_{jj})\rangle $ with centrality.

This can be compared to predictions from Hwa and
Wang~\cite{Hwa:2004zd} who assume no increase in $k_{T}$ with
centrality to reproduce the Cronin effect data at RHIC and Qiu and
Vitev \cite{Qiu:2004da} who calculate
\begin{equation}
\langle k_{T}^2\rangle_{dijet}= 2\langle k_{T}^2\rangle_{vac} +
\frac{0.72}{T_{Aminbias}}T_{A}(b)
\label{eq:QuiOriginal}
\end{equation}

To gain some insight into the magnitude of Qiu and Vitev's
predicted effect compared
to our experimental results, we recast Eq.~\ref{eq:QuiOriginal} into the
same form as Eq.~\ref{eq:centphiTAB}.
\begin{equation}
\langle k_{T}^2\rangle_{dijet}= 2\langle k_{T}^2\rangle_{vac}(1 +
\frac{0.72}{2\langle k_{T}^2\rangle_{vac}T_{Aminbias}}T_{A}(b))
\label{eq:fracQV}
\end{equation}
Hence their prediction for the fractional increase
in $\langle k_{T}^2 \rangle$ with $T_{A}(b)$ is from 0.51 to
0.72 depending on the range
$0.25 < \langle k_{T}^2 \rangle _{vac} < 0.35$ (GeV/$c$)$^2$
suggested by  Qiu and
Vitev \cite{Qiu:2004da}\footnote{For $0-88$\%
$d$ + Au collisions, $T_{A~minbias} =0.20$
mb$^{-1}$}.
Though the predicted fractional increase
is of a different quantity,
it should provide an estimate of the
magnitude of the fractional increase in $ \sin^2(\phi_{jj})\rangle$.
The prediction is slightly larger than one standard deviation
(statistical) from our experimental result, $a_{frac} =  -0.01 \pm 0.40$~mb.
If the measured value for $< \langle k_{T}^2 \rangle _{vac}$ at RHIC 
turns out to be larger
than assumed by Qiu and Vitev \cite{Qiu:2004da} then the predicted fractional
increase in Eq.~\ref{eq:fracQV} will be smaller.

Barnafoldi {\it et
al.}~\cite{Barnafoldi:2004kh} have also predicted
the increase in $k_{T}$ due to
multiple scattering in $d$ + Au collisions at RHIC.
They calculate that $\langle k_T^2
\rangle$ increases by $C=0.35$(GeV/$c$)$^2$ per collision up to
the first four collisions, then it saturates. Their prediction is
\begin{eqnarray}
\label{eq:ktFaiRelative}
\langle k_{T}^2 \rangle & = &
\langle k_{T}^2 \rangle_0(1 +\frac{0.35}{\langle k_{T}^2 \rangle_0}
                 \times(40 \times T_{A})) \text{ , }
T_{A}< 0.1 \nonumber \\
\langle k_{T}^2 \rangle & = &
\langle k_{T}^2 \rangle_0(1 +\frac{0.35}{\langle k_{T}^2 \rangle_0}
                \times(40 \times 0.1)) \text{ , }
T_{A}> 0.1 \nonumber \\
\end{eqnarray}
Barnafoldi {\it et
al.} do not provide values for $\langle k_{T}^2 \rangle_0$, however
if we use the range
$0.25 < \langle k_{T}^2 \rangle _{0} < 0.35$ (GeV/$c$)$^2$
suggested by  Qiu and
Vitev \cite{Qiu:2004da} then
the fractional increase with $T_A$ is 40 to 56 for $T_{A} < 0.1$ followed
by no further increase. This rapid increase is not observable
in our data set because the model saturates
already in the most peripheral $d$ + Au bin where
$\langle T_{A} \rangle = 0.11$ mb$^{-1}$.

\begin{figure}
\begin{center}
\includegraphics[width=1.0\linewidth]{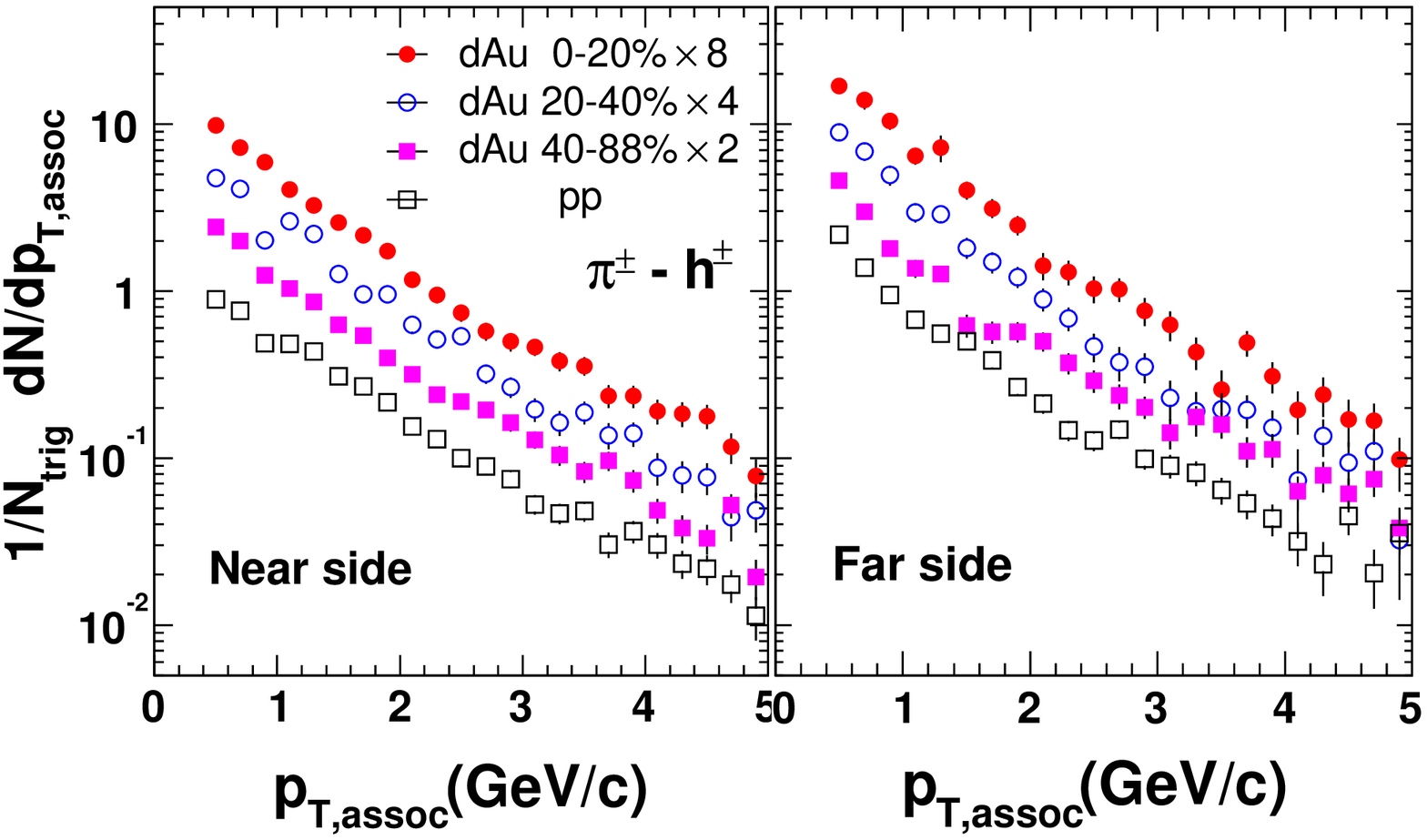}
\includegraphics[width=1.0\linewidth]{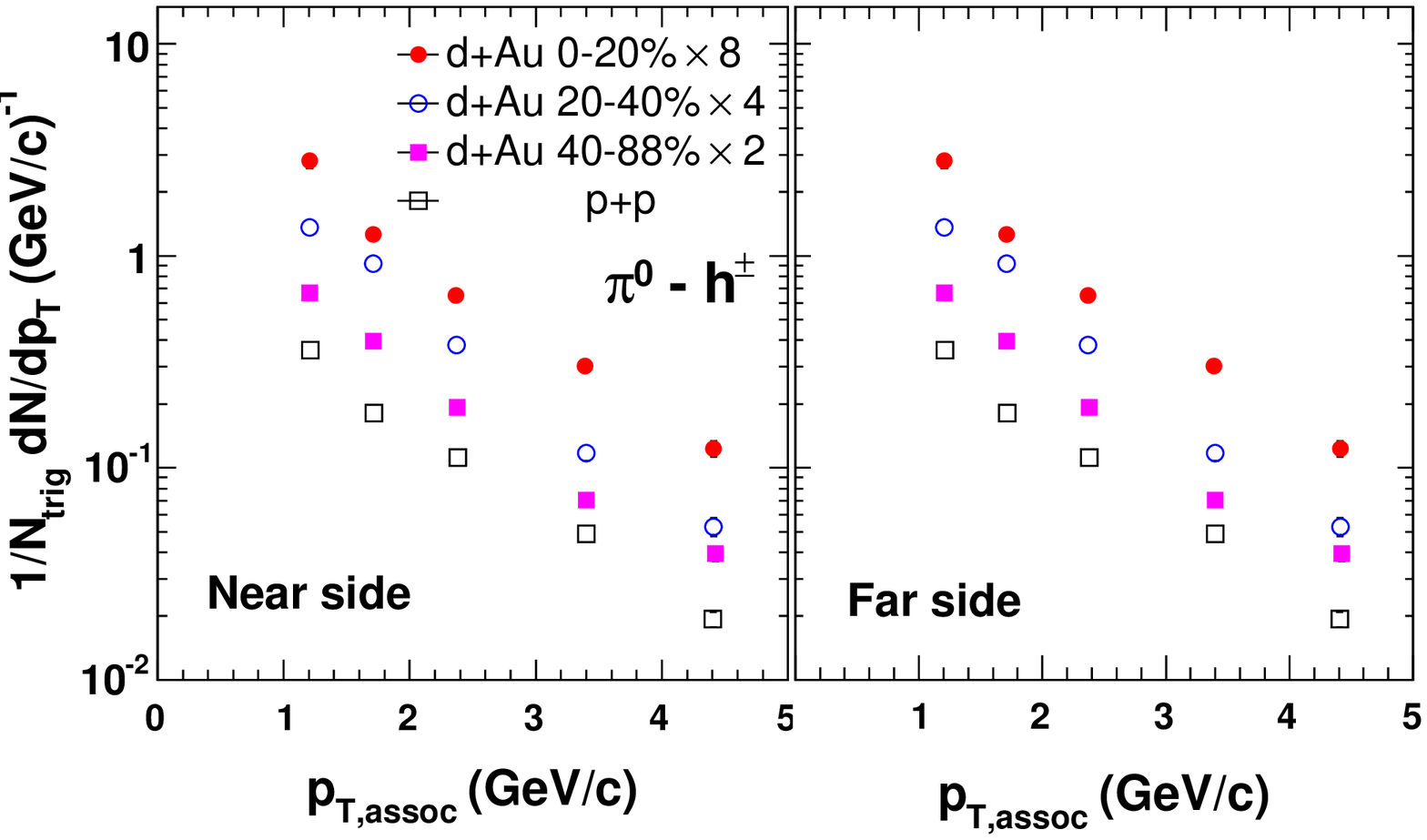}
\caption{\label{fig:ptcent} (Color online) Centrality dependent near (left) and
far (right) $CY(p_{T})$ for $\pi^{\pm}-h^{\pm}$ correlations
(upper) and $\pi^{0}-h^{\pm}$ correlations (lower).
The filled circles are from 0-20\% centrality,
the open circles are from 20-40\% centrality, the
filled squares are from 40-88\% centrality in
$d$ + Au collisions, while the open squares are
from $p+p$ collisions. Bars are statistical errors.}
\end{center}
\end{figure}

\begin{figure}
\begin{center}
\includegraphics[width=1.0\linewidth]{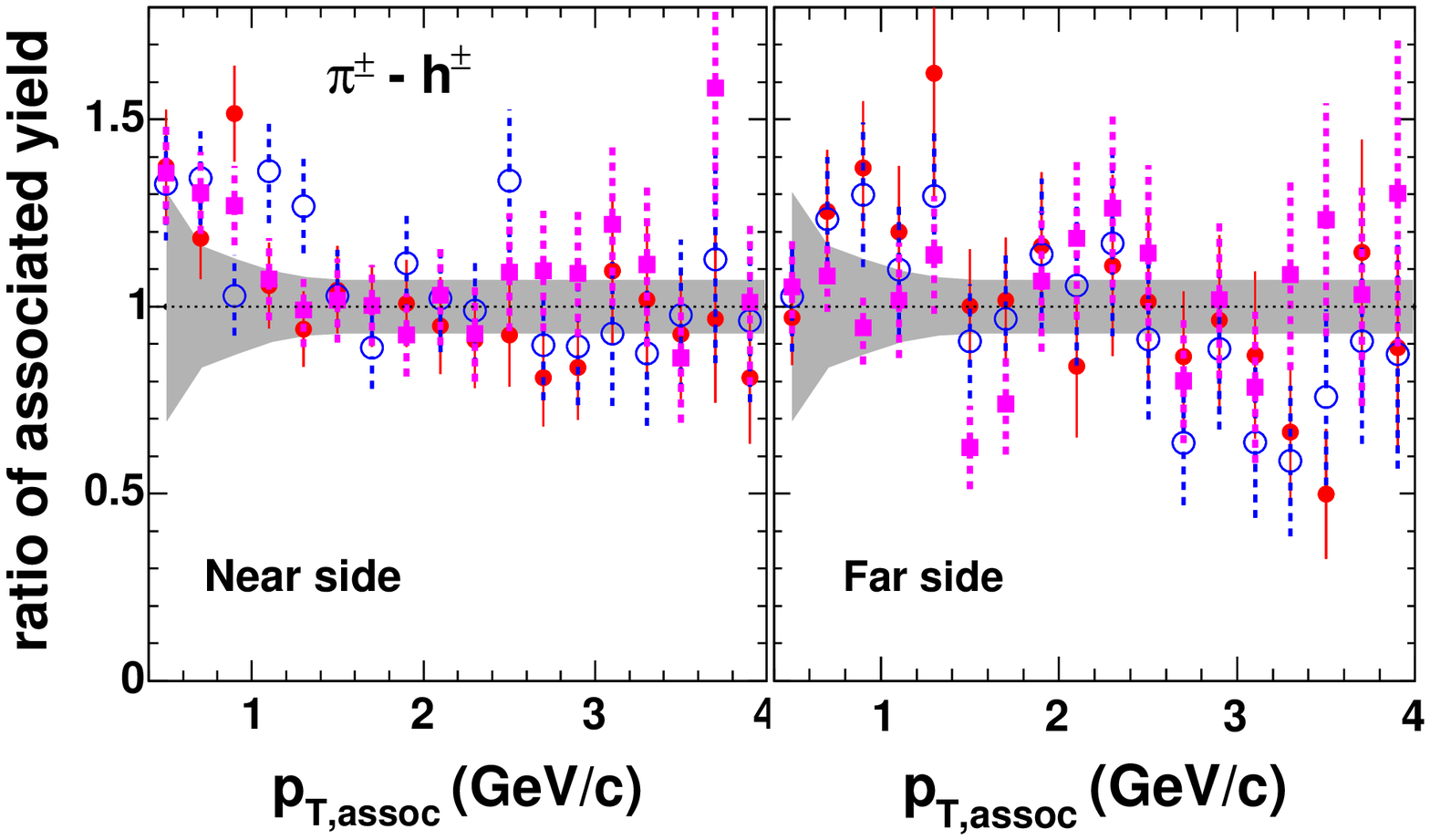}
\includegraphics[width=1.0\linewidth]{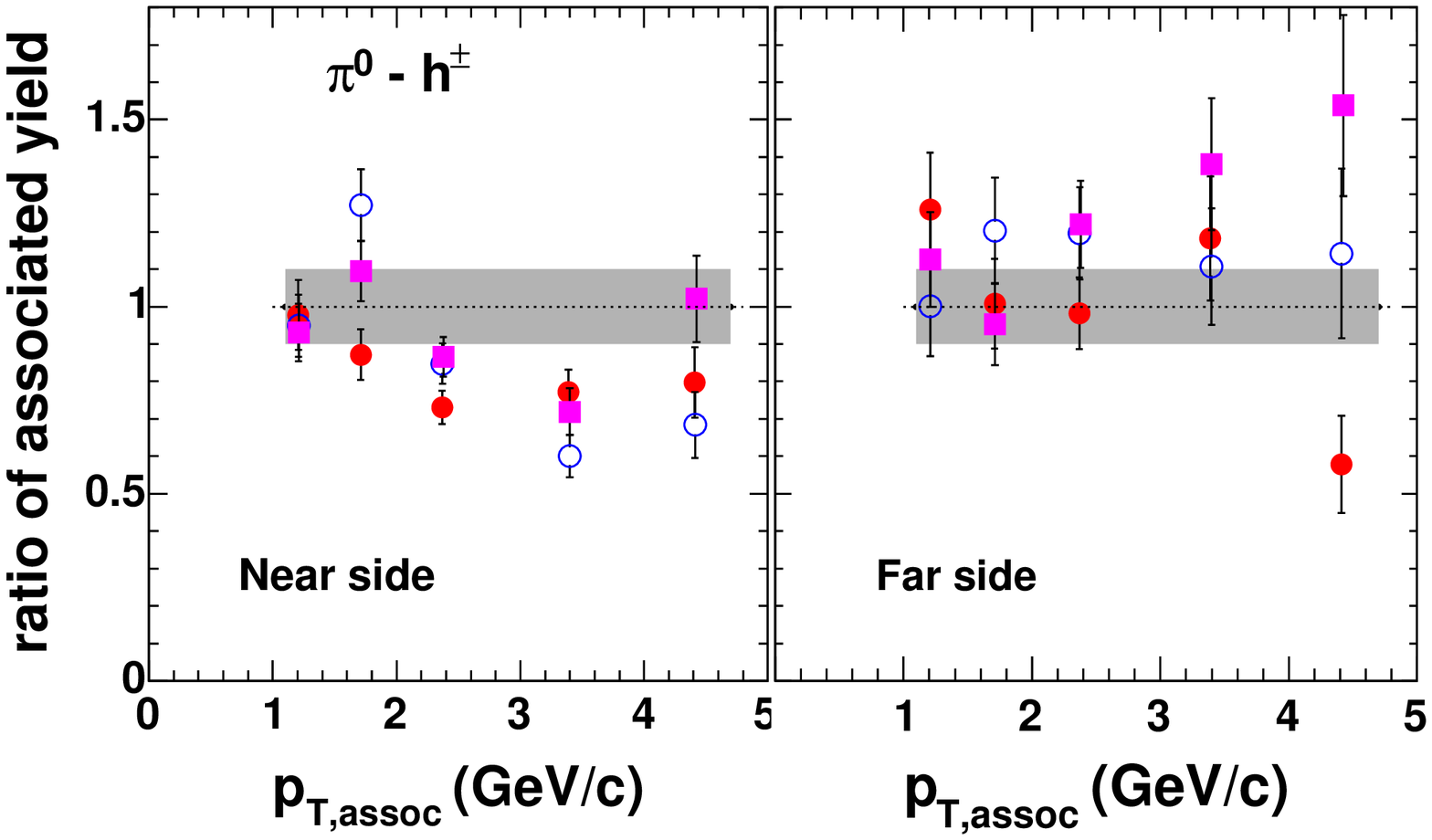}
\caption{\label{fig:ptrda} (Color online) Centrality dependence to the ratio of
near (left) and far (right) $CY(p_{T})$ for $d$ + Au to $p+p$.
$\pi^{\pm}-h^{\pm}$ correlations are the upper panels and
$\pi^{0}-h^{\pm}$ correlations are the lower panels.  The shaded
band is the systematic error due to the normalization of the
yields.
The filled circles are from 0-20\% centrality,
the open circles are from 20-40\% centrality, and the
filled squares are from 40-88\% centrality in
$d$+Au collisions. Bars are statistical errors. The shaded band represents the normalization error on the spectra.}
\end{center}
\end{figure}

As discussed in Section \ref{sec:intro}, inelastic scattering of
the hard parton in the cold-medium may also increase the
conditional yields ($CY$) of hadrons that are associated with a
high-$p_T$ trigger. Figure~\ref{fig:ptcent} shows the centrality
dependence of the extracted $CY$, together with  $CY$ from
$p+p$ collisions. The difference can be better illustrated by
taking the ratio ($d$ + Au/$p+p$) of the per trigger yield as
shown in Fig.~\ref{fig:ptrda}. There is a possible increase in
near-side particle yield for $p_T < 1$ GeV/$c$ in the $d$ + Au
collisions. In other momentum ranges there is no consistent
difference between yields in $d$ + Au and $p+p$ collisions.
The recombination model of Ref.~\cite{Hwa:2004} predicted a factor
of two increase in CY peripheral to central $d$ + Au collisions
which is much larger than observed in the data. A later
recombination model by the same authors\cite{Hwa:2005ui}
postdicted only a 30\% increase in CY for associated particles at
$p_T = 2$ GeV/$c$, which is comparable or perhaps slightly larger
than is observed in the data.

\section{Conclusions\label{sec:concl}}

We have measured several properties of jet fragmentation and
dijet correlations using two-particle correlations with three
different particle combinations: $h^{\pm}-h^{\pm}$, $\pi^0-h^{\pm}$, and
$\pi^{\pm}-h^{\pm}$. From the correlation functions we have extracted
the widths of the near- and far-angle correlations as a function
of the momentum of the two hadrons, $p_{T,\rm{trig}}$ and
$p_{T,\rm{assoc}}$. These widths decrease as a function of both
the trigger and associated particle's
momenta. From the near-angle widths we calculate
$\sqrt{\langle j_T^2 \rangle}$, the RMS of
the transverse momentum of fragmented
hadrons with respect to the hard parton. The value of
$\sqrt{\langle j_T^2 \rangle}$ saturates at
$0.64\pm0.02$~(stat)~$\pm0.04$~(sys) GeV/$c$
for $p_{T,\rm{assoc}}>2$
GeV/$c$ and is consistent with being
independent of $p_{T,\rm{trig}}$ and trigger
species. The $\sqrt{\langle j_T^2 \rangle}$ is similar for $d$ + Au and
$p+p$ collisions 
consistent with the fragmentation
process not being affected by the presence of the cold nuclear
medium.

We have also compared the measured $x_E$ distributions in $d$ + Au collisions
to the baseline distributions from $p+p$ collisions. The $x_E$
distributions extracted from the far-angle correlations provide
information on the fragmentation of a back-to-back parton
triggered on a high-$p_T$ hadron in the opposite hemisphere. The
measured $dN_{far}/dx_E$ distributions in $d$ + Au are
approximately independent of $p_{T,\rm{trig}}$, {\it i.e.} they
scale. We have quantified the level of scaling violation by
extracting the slope $\beta=d(dN/dx_E)/dp_{T,\rm{trig}}$ for
different ranges of $x_E$. The slopes are consistent with zero for
$d$ + Au collisions, {\it i.e.} there is no significant
scaling violation. Point-by-point the scaling-violation slopes for
$p+p$ collisions are not significantly different than the $d$ + Au
data. This suggests that if there is any additional
gluon radiation in $d$ + Au reactions due to multiple scattering,
then this has little observable influence on the fragmentation of
the hard parton.

We observe no centrality dependence of the conditional yield in
$d$ + Au  and these yields are very similar to those from $p+p$
collisions. The recombination model of Ref.~\cite{Hwa:2005ui}
postdicted a 30\% increase in conditional yield between $d$ + Au
and $p+p$, which is perhaps slightly larger than is observed in
the data.

We have extracted the dijet acoplanarity $\langle
\sin^2(\phi_{jj})\rangle$ from the widths of the back-to-back
correlations in $d$ + Au and $p+p$ collisions. In collisions
involving nuclei, multiple interactions within the nucleus would
tend to increase the parton transverse momentum which would be
observable as a larger dijet acoplanarity, {\it i.e.}, the back-to-back
distribution of jets should broaden. However, in $d$ + Au
collisions the extracted values of $\langle
\sin^2(\phi_{jj})\rangle $ are very similar to those observed in
$p+p$ collisions. Indeed, the quadrature difference ($\Delta$)
between $\langle \sin^2(\phi_{jj})\rangle $ in $d$ + Au and $p+p$
is consistent with zero, $\Delta\langle \sin^2(\phi_{jj})\rangle =
+0.004\pm0.008$(stat)$\pm$ 0.000(sys). The extracted $\langle
\sin^2(\phi_{jj})\rangle $ is also measured to be independent of
the nuclear thickness function $T_A(b)$, which is in contrast to
the strong A-dependence of $k_T$ observed at lower beam
energies~\cite{Stewart:1990wa,Corcoran:1990vq,Naples:1994uz}.

We have compared the centrality dependence of the extracted
$\langle \sin^2(\phi_{jj})\rangle$
with the multiple scattering model of Qiu and
Vitev~\cite{Qiu:2004da}. This model reproduces the measured
Cronin effect of single-particle spectra at
RHIC\cite{Adler:2003ii} \cite{Adams:2003im} \cite{Back:2003ns} and
predicts a finite increase of $k_T$ data with nuclear thickness
function. When converted
to a fractional increase, the
prediction is at a level that is within the experimental
uncertainty of the current data. Hence our present data on
$\langle \sin^2(\phi_{jj})\rangle $
are not inconsistent with the level of multiple scattering deduced
from the single-particle Cronin effect.

Taken together, we observe no change in fragmentation and no indication of
the effects of multiple-scattering, {\it i.e.} the
jet-structures are very similar in $d$+Au and $p+p$ collisions
at RHIC energies.
Our measurements also provide a critical
baseline for jet measurements in Au+Au collisions at RHIC.

\section{Acknowledgements}


We thank the staff of the Collider-Accelerator and Physics
Departments at Brookhaven National Laboratory and the staff
of the other PHENIX participating institutions for their
vital contributions.  We acknowledge support from the
Department of Energy, Office of Science, Nuclear Physics
Division, the National Science Foundation, Abilene Christian
University Research Council, Research Foundation of SUNY, and
Dean of the College of Arts and Sciences, Vanderbilt
University (U.S.A), Ministry of Education, Culture, Sports,
Science, and Technology and the Japan Society for the
Promotion of Science (Japan), Conselho Nacional de
Desenvolvimento Cient\'{\i}fico e Tecnol{\'o}gico and Funda\c
c{\~a}o de Amparo {\`a} Pesquisa do Estado de S{\~a}o Paulo
(Brazil), Natural Science Foundation of China (People's
Republic of China), Centre National de la Recherche
Scientifique, Commissariat {\`a} l'{\'E}nergie Atomique,
Institut National de Physique Nucl{\'e}aire et de Physique
des Particules, and Association pour la Recherche et le
D{\'e}veloppement des M{\'e}thodes et Processus Industriels
(France), Ministry of Industry, Science and Tekhnologies,
Bundesministerium f\"ur Bildung und Forschung, Deutscher
Akademischer Austausch Dienst, and Alexander von Humboldt
Stiftung (Germany), Hungarian National Science Fund, OTKA
(Hungary), Department of Atomic Energy and Department of
Science and Technology (India), Israel Science Foundation
(Israel), Korea Research Foundation and Center for High
Energy Physics (Korea), Russian Ministry of Industry, Science
and Tekhnologies, Russian Academy of Science, Russian
Ministry of Atomic Energy (Russia), VR and the Wallenberg
Foundation (Sweden), the U.S. Civilian Research and
Development Foundation for the Independent States of the
Former Soviet Union, the US-Hungarian NSF-OTKA-MTA, and the
US-Israel Binational Science Foundation.

\bibliography{ppg039}


\end{document}